\documentclass[cernpreprint, backref=false, texlive=2017, USenglish, dvipsnames, block=none, texmf]{atlasdoc}
 
\usepackage{atlaspackage}
\usepackage{multirow}
\usepackage{adjustbox}
\usepackage{atlasbiblatex}
 
\usepackage{atlasphysics}
 
\graphicspath{{logos/}{figures/}}
 
\usepackage{ANA-SUSY-2018-41-PAPER-defs}

\AtlasTitle{Search for charginos and neutralinos in final states with two boosted hadronically decaying bosons and missing transverse momentum in $pp$ collisions at $\sqrt{s}=13$~TeV with the ATLAS detector}
\PreprintIdNumber{CERN-EP-2021-127}
 
\AtlasAbstract{
A search for charginos and neutralinos at the Large Hadron Collider using fully hadronic final states and missing transverse momentum is reported.
Pair-produced charginos or neutralinos are explored, each decaying into a high-$p_{\text{T}}$ Standard Model weak boson.
Fully hadronic final states are studied to exploit the advantage of the large branching ratio,
and the efficient rejection of backgrounds by identifying the high-$p_{\text{T}}$ bosons using large-radius jets and jet substructure information.
An integrated luminosity of 139~fb$^{-1}$ of proton--proton collision data collected by the ATLAS detector at a center-of-mass energy of 13~TeV is used.
No significant excess is found beyond the Standard Model expectation.
Exclusion limits at the 95$\%$ confidence level are set on wino or higgsino production with various assumptions about the decay branching ratios and the type of lightest supersymmetric particle.
A wino (higgsino) mass up to 1060 (900)~GeV is excluded when the lightest SUSY particle mass is below 400 (240)~GeV and the mass splitting is larger than 400 (450)~GeV.
The sensitivity to high-mass winos and higgsinos is significantly extended relative to previous LHC searches using other final states.
}
 
\author{The ATLAS Collaboration}
 
\AtlasRefCode{SUSY-2018-41}
\AtlasJournalRef{Phys. Rev. D 104 (2021) 112010}
\AtlasDOI{10.1103/PhysRevD.104.112010}
\hypersetup{pdftitle={ATLAS document},pdfauthor={The ATLAS Collaboration}}
 
\begin{document}
 
\maketitle
 
\tableofcontents

\section{Introduction}
Supersymmetry (SUSY)~\cite{Golfand:1971iw,Volkov:1973ix,Wess:1974tw,Wess:1974jb,Ferrara:1974pu,Salam:1974ig}
is a theoretical framework that extends the Standard Model (SM) by introducing new particles (``superpartners'') that have the same quantum numbers as the SM particles except for their spins.
In the Minimal Supersymmetric Standard Model (MSSM)~\cite{Fayet:1976et,Fayet:1977yc},
the bino (\bino), wino (\wino) and higgsino (\hino) are the superpartners of the U(1)$_\text{Y}$ and SU(2) gauge fields and the Higgs field, respectively.
These are collectively referred to as ``electroweakinos'' and form the chargino $(\Cone,\Ctwo)$ and neutralino $(\None,\Ntwo,\Nthr,\Nfour)$ mass eigenstates through mixing, with the subscripts indicating increasing mass.
The lightest neutralino (\None) is often considered to be the lightest SUSY particle (LSP),
since it is then a viable candidate for weakly interacting massive particle (WIMP) dark matter~\cite{Goldberg:1983nd,Ellis:1983ew} when $R$-parity conservation is assumed~\cite{Farrar:1978xj}.
Depending on the signal model, the superpartner of the graviton (gravitino, $\gravitino$) and the axion (axino, $\axino$) are alternatively considered to be the LSP and dark matter candidate in this search.
 
Electroweakinos with masses of the order of 0.1--1~\TeV\ are motivated by various phenomenological arguments: (1)
the mass of the neutralino LSP dark matter candidate is constrained to be less than a few \TeV\ by the observed relic density~\cite{HISANO200734,PhysRevD.98.035026};
(2) the higgsino mass is also motivated to be of the same order as the $Z$ boson mass by naturalness arguments~\cite{Barbieri:2009ev,Baer:2011ec,Papucci:2011wy,Baer:2012up};
(3) the MSSM parameter space explaining the discrepancy between the measured muon anomalous magnetic moment~\cite{Albahri:2021ixb} and its SM predictions~\cite{Aoyama:2020ynm}
tends to include electroweakinos with masses from $200~\GeV$ to $1~\TeV$~\cite{moroi1996muon,PhysRevLett.86.3484,endo2020muon}.
 
This search targets the pair production of electroweakinos (\chiheavy),
where each of them decays into a lighter one (\chilight)
and an on-shell $W$, $Z$ or SM Higgs boson ($h$).
A mass splitting \dm greater than 400~\GeV\ is considered in the search.
The \chiheavy can be either wino- or higgsino-like, and \chilight can be a bino-, wino-, higgsino-like chargino/neutralino, gravitino or axino as discussed more in Section~\ref{sec:models}.
Cases where both \chiheavy and \chilight are wino- or higgsino-like are not considered as they lead to very small \dm.
The \chilight is either the LSP or an electroweakino nearly degenerate with it, leading to missing transverse momentum (\ptmissvec, with magnitude \met) in the decay signature.\footnote{In the regime of large mass splitting,
the effect of chargino/neutralino mixing is highly suppressed so that the gauge eigenstates (e.g.\ bino) and the mass eigenstates (e.g.\ bino-like $\None$) can be regarded as almost identical.
Therefore, this paper does not make the distinction and ``electroweakinos'' are defined to represent both unless explicitly stated otherwise.}
 
The analysis focuses on the hadronic decay modes of the $W$, $Z$ and $h$ bosons,
namely $W \to q\bar{q}$, $Z \to q\bar{q}/b\bar{b}$ and $h \ra b\bar{b}$,
where $q$ ($\bar{q}$) represents light-flavor (anti)quarks $u,d,s,c$ ($\bar{u},\bar{d},\bar{s},\bar{c}$).\footnote{In this paper the symbols $q$ ($b$) represent both (bottom) quarks and (bottom) antiquarks unless explicitly stated otherwise, since they are not distinguished in the search. In addition, charm quarks and antiquarks are treated as light-flavor quarks and antiquarks in the analysis.}
Two fully hadronic final states are considered: the $qqqq$ final state which involves two $W/Z$ bosons decaying into two light-flavor (anti)quarks,
and the $bbqq$ final state where a $W/Z$ boson decays into two light-flavor (anti)quarks and a $Z/h$ boson decays into $b\bar{b}$,
as illustrated by the diagrams in Figure~\ref{fig:Introduction:SignalDiagram}.
While the ATLAS and CMS experiments have typically searched for pair production of electroweakinos using leptonic decay modes~\cite{SUSY-2018-32,SUSY-2019-09,SUSY-2017-03,SUSY-2018-06,SUSY-2017-01,SUSY-2018-23,SUSY-2019-08,SUSY-2018-02,CMS-SUS-17-010,CMS-SUS-20-001,CMS-SUS-19-012,CMS-SUS-16-039,CMS-SUS-20-003,CMS-SUS-16-045},
the fully hadronic final states are used here to take advantage of the larger $W$, $Z$, and $h$ hadronic branching ratios,
and thereby provide sensitivity to the production of heavier electroweakinos, despite their smaller production cross-sections.
Final states with four $b$-quarks are not considered in this search,\footnote{The contributions from signals with two $h$ bosons are included in the signal models although not targeted.
Likewise, all the decays of $W$, $Z$ and $h$ are taken into account for the signals.}
however they are targeted by dedicated searches~\cite{SUSY-2017-02,CMS-SUS-16-044} complementing the sensitivity.
 
The multijet background is highly suppressed by requiring large \met,
and the dominant backgrounds in the analysis are \Znunu, \Wlv, and diboson production.
These can be effectively suppressed when targeting large mass splittings between the produced electroweakinos and the LSP by selecting high-\pt kinematics and explicitly reconstructing the two boosted SM bosons.
The ``boson tagging'' technique developed by the ATLAS and CMS experiments~\cite{JETM-2018-03,PERF-2017-04,CMS-JME-13-006,CMS-BTV-16-002} is employed,
where a single large-radius jet is used to capture the two collimated energetic jets from each boosted SM boson decay.
Analysis of the jet substructure helps to identify the hadronic decays of $W$, $Z$, and $h$.
The introduction of this technique significantly improves upon the previous fully hadronic final-state analysis in ATLAS targeting the $bbqq$ final state~\cite{SUSY-2017-01},
and establishes the sensitivity of the $qqqq$ final state for the first time at the Large Hadron Collider (LHC),
achieving unprecedented search sensitivity to electroweakinos as heavy as 1~\TeV.
 
This search uses proton--proton collision data collected by the ATLAS detector in the years 2015--2018 at a center-of-mass energy of $\sqrt{s}=13~\TeV$, corresponding to an integrated luminosity of 139~\ifb.
The results are interpreted in terms of various models with different electroweakino types for (\chiheavy, \chilight) and their branching ratio assumptions.
While each model predicts different electroweakino branching ratios into $W$, $Z$ or $h$,
the search combines the dedicated event selections for each decay to achieve only a small model dependency.
 
\begin{figure}[t]
\centering
\subfloat[$qqqq$ final state]{\includegraphics[width=0.3\columnwidth]{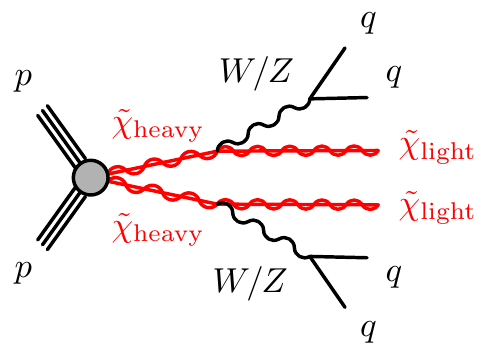}}
\hspace{3em}
\subfloat[$bbqq$ final state]{\includegraphics[width=0.3\columnwidth]{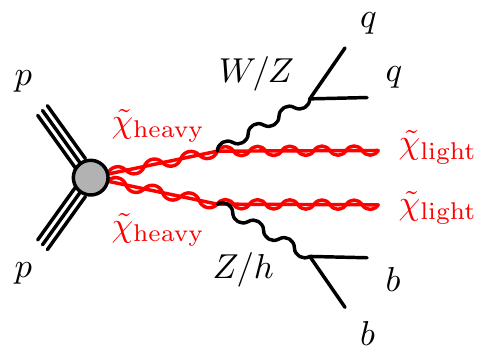}}
\caption{Diagrams for the targeted signatures and final states.
}
\label{fig:Introduction:SignalDiagram}
\end{figure}
 
\section{Target physics scenarios and the signal models}
\label{sec:models}
Three physics scenarios are considered in the analysis:
\begin{itemize}
\item a baseline MSSM scenario where bino, wino and higgsino are considered as \chiheavy or \chilight;
\item a scenario with a gravitino LSP and light higgsinos inspired by the general gauge mediation (GGM)~\cite{Meade:2008wd,Cheung:2007es,Dine:1981gu,AlvarezGaume:1981wy,Nappi:1982hm} models and naturalness;
\item a scenario with an axino LSP assuming the SM extension with a QCD axion and light higgsinos driven by naturalness.
\end{itemize}
The signal models considered in the analysis, derived from each physics scenario, are described in the following subsections.
Each model is labeled as $(A,B)$, where $A$ and $B$ represent the dominant component of \chiheavy and \chilight respectively.
All the SUSY particles other than \chiheavy and \chilight are assumed to be decoupled in mass~\cite{Giudice:2004tc,ArkaniHamed:2004fb}.
The production modes, final states, and the branching ratio assumptions for \chiheavy are summarized in Table~\ref{tab:result:models_vs_SRs}.
 
The wino-like states form a doublet consisting of one chargino and one neutralino, while the higgsino-like
triplet includes an additional neutralino.
The mass degeneracy within these wino/higgsino multiplets is dictated by the extent of mixing with the other electroweakino states, which is characterized by $\dm$.
With $\dm>400~\GeV$, the mass splittings within the wino/higgsino multiplets are typically small ($<10~\GeV$), such that the
decay products are almost never reconstructed in this analysis.
Therefore, these multiplets are treated as approximately degenerate;
the masses of the wino-like chargino and wino-like neutralino are treated as identical;
and the masses of the higgsino-like chargino and the heavier higgsino-like neutralino are set $1~\GeV$ heavier than the lighter higgsino-like neutralino.\footnote{A  mass splitting of $1~\GeV$ is used to avoid potential technical issues when setting it exactly to zero. This choice is arbitrary and has no impact on the analysis.}
Particles originating from decays within the wino doublet or higgsino triplet are ignored so that
only the decays of the form $\chiheavy \to \chilight + W/Z/h$ are taken into account in the signature.
 
\subsection{Bino/wino/higgsino LSP models: $(\wino,\bino), (\wino,\hino), (\hino,\bino), (\hino,\wino)$}
\label{sec:binoWinoHino}
Two of \bino, \wino, and \hino are assumed to be light enough to be produced at the LHC while the others are decoupled.
Four mass hierarchies are experimentally explorable under this regime: $(\wino,\bino), (\wino,\hino), (\hino,\bino), (\hino,\wino)$.
A bino is not considered as \chiheavy here because the production cross-section is negligible when sfermions and the non-SM Higgs bosons are decoupled.
 
The \winoHino, \hinoBino, and \hinoWino hierarchies are typically predicted by the MSSM parameter space that explains the muon $g-2$ anomaly with loop contributions including a wino and/or higgsino.\footnote{In this case, smuons and muon sneutralinos need to be only mildly decoupled in mass.}
The \hinoBino hierarchy is additionally motivated when the mass of a bino-like LSP is half of the $Z$ or $h$ boson mass,
where the LSP dark matter can annihilate via the $Z/h$ resonance (``$Z/h$-funnel'' dark matter).
This is a special case of \textit{well-tempered neutralino dark matter}
where a large higgsino--bino mass splitting is favored in order to realize the observed relic density~\cite{PhysRevD.47.376,PhysRevD.70.075006,han2013dark,calibbi2014lhc,chakraborti2015reduced,Hamaguchi:2015rxa,PhysRevD.93.063525,han2017unblinding,POZZO2019582}.
 
A signal model is defined for each of the four hierarchies with a set of assumptions described below.
The mass spectra, corresponding mass eigenstates, and the decays considered are illustrated for each model in Figure~\ref{fig:mass_spectra}.
A $W$ boson is generated when a chargino decays into a neutralino or vice versa;
a $Z$ or $h$ boson is emitted when a chargino decays into a chargino, or a neutralino decays into a neutralino.
 
The production modes considered in each model are shown in Table~\ref{tab:result:models_vs_SRs}.
Processes involving chargino--chargino or chargino--neutralino pair production are taken into account in the wino production models: \winoBino and \winoHino;
and neutralino--neutralino production is additionally included for the higgsino production models: \hinoBino and \hinoWino.
It should be noted that neutral wino pairs can only be produced via a $t$-channel exchange of squarks, which is prohibited when squarks are assumed to be decoupled.
 
The branching ratios of $\chiheavy$ are treated as follows.
In the \winoBino and \hinoBino models, the produced chargino ($\Cone$) decays with 100\% probability into the bino LSP ($\None$) and a $W$ boson. Meanwhile, the produced neutralino(s) can decay either into $Z$ or into $h$.\footnote{Produced heavier higgsinos are assumed to decay 100\% directly into bino rather than the lighter higgsino, which is reasonable given the much smaller higgsino mass splitting compared with the higgsino--bino mass splitting.} Given the complicated model dependency, the branching ratio of the second lightest neutralino (\Ntwo) is treated as a free parameter,\footnote{Particularly for \winoBino, the branching ratio strongly depends on the higgsino mass
as the decay has to rely on the small higgsino component, even though it is assumed to be decoupled.}
and $\mathcal{B}(\Ntwo \ra Z\None) (= 1-\mathcal{B}(\Ntwo \ra h\None))$ is scanned over 0\%, 25\%, 50\%, 75\% and 100\% in both models.
For the \hinoBino model, the branching ratios of the heavy neutral higgsino ($\Nthr$) are fixed relative to those of the light neutral higgsino ($\Ntwo$) via:
\begin{itemize}
\item $\mathcal{B}(\Ntwo \ra Z\None) + \mathcal{B}(\Nthr \ra Z\None) = 1$,
\item $\mathcal{B}(\Ntwo \ra h\None) + \mathcal{B}(\Nthr \ra h\None) = 1$.
\end{itemize}
These provide a good approximation when $\dm$ is significantly larger than $m_h$~\cite{PhysRevD.88.115010}.
 
In \winoHino and \hinoWino models, however,
the branching ratios of the produced chargino and neutralino(s) are largely dictated by the three MSSM parameters:
the wino mass parameter $M_2$, the higgsino mass parameter $\mu$, and the ratio of vacuum expectation values of the two Higgs fields, $\tan{\beta}$.
For a given set of $(M_2,\mu,\tan{\beta})$ the branching ratios are coherently derived using SOFTSUSY 4.1.7~\cite{Allanach:2001kg,Allanach:2017hcf},
with all the SUSY mass parameters except for $M_2$ and $\mu$ being set as decoupled.
The signal models are tested using various combinations of $(M_2,\mu,\tan{\beta})$
where $M_2 \in [0,1.2]~\TeV$, $\mu \in [-1.2,1.2]~\TeV$, and $\tanb=2,5,10,30$ are considered.
 
The vast majority of the previous electroweakino searches at the LHC have targeted
the simplified \winoBino model, where only a specific production channel and decay mode are considered (detailed in Section~\ref{sec:data-and-mc:signals}).
For \ConeCone production with decays into $WW$, $\mCone<400~\GeV$ is excluded for $\mNone < 200~\GeV$~\cite{SUSY-2018-32,CMS-SUS-17-010}.
For \ConeNtwo production, $\mConeNtwo<640~\GeV$ is excluded for $\mNone < 300~\GeV$ when the \Ntwo is assumed to decay into $Z$ and \None with 100\% probability~\cite{SUSY-2019-09,SUSY-2017-03,SUSY-2018-06,CMS-SUS-20-001,CMS-SUS-19-012}.
Alternatively, $\mConeNtwo<740~\GeV$ is excluded for $\mNone < 250~\GeV$~\cite{SUSY-2017-01,SUSY-2018-23,SUSY-2019-08,CMS-SUS-16-039,CMS-SUS-20-003,CMS-SUS-16-045} when the \Ntwo decays solely into $h$ and \None.
 
\begin{figure}[thb!p]
\centering
\includegraphics[width=0.98\textwidth]{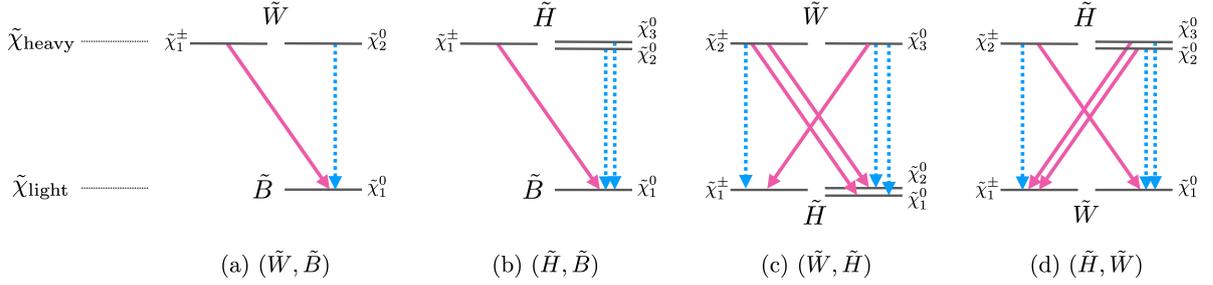}
\caption{The electroweakino mass spectra and corresponding mass eigenstates in each model in the bino/wino/higgsino LSP scenario.
The solid (dashed) arrows represent the decay modes emitting a $W$ ($Z$ or $h$) boson.
A $W$ boson is generated when a chargino decays into a neutralino or vice versa;
a $Z$ or $h$ boson is emitted when a chargino decays into a chargino or a neutralino decays into a neutralino.
}
\label{fig:mass_spectra}
\end{figure}

\subsection{GGM/naturalness-driven gravitino LSP model: $(\hino,\gravitino)$}
General gauge mediation (GGM),
a class of SUSY breaking scenarios characterized by a messenger sector
to which only SM gauge bosons can couple,
typically predicts a nearly massless gravitino (\gravitino) as the LSP.
Motivated also by the naturalness argument,
the production of a relatively light higgsino triplet (\Cone,\Ntwo,\None) decaying into a gravitino LSP
has been explored at ATLAS~\cite{SUSY-2017-02,SUSY-2018-02} and CMS~\cite{CMS-SUS-19-012},
as illustrated in Figure~\ref{fig:Introduction:SignalDiagrams_GGM}.
All of the four production modes are considered together: \ConeCone, \ConeNone, \ConeNtwo, \NoneNtwo.
A moderately small higgsino--gravitino coupling is considered in this analysis,
where the produced heavy higgsinos (\Cone/\Ntwo) always decay into a gravitino via the lightest neutral higgsino (\None),
while the \None still has a short enough lifetime to be regarded as decaying promptly.
In this model, the \None decays into a gravitino and either a $Z$ or $h$ boson,
where the branching ratio $\mathcal{B}(\None \ra Z\gravitino) (= 1-\mathcal{B}(\None \ra h\gravitino))$ is treated as a free parameter and scanned in the limit setting.
The previous searches have excluded masses of \None lighter than 650--880~\GeV\ depending on the branching ratio~\cite{SUSY-2017-02,SUSY-2018-02,CMS-SUS-19-012}.
 
\subsection{Naturalness-driven axino LSP model: $(\hino,\axino)$}
While the QCD Lagrangian generally allows for CP violation,
the absence of such observation
suggests a highly unnatural tuning of the parameters in the theory, referred to as the ``strong CP problem''.
The Peccei--Quinn mechanism aims to solve this problem by introducing an additional chiral U(1) symmetry~\cite{peccei1977cp}.
Through its spontaneous symmetry breaking, the CP-violating term vanishes dynamically,
leaving a Nambu--Goldstone boson known as the axion~\cite{weinberg1978new,wilczek1978problem}.
In the SUSY extension, the axino is introduced as the superpartner of the axion.
A model including a light higgsino triplet (\Cone,\Ntwo,\None) decaying into an axino LSP~\cite{Barenboim_axino_2014}
is proposed in the spirit of pursuing naturalness
as well as axion/axino dark matter~\cite{Covi_2009,Bae_DFSZ_Axino_2012} under $R$-parity conservation.
 
A diagram of the model is shown in Figure~\ref{fig:Introduction:SignalDiagrams_axino}.
Higgsinos are produced by each of the four modes: \ConeCone, \ConeNone, \ConeNtwo, \NoneNtwo.
The produced heavier higgsinos (\Cone/\Ntwo) are assumed to always decay into the axino via the lightest natural higgsino (\None).
This is typically valid when the wino and bino are reasonably decoupled so as to maintain approximate mass degeneracy of the higgsino triplet,
and when the conventionally motivated range of the axion coupling constant is assumed~\cite{Barenboim_axino_2014}.
A prompt \None decay into an axino and a $Z$ or $h$ boson is considered in the search.
The value of the branching ratio $\mathcal{B}(\None \ra Z\axino) (= 1-\mathcal{B}(\None \ra h\axino))$ is scanned over 25\%, 50\%, 75\%, and 100\% in the interpretation.
The model is similar to \hinoGravitino, except that the LSP can be massive.
 
\begin{figure}[thb!p]
\centering
\subfloat[\hinoGravitino]{\includegraphics[width=0.24\textwidth]{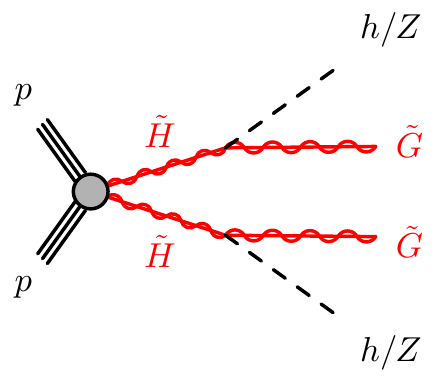} \label{fig:Introduction:SignalDiagrams_GGM}}
\hspace{3em}
\subfloat[\hinoAxino]{\includegraphics[width=0.24\textwidth]{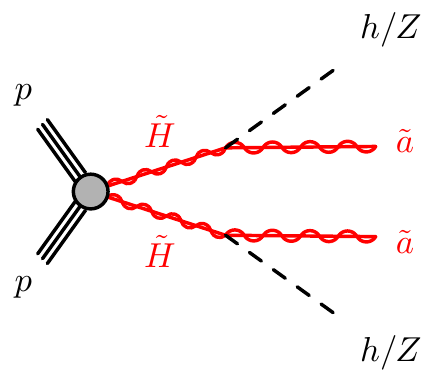} \label{fig:Introduction:SignalDiagrams_axino}}
\caption{Diagrams of signals considered in the (a) \hinoGravitino model and (b) \hinoAxino model.
In the \hinoGravitino and \hinoAxino models, the higgsino triplets ($\Cone,\Ntwo,\None$) are collectively represented by \hino.
}
\end{figure}
 
\begin{table}[thb!p]
\begin{center}
\caption{Summary of the production modes, final states, and signal regions (SRs) used for the hypothesis tests, and the branching ratio assumptions for the signal models targeted in the search.
The notation and definition of the SRs are described in Section~\ref{sec:srs}.
The \winoBino and \hinoGravitino models are used to optimize the selection, and the rest are considered in the interpretation.
The \winoBino simplified models (\winoBinoSim) discussed in Section~\ref{sec:data-and-mc:signals} are also interpreted in order to allow comparisons with the ATLAS electroweakino search results~\cite{SUSY-2018-32,SUSY-2019-09,SUSY-2017-03,SUSY-2016-24,SUSY-2019-08}.
}
\maxsizebox*{\textwidth}{\textheight}{
\begin{tabular}{l l l l l}
\toprule
\textbf{Model} & \textbf{Production} & \textbf{Final states} & \textbf{SRs simultaneously fitted} & \textbf{Branching ratio} \\
\midrule
\multirow{2}{*}{(\wino,\bino)}       & \multirow{2}{*}{\ConeCone,~\ConeNtwo}    & \multirow{2}{*}{$WW,WZ,Wh$}    & \multirow{2}{*}{4Q-VV, 2B2Q-WZ, 2B2Q-Wh}   & $\mathcal{B}(\Cone \ra W \None)=1$ \\
&                                          &                                &                                         & $\mathcal{B}(\Ntwo \ra Z\None)$ scanned \\
\midrule
\multirow{3}{*}{(\hino,\bino)}       & \multirow{3}{*}{\shortstack[l]{\ConeCone,~\ConeNtwo, \\ \ConeNthr,~\NtwoNthr}} & \multirow{3}{*}{\shortstack[l]{$WW,WZ,Wh,$ \\ $ZZ,Zh,hh$}} &
\multirow{3}{*}{4Q-VV, 2B2Q-VZ, 2B2Q-Vh} & $\mathcal{B}(\Cone \ra W \None)=1$ \\
&                      &                      &                                       & $\mathcal{B}(\Ntwo \ra Z\None)$ scanned   \\
&                      &                      &                                       & $\mathcal{B}(\Nthr \ra Z \None) = 1-\mathcal{B}(\Ntwo \ra Z\None)$ \\
\midrule
\multirow{2}{*}{(\wino,\hino)}       & \multirow{2}{*}{\CtwoCtwo,~\CtwoNthr}    & $WW,WZ,Wh,$   & \multirow{2}{*}{4Q-VV, 2B2Q-VZ, 2B2Q-Vh} & \multirow{2}{*}{Determined from $(M_2,\mu,\tan{\beta})$}  \\
&                                          & $ZZ,Zh,hh$    &                                       & \\
\midrule
\multirow{2}{*}{(\hino,\wino)}       & \CtwoCtwo,~\CtwoNtwo,                    & $WW,WZ,Wh,$   & \multirow{2}{*}{4Q-VV, 2B2Q-VZ, 2B2Q-Vh} & \multirow{2}{*}{Determined from $(M_2,\mu,\tan{\beta})$} \\
& \CtwoNthr,~\NtwoNthr                     & $ZZ,Zh,hh$    &                                       & \\
\midrule
\multirow{2}{*}{(\hino,\gravitino)}  & \ConeCone,~\ConeNone,                    & \multirow{2}{*}{$ZZ,Zh,hh$}    & \multirow{2}{*}{4Q-ZZ, 2B2Q-ZZ, 2B2Q-Zh} & \multirow{2}{*}{$\mathcal{B}(\None \ra Z \gravitino)$ scanned} \\
& \ConeNtwo,~\NoneNtwo                     &               &                                       & \\
\midrule
\multirow{2}{*}{(\hino,\axino)}      & \ConeCone,~\ConeNone,                    & \multirow{2}{*}{$ZZ,Zh,hh$}    & \multirow{2}{*}{4Q-ZZ, 2B2Q-ZZ, 2B2Q-Zh} & \multirow{2}{*}{$\mathcal{B}(\None \ra Z \axino)$ scanned} \\
& \ConeNtwo,~\NoneNtwo                     &               &                                       & \\  \midrule
\multicolumn{5}{l}{\textbf{\winoBino simplified models: \winoBinoSim}} \vspace{2mm} \\
\SMWW   & \ConeCone & $WW$     & 4Q-WW          & $\mathcal{B}(\Cone \ra W \None)=1$  \\ \midrule
\SMWZ   & \ConeNtwo & $WZ$     & 4Q-WZ, 2B2Q-WZ & $\mathcal{B}(\Cone \ra W \None)=\mathcal{B}(\Ntwo \ra Z \None)=1$  \\ \midrule
\SMWh   & \ConeNtwo & $Wh$     & 2B2Q-Wh        & $\mathcal{B}(\Cone \ra W \None)=\mathcal{B}(\Ntwo \ra h\None)=1$  \\
\bottomrule
\end{tabular}
}
\label{tab:result:models_vs_SRs}
\end{center}
\end{table}
 
 
\section{ATLAS detector}
\newcommand{\AtlasCoordFootnote}{
ATLAS uses a right-handed coordinate system with its origin at the nominal interaction point (IP) in the center of the detector and the $z$-axis along the beam pipe. The $x$-axis points from the IP to the center of the LHC ring, and the $y$-axis points upwards. Cylindrical coordinates $(r,\phi)$ are used in the transverse plane, $\phi$ being the azimuthal angle around the $z$-axis. The pseudorapidity is defined in terms of the polar angle $\theta$ as $\eta = -\ln \tan(\theta/2)$. Angular distance is measured in units of $\Delta R \equiv \sqrt{(\Delta\eta)^{2} + (\Delta\phi)^{2}}$.}
 
The ATLAS experiment~\cite{PERF-2007-01,PIX-2018-001} is a multipurpose detector with a forward--backward symmetric cylindrical geometry and nearly $4\pi$ coverage in solid angle.\footnote{\AtlasCoordFootnote}
It consists of an inner tracking detector surrounded by a thin superconducting solenoid providing a \SI{2}{\tesla} axial magnetic field, electromagnetic and hadron calorimeters, and a muon spectrometer.
 
The inner detector~(ID) consists of pixel and microstrip silicon detectors covering the pseudorapidity range $|\eta| < 2.5$ and a transition radiation tracker covering $|\eta| < 2.0$.
Outside the ID, a lead/liquid-argon (LAr) electromagnetic calorimeter (ECAL) and
a steel/scintillator-tile hadronic calorimeter cover the $|\eta| < 3.2$ and $|\eta| < 1.7$ ranges, respectively.
In the forward regions, a copper/LAr endcap calorimeter extends the coverage of hadronic measurements to $1.7 < |\eta| < 3.2$, while copper/LAr and tungsten/LAr forward calorimeters are employed for
electromagnetic and hadronic measurements in the  $3.1 < |\eta| < 4.9$ region.
The muon spectrometer (MS) surrounds the calorimeters and comprises three layers of trigger and
high-precision tracking chambers spanning $|\eta| < 2.4$ and $|\eta| < 2.7$, respectively.
A magnetic field is provided by a system of three superconducting air-core toroidal magnets with eight coils each.
The field integral of the toroids ranges between \num{2.0} and \SI{6.0}{\tesla\metre} across most of the detector.
 
Events of interest are selected and collected by the ATLAS trigger system~\cite{TRIG-2016-01},
consisting of a hardware-based first-level trigger (L1) and a software-based high-level trigger (HLT).
The L1 trigger is designed to accept events from the \SI{40}{\MHz} bunch crossings at a rate below \SI{100}{\kHz},
and the HLT reduces this to about \SI{1}{\kHz}, the rate at which events are recorded to disk.
An extensive software suite~\cite{ATL-SOFT-PUB-2021-001} is used for real and simulated data reconstruction
and analysis, for operation and in the trigger and data acquisition systems of the experiment.
 
 
\section{Data and Monte Carlo simulation}
\subsection{Data sample} \label{sec:data-and-mc:data}
The data events used in the analysis are from proton--proton collisions at $\sqrt{s}=13~\TeV$, recorded during stable beam conditions at the LHC during 2015--2018.
The collected dataset corresponds to an integrated luminosity of 139~\ifb\ after applying the data quality criteria~\cite{DAPR-2018-01}.
The primary dataset was collected by triggers targeting large missing transverse momentum~\cite{TRIG-2019-01}.
Events were accepted when the \met calculated at trigger level was greater than 70--110~\GeV, with the threshold rising with the increased instantaneous luminosity during the data-taking period.
The efficiency reaches approximately 100\% for events reconstructed offline with $\met>200~\GeV$, which is generally required in the search.
Auxiliary data samples used to validate the background estimation were selected using triggers requiring at least one isolated electron, muon or photon~\cite{TRIG-2018-05,TRIG-2018-01}.
The thresholds were $\pt=24~(26)~\GeV$ for electrons,  $\pt=20~(26)~\GeV$ for muons, and $\pt=120~(140)~\GeV$ for photons in data taken in 2015 (2016--2018).
 
\subsection{Monte Carlo simulation}
Monte Carlo~(MC) simulations are used to estimate the SM backgrounds and the signals contributing to the analysis regions.
All of the generated events were propagated through the ATLAS detector simulation~\cite{SOFT-2010-01} based on \GEANT~\cite{Agostinelli:2002hh}.
Multiple proton--proton collisions in the same and neighboring bunch crossings~(pileup) were modeled by overlaying the hard-scatter events with minimum-bias events simulated by \PYTHIA[8.186]~\cite{Sjostrand:2007gs} with a set of tuned parameters called the A3 tune~\cite{ATL-PHYS-PUB-2016-017} and \NNPDF[2.3lo] parton distribution function~(PDF) set~\cite{Ball:2012cx}.
 
The simulated events are processed with the same trigger and reconstruction algorithms as the data.
The lepton and photon trigger efficiencies in the simulation are corrected to match those in data using scale factors that depend on the \pt\ and $\eta$ of the leptons and photons, as derived from control samples~\cite{TRIG-2018-05,TRIG-2018-01}.
 
\subsubsection{Standard Model backgrounds}
Events with a leptonically decaying $W$ or $Z$ boson associated with hadron jets 
were simulated using the \SHERPA[2.2.1]~\cite{Bothmann:2019yzt} generator.
The matrix elements were calculated for up to two partons at next-to-leading-order (NLO) accuracy
and up to four jets at leading-order (LO) accuracy using the Comix~\cite{Gleisberg:2008fv} and \OPENLOOPS~\cite{Cascioli:2011va,Denner:2016kdg} generators.
The NLO matrix elements for a given jet multiplicity were matched to the parton shower
using a color-exact variant of the MC@NLO algorithm~\cite{Hoeche:2011fd}.
Different jet multiplicities were then merged
into an inclusive sample using an improved Catani-Krauss-Kuhn-Webber (CKKW) matching
procedure~\cite{Catani:2001cc,Hoeche:2009rj} which is extended to NLO
accuracy using the \MEPSatNLO prescription~\cite{Hoeche:2012yf}.
The \NNPDF[3.0nnlo] PDF sets~\cite{Ball:2014uwa} were used.
Prompt single-photon production, denoted by \gammajets, was simulated using the same configuration except that the generator version is \SHERPA[2.2.2].
The photons must be isolated according to a smooth-cone isolation criterion~\cite{Frixione:1998jh}.

Samples of \ttbar\ and single-top-quark ($t$+$X$) events were generated with \POWHEGBOX~\cite{Frixione:2007nw,Nason:2004rx,Frixione:2007vw,Alioli:2010xd}~v2 at NLO with the \NNPDF[3.0nlo] PDF sets.
The top quark mass was set to 172.5~\GeV.
For the \ttbar generation, the $h_{\mathrm{damp}}$ parameter, which controls the \pt of the first additional emission beyond the Born configuration in \textsc{Powheg},
was set to 1.5 times the top quark mass.
This is to regulate the high-\pt\ emission recoiling against the \ttbar system so as to reproduce the data~\cite{ATL-PHYS-PUB-2016-020}.
The parton shower, fragmentation, and underlying event were simulated using \PYTHIA[8.230] with the \NNPDF[2.3lo] PDF set and the A14 tune~\cite{ATL-PHYS-PUB-2014-021}.
The decays of bottom and charm hadrons were performed by \EVTGEN[1.6.0]~\cite{Lange:2001uf}.
The diagram removal scheme~\cite{Frixione:2008yi} was employed to
account for the interference between \ttbar\ and single-top $Wt$ production.
 
Events containing \ttbar\ with additional heavy particles, such as $\ttbar$+$W/Z/h$, $\ttbar$+$WW$, $t\bar{t}t$ and $t\bar{t}t\bar{t}$,
are collectively referred to as \ttbarX in this paper.
The $\ttbar{+}W/Z$ events were modeled by \textsc{MadGraph5\_aMC@NLO}\,2.3.3~\cite{Alwall:2014hca} at NLO accuracy with the \NNPDF[3.0nlo] PDF~\cite{Ball:2014uwa} set,
while $\ttbar{+}h$ events were modeled by \POWHEGBOX[2.2].
Both were interfaced with \PYTHIA[8.230] for parton showering.
The $\ttbar{+}h$ events were generated using the same setup as for \ttbar described above, with the $h$ mass set to 125~\GeV.
The $\ttbar{+}WW$, $t\bar{t}t$, and $t\bar{t}t\bar{t}$ events were generated with \textsc{MadGraph5\_aMC@NLO}\,2.2.2,
interfaced to \PYTHIA[8.210] using the A14 tune and the \NNPDF[2.3lo] PDF set.
The decays of bottom and charm hadrons were simulated using \EVTGEN[1.2.0].
 
Diboson production, including $WW$, $WZ$ and $ZZ$, is collectively denoted here by $VV$.
Events with semileptonic $VV$ decays ($\ell\nu qq$, $\ell\ell qq$ and $\nu\nu qq$) were simulated with \SHERPA[2.2.1]
using matrix elements with up to one additional parton at NLO accuracy in QCD and up to three additional parton emissions at LO accuracy.
Events with fully leptonic $VV$ decays were simulated with \SHERPA[2.2.1] ($ZZ \to \nu\nu\nu\nu$) or \SHERPA[2.2.2] ($\ell\ell\ell\ell$, $\ell\ell\ell\nu$, $\ell\ell\nu\nu$ and $\ell\nu\nu\nu$), where all the processes at orders of $(\alpha_{\textsc{ew}}^{4}, \alphas^{2})$ and $(\alpha_{\textsc{ew}}^{6}, \alphas^{0})$ are taken into account, including the off-shell contributions and those mediated by Higgs bosons.
The \NNPDF[3.0nnlo] PDF set is used.
 
The production of $Wh$ and $Zh$ (collectively denoted by $Vh$) was modeled by \POWHEGBOX[2.2] interfaced with \PYTHIA[8.186].
The $h$ mass was set to 125~\GeV.
The \NNPDF[3] PDF set and the AZNLO tune were used.
 
Triboson production, including $WWW$, $WWZ$, $WZZ$ and $ZZZ$, are collectively denoted by $VVV$.
The $V(\ra \ell\ell/\ell\nu/\nu\nu)V(\ra qq)V(\ra qq)$ processes were generated using \textsc{MadGraph5\_aMC@NLO}\,2.6.6 at LO,
interfaced to \PYTHIA[8.243] for modeling of parton showers and hadronization using the A14 tune.
The \NNPDF[3.0lo] PDF set was used.
 
Theoretical cross-sections are used to normalize the generated background samples.
The \ttbar\ sample is normalized to the cross-section predicted at NNLO
in QCD, including the resummation of next-to-next-to-leading-logarithmic (NNLL) soft-gluon terms calculated using
\TOPpp[2.0]~\cite{Beneke:2011mq,Cacciari:2011hy,Baernreuther:2012ws,Czakon:2012zr,Czakon:2012pz,Czakon:2013goa,Czakon:2011xx}.
The cross-sections of single-top-quark $t$- and $s$-channel production are calculated using the \HATHOR[2.1] program~\cite{Kant:2014oha, Aliev:2010zk}, while the $Wt$-channel calculation followed the prescriptions from Refs.~\cite{Kidonakis:2011wy, Kidonakis:2010ux}.
The $VV$ and $\ttbar{+}W/Z$ samples are normalized to the cross-sections calculated at NLO~\cite{Lazopoulos:2008de,Campbell:2012dh,Campbell:2011bn}.
The cross-sections for the \Vjets and \gammajets samples are calculated at NNLO~\cite{Anastasiou:2003ds}.
The remaining samples, including the production of $Vh$, $V\gamma$, $VVV$, $\ttbar{+}WW$, $tWZ$, $\ttbar{+}h$, $t\bar{t}t$, and $t\bar{t}t\bar{t}$, are normalized to the cross-sections calculated at LO by the generators.
 
\subsubsection{Signals}  \label{sec:data-and-mc:signals}
The \winoBino, \hinoBino, \winoHino, \hinoWino, and \hinoAxino models
discussed in Section~\ref{sec:models} were simulated
by combining the ``simplified model'' signals in which a fixed production mode and decay chain are considered.
Six sets of simplified model samples, derived as variants of the \winoBino and \hinoBino models,
were generated to cover the different final states:
\begin{itemize}
\item wino pair production, with each decaying into a bino LSP (referred to as \winoBino simplified models, or \winoBinoSim):
\begin{align}
\text{\textbf{\SMWW:}} \quad\quad  & \Cone\Cone \to WW~\None\None~~~(\mathcal{B}(\Cone \ra W\None)=100\%)  \nonumber \\
\text{\textbf{\SMWZ:}} \quad\quad  & \Cone\Ntwo \to WZ~\None\None~~~(\mathcal{B}(\Cone \ra W\None)=\mathcal{B}(\Ntwo \ra Z\None)=100\%)  \nonumber \\
\text{\textbf{\SMWh:}} \quad\quad  & \Cone\Ntwo \to Wh~\None\None~~~~(\mathcal{B}(\Cone \ra W\None)=\mathcal{B}(\Ntwo \ra h\None)=100\%),  \nonumber
\end{align}
\item neutral higgsino pair production, with each decaying into a bino LSP (denoted by \hinoBinoSim):
\begin{align}
\text{\textbf{\SMZZ:}} \quad\quad  & \Ntwo\Nthr \to ZZ~\None\None~~~(\mathcal{B}(\Ntwo \ra Z\None)=\mathcal{B}(\Nthr \ra Z\None)=100\%) \nonumber \\
\text{\textbf{\SMZh:}} \quad\quad  & \Ntwo\Nthr \to Zh~\None\None~~~(\mathcal{B}(\Ntwo \ra Z\None)=\mathcal{B}(\Nthr \ra h\None)=100\%) \nonumber \\
\text{\textbf{\SMhh:}} \quad\quad  & \Ntwo\Nthr \to hh~\None\None~~~~(\mathcal{B}(\Ntwo \ra h\None)=\mathcal{B}(\Nthr \ra h\None)=100\%). \nonumber
\end{align}
\end{itemize}
The symbols C1, N2 and N3 in the model names represent \Cone, \Ntwo and \Nthr respectively.
These samples are used to model event kinematics for each final state;
the \winoBinoSim samples are used for events containing $WW,WZ$ or $Wh$,
while \hinoBinoSim samples are employed to model events with $ZZ,Zh$ or $hh$.
The events are reweighted to account for the production cross-section
and the \chiheavy branching ratios considered in the model.
The underlying assumption for the method is that the event kinematics depend only on \mlight\ and \mheavy,
and not on the production mode, type of LSP, or other MSSM variables such as $\tan{\beta}$.
This is validated in the phase space considered in the analysis by using generator-level samples.
 
The \winoBinoSim models are also considered in the limit interpretation in order to allow comparisons with previous ATLAS searches, as these models were the most commonly studied~\cite{SUSY-2018-32,SUSY-2017-03,SUSY-2018-06,SUSY-2017-01,SUSY-2018-23,SUSY-2019-08,CMS-SUS-17-010,CMS-SUS-20-001,CMS-SUS-19-012,CMS-SUS-16-039,CMS-SUS-20-003,CMS-SUS-16-045}.
For the \hinoGravitino model, a dedicated sample was generated with
all of the production modes (\ConeCone, \ConeNone, \ConeNtwo, \NoneNtwo) included.
 
The \winoBinoSim and \hinoGravitino signal samples were simulated using LO matrix elements with up to
two extra partons in \textsc{MadGraph5\_aMC@NLO}\,2.6.2.
The events were subsequently interfaced to \PYTHIA[8.230]~\cite{Sjostrand:2014zea}
together with the A14 tune and \NNPDF[2.3lo] PDF set for simulation of parton showering and hadronization.
For \hinoBinoSim, \textsc{MadGraph5\_aMC@NLO}\,2.6.7 and \PYTHIA[8.244] were used.
The mass of the SM Higgs boson was set to 125~\GeV.
 
Decays of the produced electroweakinos are simulated using \PYTHIA.
As the performance of the boosted $W/Z$ boson tagging is known to be sensitive to the boson polarization~\cite{JETM-2018-03},
$W/Z$ bosons from the electroweakino decays were carefully modeled either by using \MADSPIN[2.7.3]~\cite{Artoisenet:2012st,Frixione:2007zp}
or by reweighting the helicity angle distribution such that the overall cross-section remains unchanged.
The $W/Z$ bosons are typically longitudinally polarized when $\dm>400~\GeV$ is considered.
The decays of bottom and charm hadrons were performed by \EVTGEN[1.2.0].

The signal cross-sections are computed at NLO in the strong coupling constant,
adding the resummation of soft-gluon emission at next-to-leading-logarithm (NLL) accuracy~\cite{Beenakker:1999xh,Debove:2010kf,Fuks:2012qx,Fuks:2013vua,Fiaschi:2018hgm,Borschensky:2014cia}.
The \PDFforLHC[15\_mc] PDF set is used following the recommendations in Ref.~\cite{Butterworth:2015oua}.
Assuming a mass of 800~\GeV, the calculated cross-section for wino-like (higgsino-like) chargino pair production
is 2.21~(0.63)~fb, and 4.76~(1.12)~fb for wino-like (higgsino-like) chargino--neutralino production.
For neutral higgsino pair production, the corresponding cross-section is 0.59~fb.

 
\section{Event reconstruction}
\label{sec:object}
The primary reconstructed objects used in the analysis are large-radius (large-$R$) jets, denoted by $J$.
These are reconstructed from locally calibrated topo-clusters~\cite{PERF-2014-07} using the anti-$k_t$ algorithm~\cite{Cacciari:2008gp} implemented in the \textsc{FastJet} package~\cite{Cacciari:2011ma} with a radius parameter $R=1.0$.
A trimming algorithm~\cite{Krohn:2009th} is applied to mitigate the effects of pileup and soft radiation.
The constituents of each jet are reclustered with the $k_t$ algorithm~\cite{Ellis:1993tq} into $R = 0.2$ subjets,
and the subjets are removed if $p_{\mathrm{T}}^{\mathrm{subjet}} / \ptJ < 0.05$,
where $p_{\mathrm{T}}^{\mathrm{subjet}}$ and \ptJ\ are the transverse momenta of the subjet and the large-$R$ jet, respectively.
The jet mass, \mJ, is calculated according to the combined mass prescription~\cite{ATLAS-CONF-2016-035} in order to achieve the best mass resolution,
which is given by the weighted sum of masses computed using only the calorimeter information and with tracking information included.
The \pt\ and mass scales are calibrated using simulation, followed by an in situ calibration~\cite{JETM-2018-02} to correct for residual differences between data and MC simulation.
Large-$R$ jets used in the analysis are selected with $\pt>200$~\GeV, $|\eta|<2.0$, and $\mJ > 40~\GeV$.
 
Track-jets are used to identify large-$R$-jet subjets that contain $b$-hadrons.
They are reconstructed from ID tracks by using the anti-$k_t$ algorithm with a sliding radius parameter $R=30~\GeV/\pt$ truncated at 0.02 and 0.4~\cite{ATL-PHYS-PUB-2017-010}.
The MV2c10 $b$-tagging algorithm~\cite{FTAG-2018-01} is applied to track-jets satisfying $\pt>20$~\GeV\ and $|\eta|<2.5$.
This algorithm is a multivariate discriminator that utilizes track impact parameters, the presence of secondary vertices,
and the trajectories of $b$- and $c$-hadrons inside the jet.
A working point is chosen such that $b$-jets from simulated \ttbar events are identified
with 85\% efficiency, with rejection factors of 3 against $c$-jets and 33 against jets originating from other light-flavor quarks or gluons~\cite{FTAG-2018-01}.
Efficiency correction factors are applied to simulated samples to
account for the efficiency difference observed between MC and data events in dedicated measurement regions in which
\ttbar samples~\cite{FTAG-2018-01,ATLAS-CONF-2018-045} are used for $b$-jets and mistagged $c$-jets, and \Zjets samples are used for the light-flavor mistagged jets~\cite{ATLAS-CONF-2018-006}.
 
The $b$-jet multiplicity of each large-$R$ jet is defined by the number of $b$-tagged track-jets it contains.
A $b$-tagged track-jet is contained in a large-$R$ jet if the two jet axes have an angular separation $\Delta R<1.0$.
 
Two types of boosted boson tagging are employed for the preselected large-$R$ jets to identify the SM boson decays:
$\Wqq(\Zqq)$-tagging targeting $W(Z) \to qq$, and $\Zbb(\hbb)$-tagging targeting $Z(h) \to bb$.
The $\Wqq(\Zqq)$-tagging utilizes cuts on \mJ, the energy correlation function $D_2$, and the track multiplicity \ntrk~\cite{JETM-2018-03,ATL-PHYS-PUB-2020-017}.
In order to maintain orthogonality with the $\Zbb(\hbb)$-tagging,
the $b$-jet multiplicity of the large-$R$ jet is required to be less than two.
The $D_2$ variable is defined as a ratio of three-point to two-point energy correlation functions~\cite{Larkoski:2013eya,Larkoski:2014gra} based on the energies and pairwise angular separations of particles within a jet, and
\ntrk is the number of tracks matched to the large-$R$ jet by ghost association~\cite{Cacciari:2008gn} before trimming is applied.
While the upper bound on \ntrk is fixed, the cut values applied to \mJ and $D_2$ are shifted smoothly as a function of \pt to maximize the rejection for typical single-parton-initiated jets,
maintaining a constant efficiency for signal jets that contain the decay products of $W(Z) \to qq$.
The selections are optimized separately for targeting $W$ and $Z$,
and \Vqq-tagging is defined by a selection satisfying either the $\Wqq$- or $\Zqq$-tagging.
The cut values applied to \mJ and $D_2$ are the same as for the ``50\% efficiency $W/Z$ tagger'' in Ref.~\cite{ATL-PHYS-PUB-2020-017};
however, a loosened \ntrk cut, from $\le 26$ to $\le 32~(34)$ for the $\Wqq(\Zqq$)-tagging,
is applied to achieve the optimum sensitivity for the analysis and better modeling in MC simulation.
The efficiency correction factors are rederived according to this refinement,
using a methodology similar to that described in Ref.~\cite{ATL-PHYS-PUB-2020-017}.
A correction factor of 0.85--1.05 is typically obtained.
The performance of the \Wqq- and \Zqq-tagging is summarized in Figures~\ref{fig:boson_tagger_WZqq_eff} and~\ref{fig:boson_tagger_WZqq_rej}.
For a sample of preselected large-$R$ jets ($\pt>200~\GeV, |\eta|<2.0, \mJ > 40~\GeV$),
the tagging efficiency is about 50\% for the signal jets originating from electroweakino decays,\footnote{The corresponding efficiency for the nominal working point (``50\% efficiency $W/Z$ tagger'' in Ref.~\cite{ATL-PHYS-PUB-2020-017}) is 35\%--40\% for this particular set of preselected jets.}
while the background rejection is typically about 10~(40) at $\pt=200~(1000)~\GeV$ per jet in \Znunu events.

The $\Zbb(\hbb)$-tagging is applied to large-$R$ jets containing exactly two $b$-jets (denoted by \Jbb)
by applying a jet mass window cut that selects the peak consistent with $Z(h)$ bosons.
The jet mass is corrected by adding the momentum of the highest-\pt muon identified inside the large-$R$ jet in order to improve the resolution of the mass peak.
The $\Zbb(\hbb)$-tagging requires the jet mass to satisfy $70~(100)~\GeV < \mJ < 100~(135)~\GeV$.
The performance of the \Zbb- and \hbb-tagging is summarized in Figures~\ref{fig:boson_tagger_Zhbb_eff} and~\ref{fig:boson_tagger_Zhbb_rej}.
 
\begin{figure}[thb!p]
\centering
\subfloat[\Wqq/\Zqq-tagging efficiency]{\includegraphics[width=0.485\textwidth]{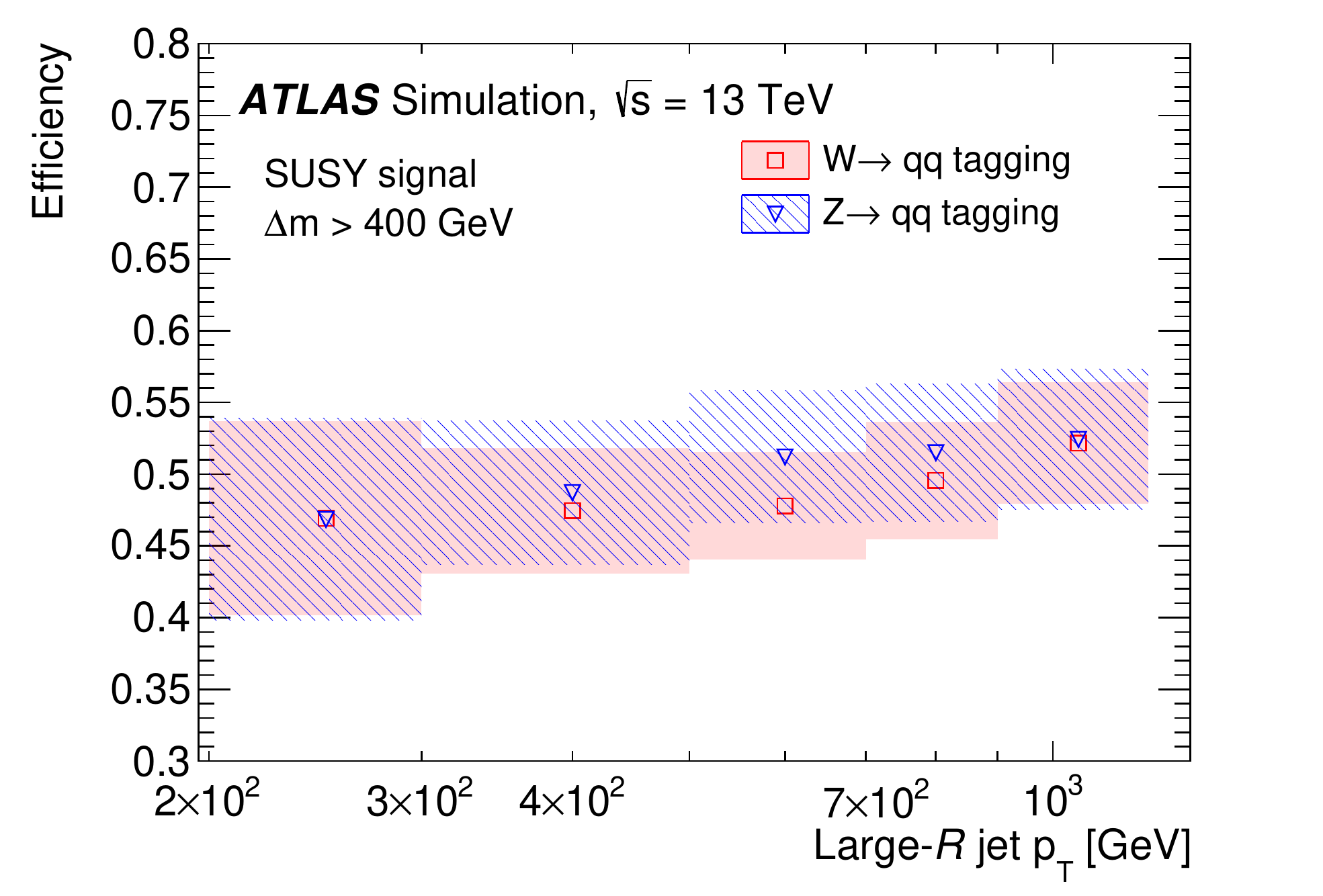} \label{fig:boson_tagger_WZqq_eff}}
\subfloat[\Wqq/\Zqq-tagging rejection]{\includegraphics[width=0.485\textwidth]{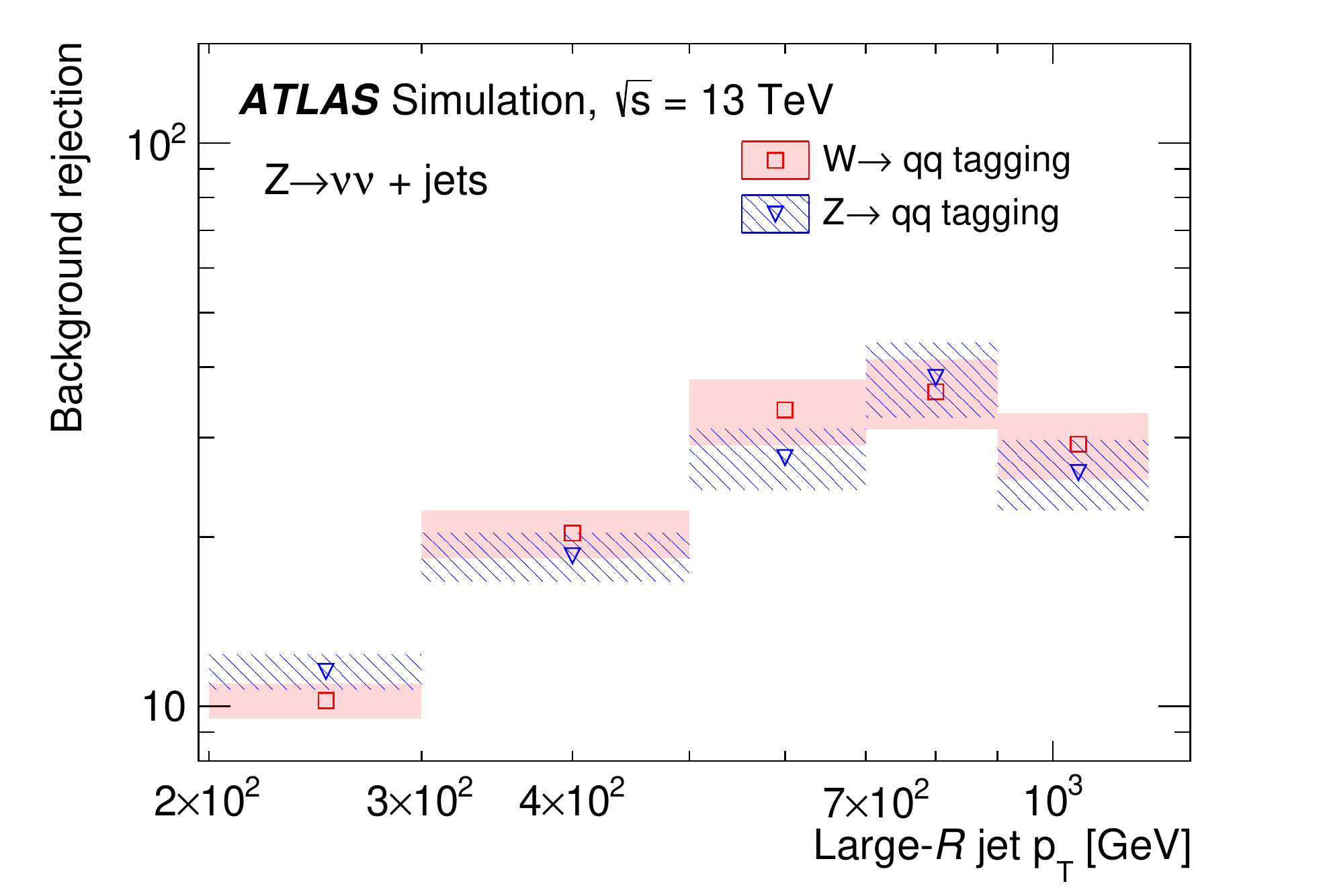} \label{fig:boson_tagger_WZqq_rej}} \\
\subfloat[\Zbb/\hbb-tagging efficiency]{\includegraphics[width=0.485\textwidth]{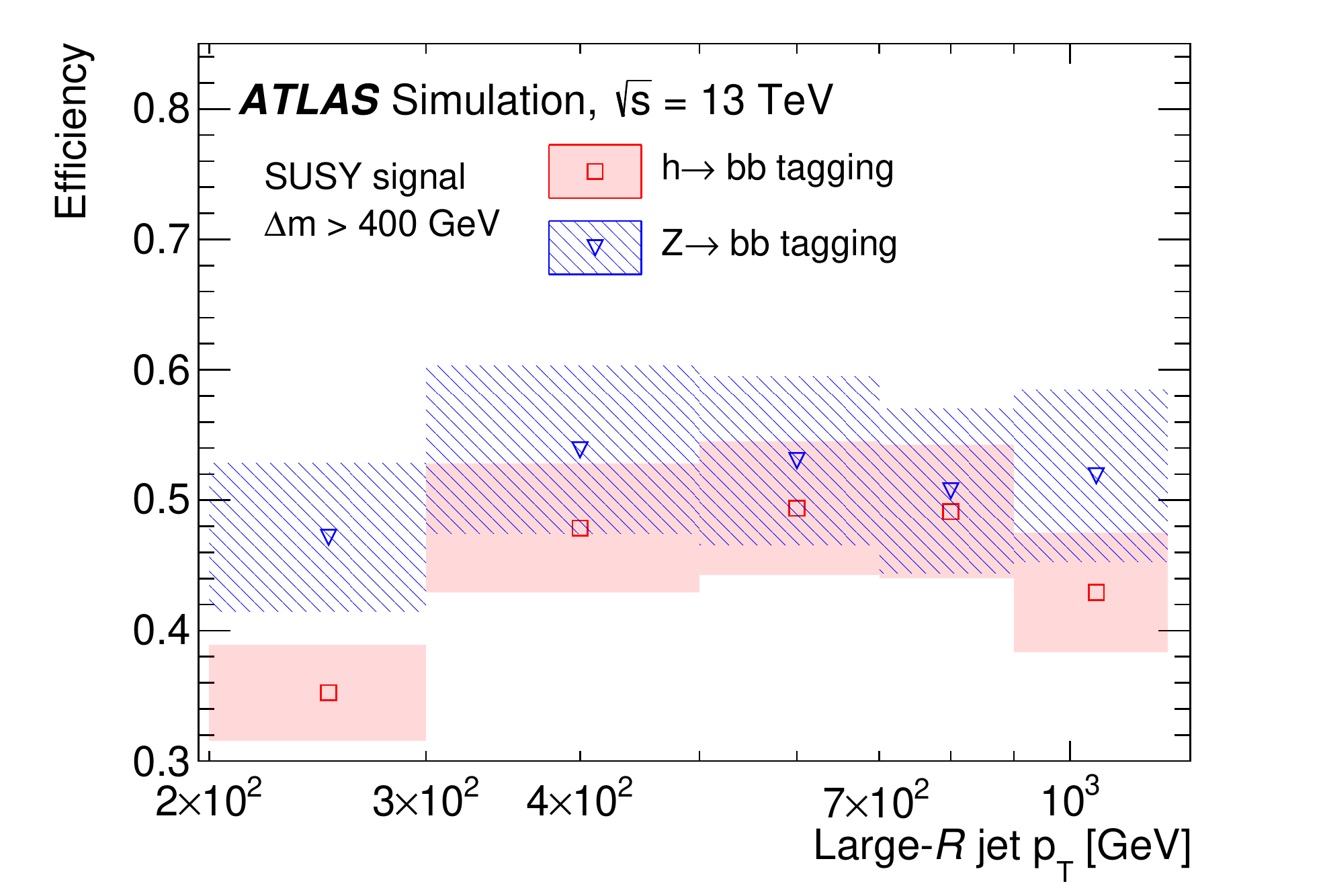} \label{fig:boson_tagger_Zhbb_eff}}
\subfloat[\Zbb/\hbb-tagging rejection]{\includegraphics[width=0.485\textwidth]{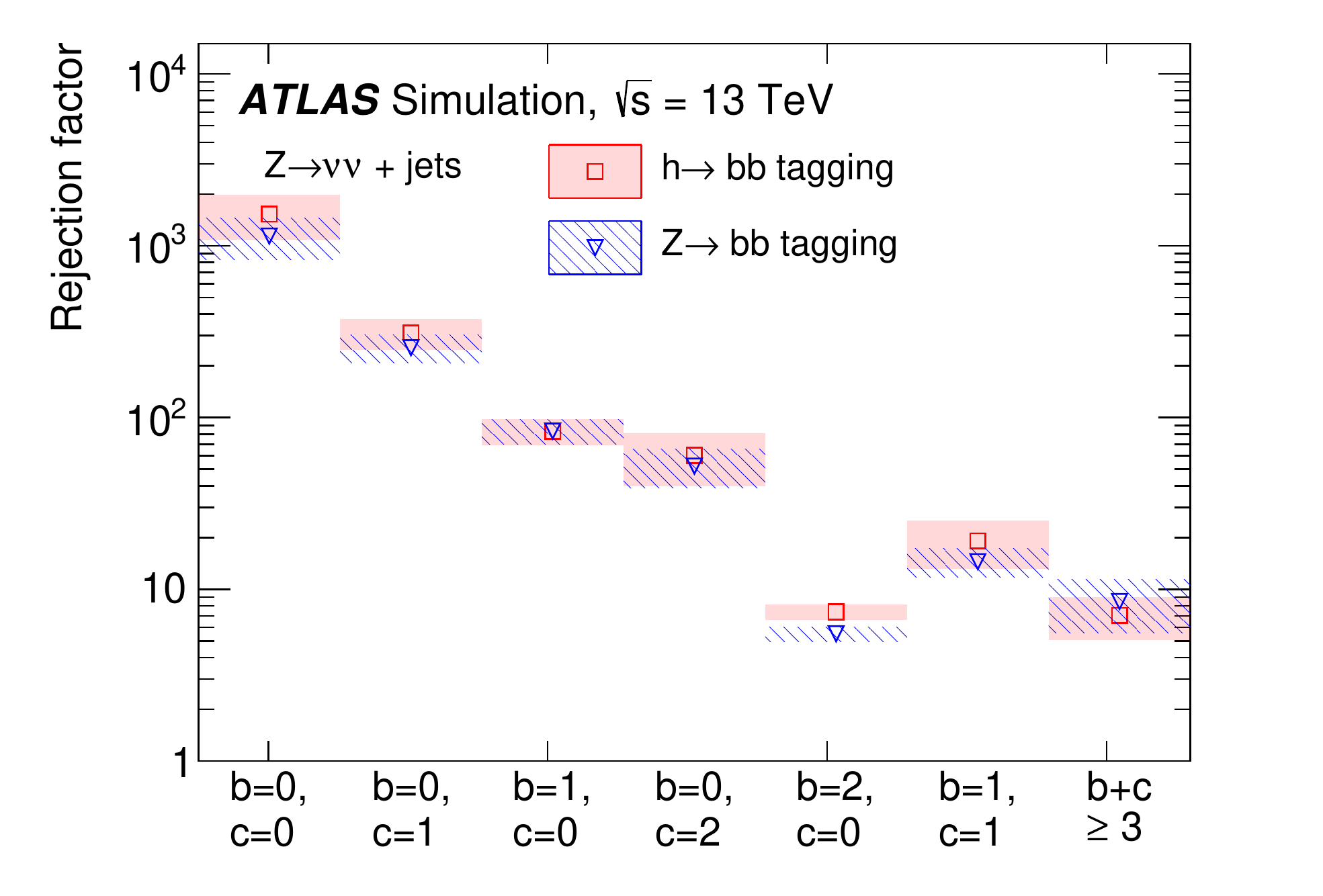} \label{fig:boson_tagger_Zhbb_rej}}
\caption{(a,c) The boson-tagging efficiency for jets arising from $W/Z/h$ bosons decaying into $q\bar{q}$ or $b\bar{b}$ (signal jets)
and (b,d) the rejection factor (inverse of the efficiency) for jets that have other origins (background jets) are shown.
The signal jet efficiency of \Wqq/\Zqq-tagging (\Zbb/\hbb-tagging) is evaluated using a sample of preselected large-$R$ jets ($\pt>200~\GeV, |\eta|<2.0, \mJ > 40~\GeV$)
in the simulated \winoBinoSim signal events with $\dm \ge 400~\GeV$.
The jets are matched with generator-level $W/Z$~($Z/h$) bosons within $\Delta R<1.0$ which decay into $q\bar{q}$ ($b\bar{b}$).
The background jet rejection factor is calculated using preselected large-$R$ jets in the sample of simulated \Znunu events, dominated by initial-state radiation jets.
As in the \Zbb/\hbb-tagging, the rejection factor is shown as a function of the number of $b$- or $c$-quarks contained in the large-$R$ jet within $\Delta R<1.0$.
The efficiency correction factors are applied to the signal efficiency and background rejection for the \Wqq/\Zqq-tagging.
The uncertainty is represented by the hashed bands, which includes the MC statistical uncertainty and the systematic uncertainties discussed in Section~\ref{sec:systematics:exp}.
}
\label{fig:boson_tagger}
\end{figure}

The analysis also uses reconstructed electrons and muons (collectively referred to as ``leptons''), as well as
photons and small-radius (small-$R$) jets, for kinematic selection, validation of the background estimation, and the \met computation.
Electron candidates are reconstructed from energy clusters that are consistent with electromagnetic showers in the ECAL and are matched to tracks in the ID, which are calibrated in situ using $Z \to ee$ samples~\cite{EGAM-2018-01}.
Muon candidates in the detector are typically reconstructed by matching tracks in the MS to tracks in the ID, and they are calibrated in situ using $Z \to \mu\mu$ and $J/\psi \to \mu\mu$ samples~\cite{MUON-2018-03_tmp}.
Small-$R$ jet candidates are reconstructed from particle-flow objects~\cite{PERF-2015-09} calibrated at the electromagnetic scale using the anti-$k_t$ algorithm with a radius parameter of $R=0.4$.
After subtracting the expected energy contribution from pileup using the jet area technique~\cite{PERF-2014-03},
the jet energy scale and resolution are corrected to particle level using MC simulation
as well as by in situ calibration using \Zjets, \gammajets, and multijet events~\cite{PERF-2016-04}.
Photon candidates are reconstructed either as electromagnetic clusters with no matching ID track or as $e^{+}e^{-}$ pairs from photon conversions in the ID material~\cite{EGAM-2018-01}.
 
Reconstructed electrons, muons, small-$R$ jets, and photons are subject to two sets of identification criteria: the looser ``baseline'' criteria and the tighter ``signal'' criteria.
The baseline objects are used for the \met computation, event cleaning, and the overlap removal procedure that resolves ambiguities between reconstructed objects as described below.
Baseline electrons are required to have $\pt>4.5$~\GeV\ and $|\eta|<2.47$, and meet the \emph{Loose}
criteria of the likelihood-based identification~\cite{EGAM-2018-01}.
Baseline muons are required to have $\pt>3$~\GeV\ and $|\eta|<2.7$, and to meet the \emph{Medium} identification criteria defined in Ref.~\cite{MUON-2018-03_tmp}.
To suppress the contributions from pileup, baseline leptons are also required to have a trajectory consistent with the primary vertex,
i.e.\ $|z_0\sin\theta|<0.5$~mm.\footnote{The transverse impact parameter, $d_0$, is defined as the distance of closest approach in the transverse plane between a track and the beam-line. The longitudinal impact parameter, $z_0$, corresponds to the $z$-coordinate distance between the point along the track at which the transverse impact parameter is defined and the primary vertex.}
Baseline small-$R$ jets must have $\pt > 30~\GeV$ and $|\eta|<4.5$.
Baseline photons must meet the \emph{Tight} identification criteria~\cite{EGAM-2018-01} in addition to satisfying $\pt > 50~\GeV$ and $\left| \eta \right| < 2.4$.
 
To prevent the reconstruction of a single particle as multiple objects,
an overlap removal procedure is applied to the baseline leptons, photons, and jets in the following order.
Any electron sharing an ID track with a muon or other electrons is removed,
such that only the highest-\pt electron is kept when multiple electrons share the same ID track.
Photons around the remaining electrons and muons are removed if the photon--lepton separation is $\Delta R<0.4$.
Next, small-$R$ jets are removed if they are $\Delta R<0.2$ from a remaining electron or photon,
or are $\Delta R<0.4$ from a muon and the jet has fewer than three associated tracks with $\pt>500~\MeV$.
Leptons (photons) are removed if they are separated from a remaining small-$R$ jet by $\Delta R < \min(0.4, 0.04+10~\GeV/p_{\mathrm{T},\ell})$ ($\Delta R<0.4$).
Finally, large-$R$ jets are removed if they are separated by $\Delta R < 1.0$ from any remaining electrons.
 
The missing transverse momentum, with magnitude \met,
is calculated as the negative vectorial sum of the transverse momenta of all baseline leptons, photons and small-$R$ jets calibrated to their respective energy scales,
and an additional ``soft term'' constructed from tracks originating from the primary vertex but not associated with any of the baseline objects~\cite{PERF-2016-07}.
 
Signal objects are defined by applying additional stringent criteria in event selections to ensure a high selection purity.
Signal leptons and photons are used only for validating the background estimation.
Signal electrons and muons must satisfy $\pt>10$~\GeV.
Additionally, the \emph{Tight} identification criteria are imposed for signal electrons.
To reduce the contribution from nonprompt decays of heavy-flavor hadrons,
the significance of the transverse impact parameter is required to satisfy $|d_0/\sigma(d_0)| < 5~(3)$ for signal electrons (muons).
An isolation selection is imposed to further suppress the residual misidentified leptons originating from jets.
The \emph{Tight} (\emph{HighPtCaloOnly}) working point defined in Ref.~\cite{EGAM-2018-01} is used for signal electrons with $\pt<200~\GeV$ ($\pt>200~\GeV$),
and the \emph{Tight} working point defined in Ref.~\cite{MUON-2018-03_tmp} is applied for signal muons.
Signal small-$R$ jets are selected within $|\eta|<2.8$, and those with $\pt < 120~\GeV$ and $|\eta| < 2.5$ must satisfy the \emph{Tight} quality criteria
of the track-based jet vertex tagger~\cite{PERF-2014-03,ATLAS-CONF-2014-018} in order to suppress jets originating from pileup.
Signal photons are defined as baseline photons with $\pt>200~\GeV$.
 
\section{Event selection}
\subsection{Common preselection}
As described below, a common preselection is used when defining the signal regions (SRs),
as well as the control regions (CRs) and validation regions (VRs) used for background estimation.
 
After the trigger requirement discussed in Section~\ref{sec:data-and-mc:data},
both the data and MC events are required to have at least one reconstructed vertex that is associated with two or more tracks with $\pt > 500~\MeV$.
The primary vertex of each event is selected as the vertex with the largest $\sum\pt^2$ of associated tracks~\cite{ATL-PHYS-PUB-2015-026}.
 
A set of event cleaning criteria are applied to ensure the quality of the measurements as well as to veto noncollision backgrounds.
Events with baseline muons consistent with cosmic muons ($|z_{0}| > 1$~mm or $|d_{0}| > 0.2$~mm) are removed.
To avoid pathological \met reconstruction,
events are also vetoed if any baseline jet points to a module of the tile hadronic calorimeter that was not operational during the data taking,
or if any baseline muon suffers from a poor momentum measurement ($\sigma(q/p)/(q/p) > 0.4$ where $q/p$ is the measured charge divided by momentum and $\sigma(q/p)$ is its uncertainty).
Track-jets can overlap due to their \pt-dependent variable radius.
To avoid the ambiguous cases of concentric jets,
events with a track-jet overlapping with another track-jet are removed.
An overlap is defined by $\Delta R < R_{\text{min}}$ where $\Delta R$ is the angular distance between a given pair of track-jets and $R_{\text{min}}$ is the smaller of their radii.
 
Beam-induced background is one of the major noncollision backgrounds.
Particles (typically muons) generated by interactions between the beam and the upstream collimators may directly hit the detector material.
Particular care is needed for this background since it can cause high-energy jet-like signatures and, correspondingly, large spurious \met~\cite{ATLAS-CONF-2011-137}.
Events with baseline jets failing the \emph{Loose} cleaning~\cite{ATLAS-CONF-2015-029} are removed.
Further cleaning is applied to events with no baseline leptons and photons
by requiring consistency between the \met and the alternative \metTrack, which is computed using only the good-quality ID tracks associated with the primary vertex.
While those events satisfy $\met>200~\GeV$ due to the trigger requirements, $\metTrack > 75~\GeV$ and $\dphimetMetTrack < 2.0$ are required in the cleaning procedure.
 
Finally, events are required to contain at least two large-$R$ jets.
Events with three or more large-$R$ jet are not vetoed in the analysis, in order to be as inclusive as possible in model coverage;
however, the applied event selection is always based on the two highest-\pt large-$R$ jets.
 
\subsection{Signal region selection} \label{sec:srs}
 
After the preselection, events with no baseline leptons are selected.
Two orthogonal signal region categories, 4Q and 2B2Q, based on the absence or presence of a large-$R$ jet containing
exactly two $b$-tagged track-jets (\Jbb) are defined in order to target the $qqqq$ and $bbqq$ final states respectively.
 
Boson tagging is required for the two leading large-$R$ jets.
The two leading large-$R$ jets must pass the \Vqq-tagging ($\nV=2$) in SR-4Q.
On the other hand, SR-2B2Q requires \Jbb to satisfy the \Zbb- or \hbb-tagging, while the
other jet (denoted by \Jqq) must satisfy the \Vqq-tagging criteria.
Multiple SRs are defined in each SR category to target the different final states of the signal processes.
The WW\,(ZZ) region in SR-4Q requires both leading large-$R$ jets to pass the $\Wqq(\Zqq)$-tagging,
while the WZ region is defined to contain events with at least one \Wqq-tagged jet and at least one \Zqq-tagged jet.
The inclusive bin VV, the logical union of WW, WZ and ZZ,
is designed to cover the signals that can have various \chiheavy branching ratios into $W$ relative to $Z$.
Similarly, the 2B2Q category accommodates the WZ/ZZ/Wh/Zh region, which varies by \Wqq- versus \Zqq-tagging or
\Zbb- versus \hbb-tagging.
The VZ (Vh) region serves as the inclusive SR defined by the logical union of WZ and ZZ (Wh and Zh).
The SR segmentation is summarized in Table~\ref{tab:selection:SRoverview}
and illustrated schematically in Figure~\ref{fig:SRsegmentation}.
In total, four and six SRs are defined in SR-4Q and SR-2B2Q, respectively.
SR-4QXX, SR-2B2QXZ and SR-2B2QXh (where X = W/Z/V) are mutually orthogonal,
and are statistically combined according to the models considered in Table~\ref{tab:result:models_vs_SRs}.

\begin{table}[htb]
\begin{center}
\caption{Definition of each SR in the 4Q and 2B2Q categories, where \nW, \nZ, \nV, \nZbb, and \nH are respectively the number of large-$R$ jets passing the \Wqq-, \Zqq-, \Vqq-, \Zbb-, and \hbb-tagging of the two leading large-$R$ jets.
SR-4Q-WZ requires $\nW,\nZ \ge 1$ instead of $\nW=\nZ=1$ because the selections in \Wqq- and \Zqq-tagging are not exclusive.
The overlap and the segmentation between the SRs are illustrated in Figure~\ref{fig:SRsegmentation}.}
\begin{tabular}{lccccc}
\toprule
& $\nW$  & $\nZ$   & $\nV$ & $\nZbb$ & $\nH$ \\
\midrule
4Q-WW   & $=2$   & -       & $=2$  &  $=0$   & $=0$  \\
4Q-WZ   & $\ge1$ & $\ge1$  & $=2$  &  $=0$   & $=0$  \\
4Q-ZZ   & -      & $=2$    & $=2$  &  $=0$   & $=0$  \\
4Q-VV   & -      & -       & $=2$  &  $=0$   & $=0$  \\
\midrule
2B2Q-WZ & $=1$   & -       & $=1$  &  $=1$   & $=0$  \\
2B2Q-ZZ & -      & $=1$    & $=1$  &  $=1$   & $=0$  \\
2B2Q-Wh & $=1$   & -       & $=1$  &  $=0$   & $=1$  \\
2B2Q-Zh & -      & $=1$    & $=1$  &  $=0$   & $=1$  \\
2B2Q-VZ & -      & -       & $=1$  &  $=1$   & $=0$  \\
2B2Q-Vh & -      & -       & $=1$  &  $=0$   & $=1$  \\
\bottomrule
\end{tabular}
\label{tab:selection:SRoverview}
\end{center}
\end{table}
 
\begin{figure}[h]
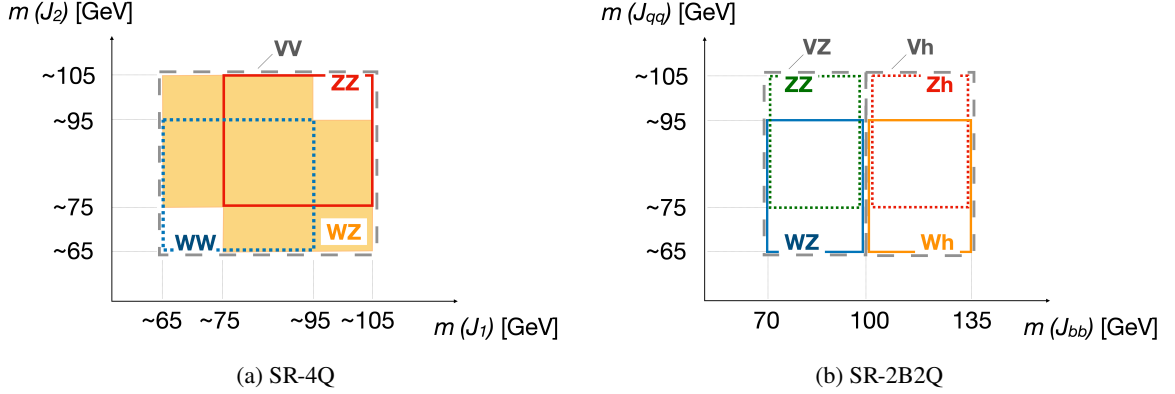

\centering
\subfloat[SR-4Q]{\includegraphics[width=0.485\textwidth]{fig_05a.pdf}}
\subfloat[SR-2B2Q]{\includegraphics[width=0.485\textwidth]{fig_05b.pdf}}
\caption{The SR segmentation illustrated as a function of the masses of the two leading large-$R$ jets.
(a) In SR-4Q, both jets are required to pass the \Wqq- or \Zqq-tagging.
(b) In SR-2B2Q, one of the two jets is required to contain exactly two $b$-tagged track-jets (\Jbb) while the other (\Jqq) has at most one.
The mass of \Jbb is required to be consistent with a $Z$ boson (70--100~\GeV) or an $h$ boson (100--135~\GeV),
while \Jqq is required to pass the \Wqq- or \Zqq-tagging.
The mass window cuts of the \Wqq/\Zqq-tagging shown in the plot only indicate the typical values,
while variable cut values along \pt are applied in the analysis.
The inclusive SRs, defined by the logical union of a few mutually overlapping SRs,
are indicated by the gray dashed lines.
}
\label{fig:SRsegmentation}
\end{figure}
 
Finally, a series of background rejection cuts are applied to further suppress the main SM backgrounds at this stage: \Znunu, \Wlv, and \ttbar.
To reject backgrounds including top quarks,
the $b$-jets that do not originate from boosted boson candidates are vetoed.
This is done by requiring $\nbOutJ=0$, where $\nbOutJ$ denotes the number of $b$-tagged track-jets that are not matched with any of the two leading large-$R$ jets by the $\Delta R<1.0$ criterion.
For the 4Q category, the total number of $b$-tagged track-jets in the event (\nb) must be less than two in order to further suppress the \ttbar background.
The effective mass variable, \meff, defined as the scalar sum of the \pt of the two leading large-$R$ jets and \met, is used to select events with hard kinematics together with \met.
Selections of $\met>300~(200)~\GeV$ and $\meff>1300~(1000)~\GeV$ are applied in the 4Q (2B2Q) regions.
Event shape information is also useful in distinguishing the signals from the backgrounds.
Signal events have a relatively spherical shape, where the jets tend to be isolated from \met, indicating that the heavy electroweakinos are produced nearly at rest.
Background events, meanwhile, have a higher chance of containing a jet aligned with a $W/Z$ boson, due to boosted top decays in \ttbar events or the emission of collinear radiation in \Zjets and \Wjets events.
The minimum azimuthal angle separation between \met and any signal small-$R$ jets must satisfy $\dphimin > 1.0$ for both SR categories.
The use of small-$R$ jets is motivated by the need to effectively identify low-\pt jets and it also provides better resolution in terms of jet alignment with \met.
In 2B2Q, the stransverse mass variable \mTtwo~\cite{Lester:1999tx,Barr:2003rg} is also used,
constructed by assigning each of the two leading large-$R$ jets to the visible particle legs.
A selection of $\mTtwo>250~\GeV$ is found to effectively suppress the SM backgrounds, particularly \ttbar which exhibits a kinematic cutoff at $\mTtwo \sim 200~\GeV$, driven by the top quark mass constraint.\footnote{The hypothetical missing-particle mass is set to $100~\GeV$ and this offset is subtracted from the calculated \mTtwo, although the dependency on the choice of missing-particle mass is very small.
}
 
The cut values of the kinematic selection are equivalent within the SR-4Q and SR-2B2Q categories.
The selection criteria that define the SR-4Q, SR-2B2Q-Wh, and SR-2B2Q-Vh regions are obtained by optimizing the sensitivity to the \winoBino model with $(\mCone,\mNone)\approx(800,100)~\GeV$,
while those for SR-2B2Q-WZ, SR-2B2Q-ZZ and SR-2B2Q-VZ are determined by optimizing the sensitivity to the \hinoGravitino model with $m_{\None} \approx 800~\GeV$.
The discovery significance is used as the metric of sensitivity.
The obtained cuts are also found to be nearly optimal for the other signal models.
 
The acceptance times efficiency for signal events ranges from 1\% to 4\% depending on \dm and the SR.
For example, it is about 1\%--2\% (1.5\%) in SR-4Q-VV (SR-2B2Q-Vh) for the \SMWZ (\SMWh) signals with $\dm = 600~\GeV$,
and 3\%--4\% (2\%--3\%) with $\dm=1~\TeV$.
 
\section{Background estimation}
The main SM background process in the SRs is \Znunu (${\sim}50\%$), followed by \Wlv (15\%--20\%), and $VV$ (10\%--20\%).
The rest consists of $VVV$ events in SR-4Q (5\%--10\%),
or \ttbar, single-top and \ttbarX events in SR-2B2Q (10\%--20\%).
 
The estimation strategy varies between the ``reducible'' and ``irreducible'' backgrounds.
The irreducible backgrounds in this search are due to SM events including at least two hadronic $W/Z/h$ decays and large \met from high-\pt neutrinos.
These consist of $VVV$ and \ttbarX, and are estimated using MC simulation.
The contribution from fully-hadronic $VV$ and \ttbar are negligible due to the stringent \met requirement.
 
The reducible backgrounds are all that remain, including the dominant \Znunu production.
These backgrounds are characterized by the presence of at least
one large-$R$ jet
that originates from a process other than a $W/Z \ra qq$ decay and is referred to as a ``fake boson jet''.
The fake boson jets are typically caused by two collimated high-\pt initial-state radiation (ISR) jets that are clustered together as a single large-$R$ jet.
A partly data-driven method is used to estimate the reducible backgrounds;
a control region (CR0L)\footnote{``0L'' stands for regions with no baseline leptons.}
is defined in the phase space adjacent to a SR,
where the MC sample is normalized to the CR data.
The SR expectation is obtained using the normalized MC sample,
assuming the modeling of the SR/CR0L yield ratio (``0L transfer factor'') is reliable.
This assumption is tested in a number of validation regions (VR) in data.
A CR0L is defined for each 4Q and 2B2Q category
by reversing the $\Vqq$-tagging requirement
on one of the two leading large-$R$ jets in the SR,
and is denoted by CR0L-4Q and CR0L-2B2Q, respectively.
The multijet and noncollision background contributions are found to be negligible
using the data-driven methods or from the estimation in a similar phase space carried out in Ref.~\cite{SUSY-2018-22}.
 
The following subsections discuss the methodology and results of the irreducible and reducible background estimations.

\subsection{Irreducible background estimation}
\label{sec:VRTTX}
The $VVV$ (\ttbarX) events account for at most 10\% of the total background in SR-4Q (SR-2B2Q),
and are negligible in SR-2B2Q (SR-4Q).
Given their minor contribution, these backgrounds are estimated directly from the MC predictions and assigned conservative uncertainties.
 
The dominant \ttbarX component in SR-2B2Q is $\ttbar(\ra bqqbqq)+Z(\ra \nu\nu)$.
To validate the MC modeling, a dedicated validation region VRTTX is defined in a 3-lepton region populated by
$\ttbar(\ra b\ell\nu bqq)+Z(\ra \ell\ell)$.
The selections are summarized in Table~\ref{tab:cutsummary}.
Exactly three baseline and signal leptons are required,
with the leading lepton of $\pt>30~\GeV$ firing the single-lepton trigger.
At least one large-$R$ jet is required in the event,
and it must contain exactly two $b$-tagged track-jets.
No further kinematic cuts are applied, in order to maintain a sufficient data sample size in the region.
The \ttbarX purity in this region is about 70\%.
Sixty-eight data events are observed in VRTTX, while $46.5$ events are predicted by MC simulation (31.5 from \ttbarX, 12.1 from $VV$, and 2.9 from the others).
A 70\% uncertainty is therefore assigned to the \ttbarX normalization to fully cover the observed discrepancy;
this has only a small impact on the total background estimation, given the minor contribution in the SRs.
 
The $VVV$ contribution to SR-4Q is mainly from $VV(\to qqqq)+Z(\to \nu\nu)$ processes.
The $VVV$ process has only recently been observed at the LHC~\cite{STDM-2017-22,CMS-SMP-19-014}.
As the data sample size in that phase-space region is insufficient and loosening the selection leads to poor $VVV$ purity,
no control or validation regions were designed and this background is estimated directly from the MC prediction.
A 50\% uncertainty is assigned for the normalization based on the precision of the $WWZ$ production cross-section measurement performed by the CMS experiment~\cite{CMS-SMP-19-014}.
 
\subsection{Reducible background estimation}
\label{sec:CR_definitions}
The CR0L-4Q is defined by the same selection as SR-4Q-VV except that one of the two leading large-$R$ jets must fail the \Vqq-tagging,
and the CR0L-2B2Q is constructed as the logical union of SR-2B2Q-VZ and SR-2B2Q-Vh with the \Jqq failing the \Vqq-tagging.
For the reducible backgrounds, the extrapolation from a CR0L to the SR is mainly characterized by the \Vqq-tagging response for a fake boson jet.
In order to maintain a sufficient data sample size and to suppress signal contamination, the \meff selection and the \Jbb mass window cut for CR0L-2B2Q are loosened
relative to the SR, from $\meff > 1000~\GeV$ to $\meff > 900~\GeV$ and from $70 < \mJbb < 135~\GeV$ to $70 < \mJbb < 150~\GeV$.
The signal contamination in CR0L-4Q and CR0L-2B2Q is evaluated using the \winoBinoSim and \hinoGravitino samples.
While a contribution of up to 15\% (24\%) of the expected backgrounds in CR0L-4Q (CR0L-2B2Q) can be caused by the non-excluded signals,
the introduced bias in the estimate is still smaller than the total uncertainty in the SRs and VRs,
and therefore has only a small impact on the final sensitivity.
The selections are summarized in Table~\ref{tab:cutsummary},
and the relation with the SR is illustrated in Figure~\ref{fig:CR_strategy}.
 
\begin{figure}[htb!p]
\centering
\subfloat[4Q]{\includegraphics[width=0.48\textwidth]{fig_06a.pdf}}
\subfloat[2B2Q]{\includegraphics[width=0.48\textwidth]{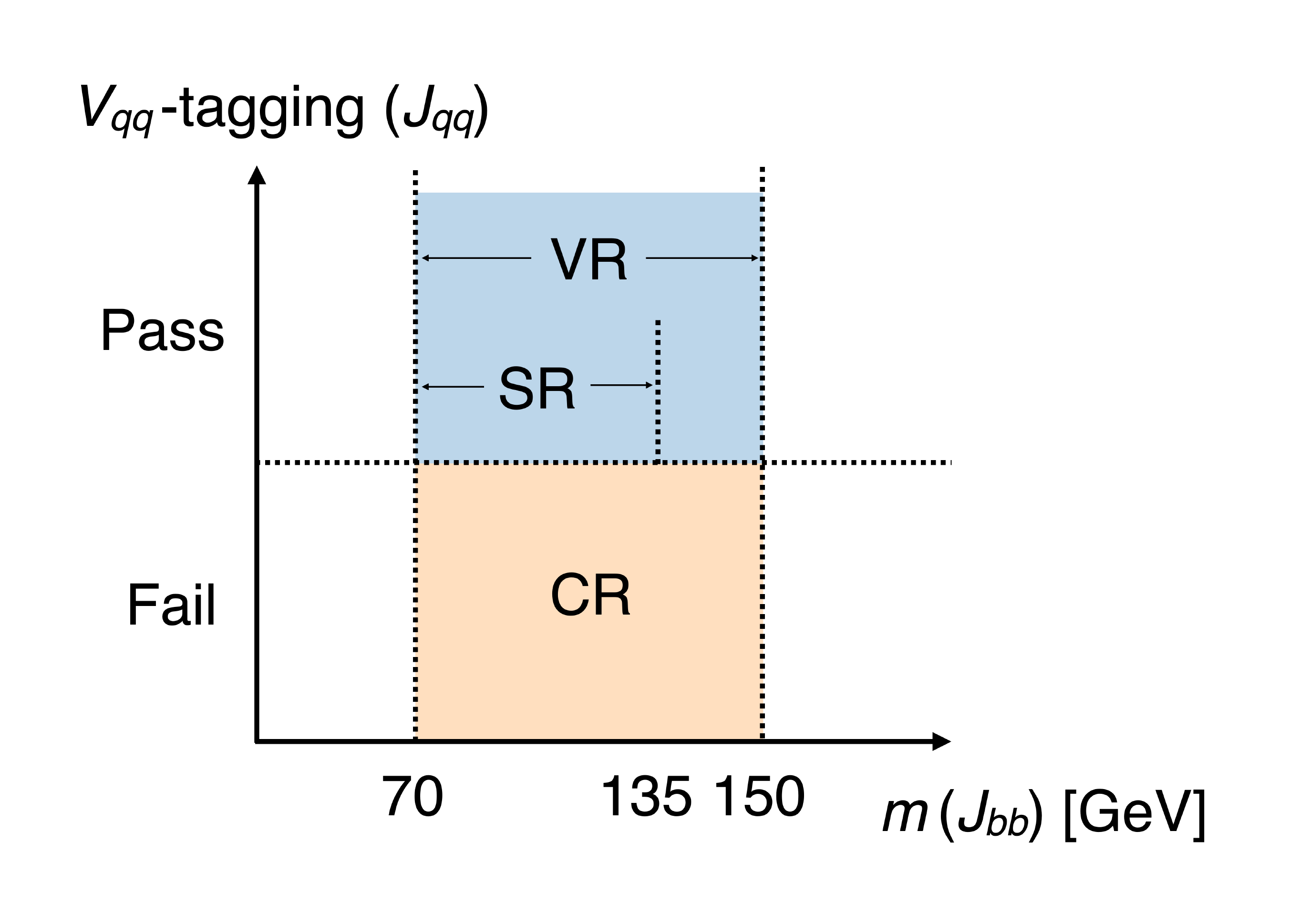}}
\caption{
Schematics illustrating the relation between
the signal regions (SRs), validation regions (VRs) and control regions (CRs)
in the (a) 4Q and (b) 2B2Q categories.
A CR is constructed by inverting the \Vqq-tagging requirement for one of the two leading large-$R$ jets in the corresponding SR (VR).
In 2B2Q, the \Jbb mass window cut in the CRs and VRs is loosened.
See Table~\ref{tab:cutsummary} for details.
}
\label{fig:CR_strategy}
\end{figure}
 
One advantage of this CR0L definition is that
all physics processes contributing to the background (\Zjets, \Wjets, \ttbar, etc.) have comparable 0L transfer factors.
This is because the fake boson jets have similar origins and kinematics,
and therefore the same boson tagging efficiency, confirmed by the simulation.
Consequently, they can be treated as a single combined component
with a common normalization factor assigned to correct their normalizations at once.
The normalization is performed in 4Q and 2B2Q separately with an independent normalization factor,
based on the ``background-only fits'' described in Section~\ref{sec:stat}.
 
Good MC modeling of the 0L transfer factor is essential for the estimation.
Data events with exactly one lepton or one photon are utilized for the validation.
This is motivated by the fact that in these regions the main backgrounds,
\Znunu in SR/CR0L, \Wlv in the 1-lepton region and \gammajets in the 1-photon region, have similar ISR jet kinematics
and, relative to the minor backgrounds, contribute similarly to the respective regions
when a compatible kinematic phase space is chosen.
The 1-lepton (1-photon) regions corresponding to CR0L and the SR are constructed,
denoted by CR1L(1Y) and VR1L(1Y) respectively.
The level of data-vs-MC agreement in the ratio VR1L(1Y)/CR1L(1Y) (``1L(1Y) transfer factor'')
is validated as a proxy for the MC modeling of the 0L transfer factor.
 
The selections applied for VR(CR)1L and VR(CR)1Y are listed in Table~\ref{tab:cutsummary}.
The VR(CR)1L is defined by requiring exactly one baseline lepton and a signal lepton with $\pt>30~\GeV$ that fires the single-lepton trigger.
In addition, $\met>50~\GeV$ is required to suppress contributions from QCD multijet events with poorly measured \met.
The VR(CR)1Y requires exactly one baseline and signal photon to fire the single-photon trigger,
and vetoes events with at least one baseline lepton.
For both VR(CR)1L and VR(CR)1Y, at least two large-$R$ jets are required,
and the events are separated into the 4Q and 2B2Q categories, based on the absence or presence of \Jbb, respectively
A $b$-jet veto strategy similar to that for the SRs and the CR0L bins is applied:
$\nbOutJ=0$ in all the regions, $\nb \le 1$ in the 4Q regions.
A stricter veto of $\nb=0$ is applied in CR(VR)1L-4Q to suppress the large single-top contribution relative to the other 4Q regions.
 
The kinematic selection in the 1L (1Y) regions is applied with \met replaced by \ptW (\ptY) in the variables.
This is to ensure that compatible phase spaces are probed for
\Znunu in SR/CR0L, \Wlv in VR1L/CR1L, and \gammajets in VR1Y/CR1Y,
since \met in \Znunu events typically represents the \pt of the $Z$ boson.
The \ptW variable is defined as the vector sum of the \pt of the lepton and \met in the 1L regions.
The \met, \ptW and \ptY are collectively denoted by \ptV.
The \dphimin variable, as well as those analogously derived from \ptW and \ptY, are collectively denoted by \dphiminV.
The cuts in \ptV and \meff are moderately loosened relative to the SRs to ensure a data sample size sufficient for the validation. For 2B2Q regions, the \Jbb mass window cut for the \Zbb/\hbb-tagging is loosened to $70 < \mJ < 150~\GeV$ to match CR0L-2B2Q, further increasing the data sample size.
The contributions from events containing fake leptons and fake photons are estimated by data-driven methods~\cite{STDM-2011-24,STDM-2010-08} and are found to be negligible.
 
The same estimation procedure used to estimate the backgrounds in the SRs is employed to predict the backgrounds in the VRs.
A background-only fit, described in Section~\ref{sec:stat}, is performed,
where the sum of the reducible backgrounds in MC samples is normalized to the data in each CR1L~(CR1Y).
The modeling of the transfer factor is examined using the level of agreement between the data and the post-fit background expectation in the corresponding VR1L~(VR1Y).
Four sets of fits are performed in the 1L-4Q, 1L-2B2Q, 1Y-4Q and 1Y-2B2Q categories separately.
Including all the systematic uncertainties discussed in Section~\ref{sec:systematics},
the observed data event yields and the post-fit background expectations in the VRs are summarized in Table~\ref{tab:VRs_yields}, and compared graphically in Figure~\ref{fig:VRfit}.
The distributions of \meff\ in VR1L(1Y)-4Q and \mTtwo\ in VR1L(1Y)-2B2Q are shown in Figure~\ref{fig:N-1plots_VR}.
The data agree reasonably well with the estimated backgrounds across the VRs.
The largest disagreement is found in VR1L-4Q, corresponding to a statistical significance of $1.8\sigma$.

\begin{table}
\centering
\caption{Summary of selections for the SRs, CRs, and VRs.
\nJTagged (\nJFailed) represents the number of large-$R$ jets passing (failing) the \Vqq-tagging of the two highest-\pt large-$R$ jets.
The same selection is applied to the SR~(VR) and CR in the same category except for the \Vqq-tagging and some kinematic selections
that are explicitly indicated in parentheses.
The trigger selection and event cleaning described in Section~\ref{sec:object} are also applied.
VRTTX is a validation region used to validate the \ttbarX modeling as described in Section~\ref{sec:VRTTX}.
\ptW is the vector sum of the \pt of the lepton and \met in the 1L regions.
In the 1L (1Y) regions, \met is replaced by \ptW (\ptY) when calculating the kinematic variables \meff, \dphimin and \mTtwo.
Details are given in Section~\ref{sec:CR_definitions}.
}
\label{tab:cutsummary}
\maxsizebox*{\textwidth}{\textheight}{
\begin{tabular}{l|cc|cc|cc|c}
\toprule
& \multicolumn{2}{c|}{SR(CR0L)}            & \multicolumn{2}{c|}{VR(CR)1L}             & \multicolumn{2}{c|}{VR(CR)1Y}  & \multirow{2}{*}{VRTTX} \\
& 4Q         & 2B2Q                & 4Q         & 2B2Q                         & 4Q         &  2B2Q             &             \\
\midrule
\nJ                & \multicolumn{2}{c|}{$\ge 2$}     & \multicolumn{2}{c|}{$\ge 2$}              & \multicolumn{2}{c|}{$\ge 2$}   & $=1$        \\
\nL                & \multicolumn{2}{c|}{$=0$}        & \multicolumn{2}{c|}{$=1$}                 & \multicolumn{2}{c|}{$=0$}      & $=3$        \\
$\pt(\ell_1)$ [GeV]& \multicolumn{2}{c|}{-}           & \multicolumn{2}{c|}{$>30$~~~}                & \multicolumn{2}{c|}{-}         & $>30$~~  \\
\nPhoton           & \multicolumn{2}{c|}{-}           & \multicolumn{2}{c|}{-}                    & \multicolumn{2}{c|}{$=1$}      & -      \\
\midrule
\nJTagged          & $=2~(=1)$ & $=1~(=0)$            & $=2~(=1)$ & $=1~(=0)$                     & $=2~(=1)$ & $=1~(=0)$          & - \\
\nJFailed          & $=0~(=1)$ & $=0~(=1)$            & $=0~(=1)$ & $=0~(=1)$                     & $=0~(=1)$ & $=0~(=1)$          & - \\
\midrule
$\nJTwoB$          & $=0$ & $=1$                      & $=0$ & $=1$                               & $=0$ & $=1$                    & $=1$ \\
$\mJbb$~[GeV]      & -     & $\in[70,135~(150)]$      & - & $\in[70,150]$                         &   - & $\in[70,150]$            & - \\
\midrule
$\nbOutJ$          & \multicolumn{2}{c|}{$=0$}        & \multicolumn{2}{c|}{$=0$}                 & \multicolumn{2}{c|}{$=0$}      & -      \\
$\nb$              & $\leq 1$ & -                     & $=0$ & -                                  & $\leq 1$ & -                   & -      \\
\midrule
\met\ [GeV]        & $>300$  & $>200$                 & \multicolumn{2}{c|}{$>50$}                & \multicolumn{2}{c|}{$<200$}    & - \\
\ptW\ [GeV]        & \multicolumn{2}{c|}{-}           & \multicolumn{2}{c|}{$>200$~~~}               & \multicolumn{2}{c|}{-}         & - \\
\ptY\ [GeV]        & \multicolumn{2}{c|}{-}           & \multicolumn{2}{c|}{-}                    & \multicolumn{2}{c|}{$>200$}    & - \\
\meff\ [GeV]       & $>1300$ & $>1000~(>900)$         & $>1000$ & $>900$                          & $>1000$ & $>900$               & - \\
\dphimin          & \multicolumn{2}{c|}{$>1.0$}      & \multicolumn{2}{c|}{$>1.0$~\,}               & \multicolumn{2}{c|}{\,$>1.0$}    & - \\
\mTtwo\ [GeV]      & -       & $>250$                 & -       & $>250$                          & -       & $>250$               & - \\
\bottomrule
\end{tabular}
}
\end{table}

 
\begin{figure}[thb!p]
\centering
\includegraphics[width=0.85\textwidth]{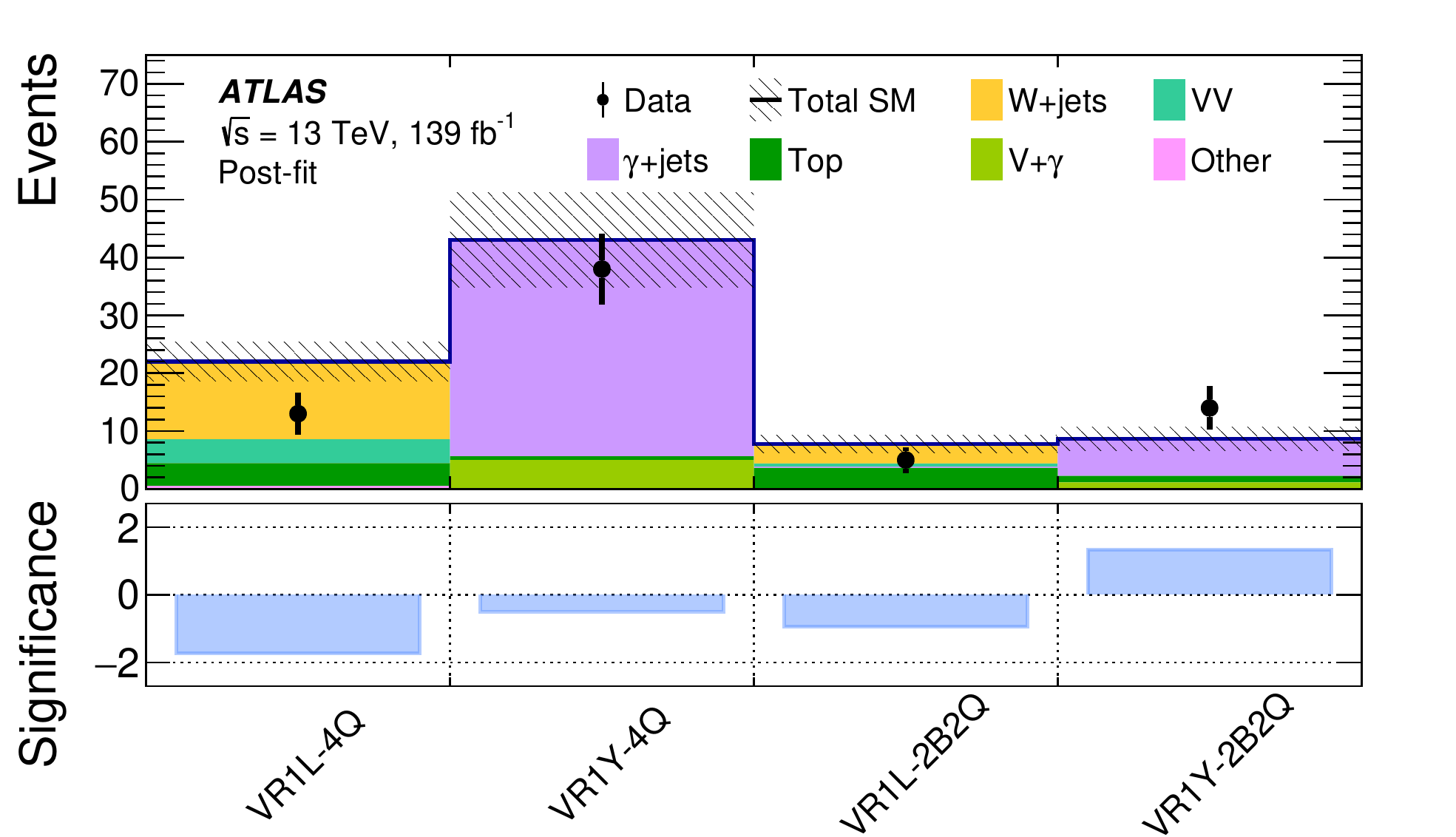}
\caption{Comparison between the observed data and the post-fit SM background prediction in VR1L and VR1Y.
The background predictions are obtained using the background-only fit described in Section~\ref{sec:stat}.
``Top'' includes \ttbar, $t{+}X$ and \ttbarX production.
The total systematic uncertainty of the background prediction is shown by the hatched band.
The bottom panel shows the statistical significance of the discrepancy between the observed number of events and the SM expectation, following the prescription described in Ref.~\cite{Cousins:2007bmb}.
}
\label{fig:VRfit}
\end{figure}
 
\begin{table}
\centering
\caption{Number of observed data events and the post-fit SM background prediction in the VR1L~(1Y) bins and the corresponding CR1L~(1Y) bins.
Negligibly small contributions are indicated by ``-''. Within each CR the reducible backgrounds have the same relative uncertainty in their expected yield because a common
normalization factor is assigned to all of them in the fit.}
\label{tab:VRs_yields}
{\small
\begin{tabular}{l cccc}
\toprule
Region          & CR1L-4Q	& VR1L-4Q	& CR1L-2B2Q	  & VR1L-2B2Q	   \\[-0.05cm]
\midrule
Observed        & $439$	& $13$	& $96$	  & $5$  \\
\midrule
Post-fit        & $439 \pm 21~~$          & $22.0  \pm 3.4~~$       & $96 \pm 10$           & $7.8 \pm 1.5$                \\
\midrule
\Wjets   	& $325    \pm 16~~$       & $13.4  \pm 2.2~~$       & $48     \pm 5~~$        & $3.4   \pm 0.7$              \\
\Zjets   	& $4.45   \pm 0.21$    	& $0.198 \pm 0.035$     & $0.58   \pm 0.06$     & $0.044 \pm 0.012$            \\
\gammajets	& $<1$      		& -         		& $0.57   \pm 0.06$     & $0.22  \pm 0.10$             \\
$VV$	 	& $65.4   \pm 3.1~~$      & $4.1   \pm 0.8$       & $6.9    \pm 0.7$      & $0.55  \pm 0.15$             \\
$V\gamma$	& $<1$      		& -         		& $<0.1$      		& -                \\
$VVV$	        & $1.3    \pm 0.6$     	& $0.52  \pm 0.28$      & $0.14   \pm 0.08$     & $0.09  \pm 0.05$             \\
\ttbar	 	& $30.4   \pm 1.5~~$     	& $2.7   \pm 0.4$       & $24.0   \pm 2.5~~$      & $1.8   \pm 0.4$                            \\
$t{+}X$	 	& $11.0   \pm 0.5~~$     	& $0.91  \pm 0.21$      & $13.2   \pm 1.4~~$      & $1.27  \pm 0.34$                           \\
\ttbarX	 	& $1.5    \pm 1.2$      & $0.16  \pm 0.12$      & $1.5    \pm 1.1$      & $0.4 \pm 0.4$  \\
$Vh$	 	& $<0.1$    		& $<0.001$  		& $0.69   \pm 0.07$     & $0.046 \pm 0.009$                          \\
\midrule
\midrule
Region     & CR1Y-4Q    & VR1Y-4Q         & CR1Y-2B2Q	 & VR1Y-2B2Q	   \\[-0.05cm]
\midrule
Observed   & $1001$    & $38$           & $127$	 & $14$	       \\
\midrule
Post-fit   & $1001  \pm 32~~~~$     & $43    \pm 8~~$         & $127 \pm 11~~$          & $8.6 \pm 2.0$              \\
\midrule
\Wjets     & $2.59  \pm 0.08$   & $<0.1$         	& $<0.1$    		& -         	 \\
\Zjets     & $<1$      		& -   	        	& $<0.01$   		& -         	 \\
\gammajets & $856   \pm 28~~$     & $37    \pm 7~~$  	& $107   \pm 11~~$        & $6.4  \pm 1.6$         	 \\
$VV$       & $<1$      		& -  	        	& -         		& -         	 \\
$V\gamma$  & $131   \pm 4~~~~$      & $5.0   \pm 0.9$       & $12.6  \pm 1.3~~$       & $1.13 \pm 0.27$         	 \\
$VVV$	   & $<0.1$    		& $<0.01$        	& -         		& -         	 \\
\ttbar	   & $1.28  \pm 0.04$   & -              	& $0.57  \pm 0.06$      & $0.28 \pm 0.18$         	 \\
$t{+}X$	   & $<1$      		& -              	& $<0.1$    		& -         	 \\
\ttbarX	   & $9     \pm 6$      & $0.6   \pm 0.5$       & $7     \pm 5$         & $0.8  \pm 0.6$         	 \\
$Vh$	   & $<0.001$  		& -              	& $<0.01$   		& -        	 \\
\bottomrule
\end{tabular}
}
\end{table}
 
\begin{figure}[h]
\centering
\subfloat[VR1L-4Q]{\includegraphics[width=0.48\textwidth]{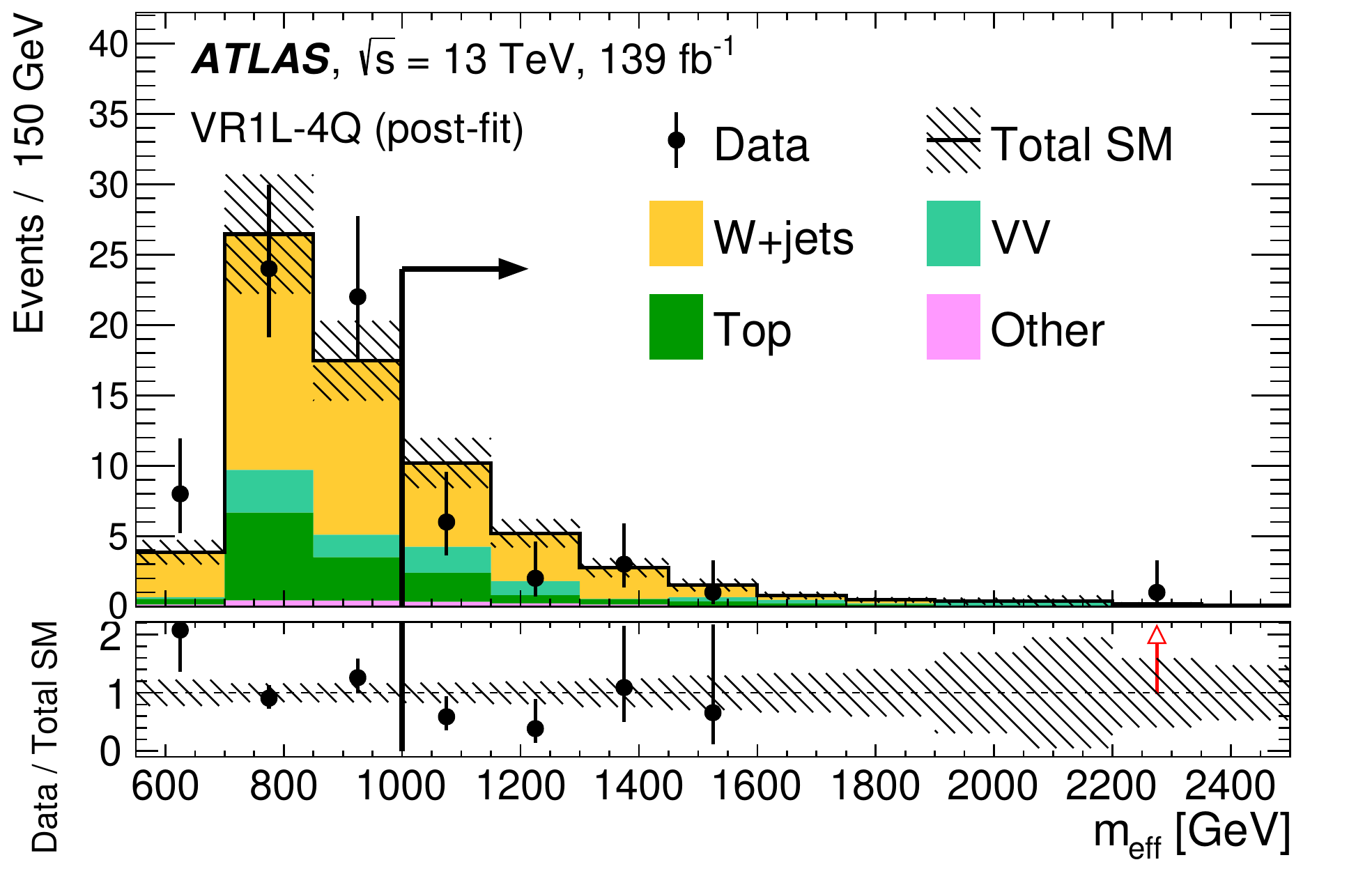}}
\subfloat[VR1Y-4Q]{\includegraphics[width=0.48\textwidth]{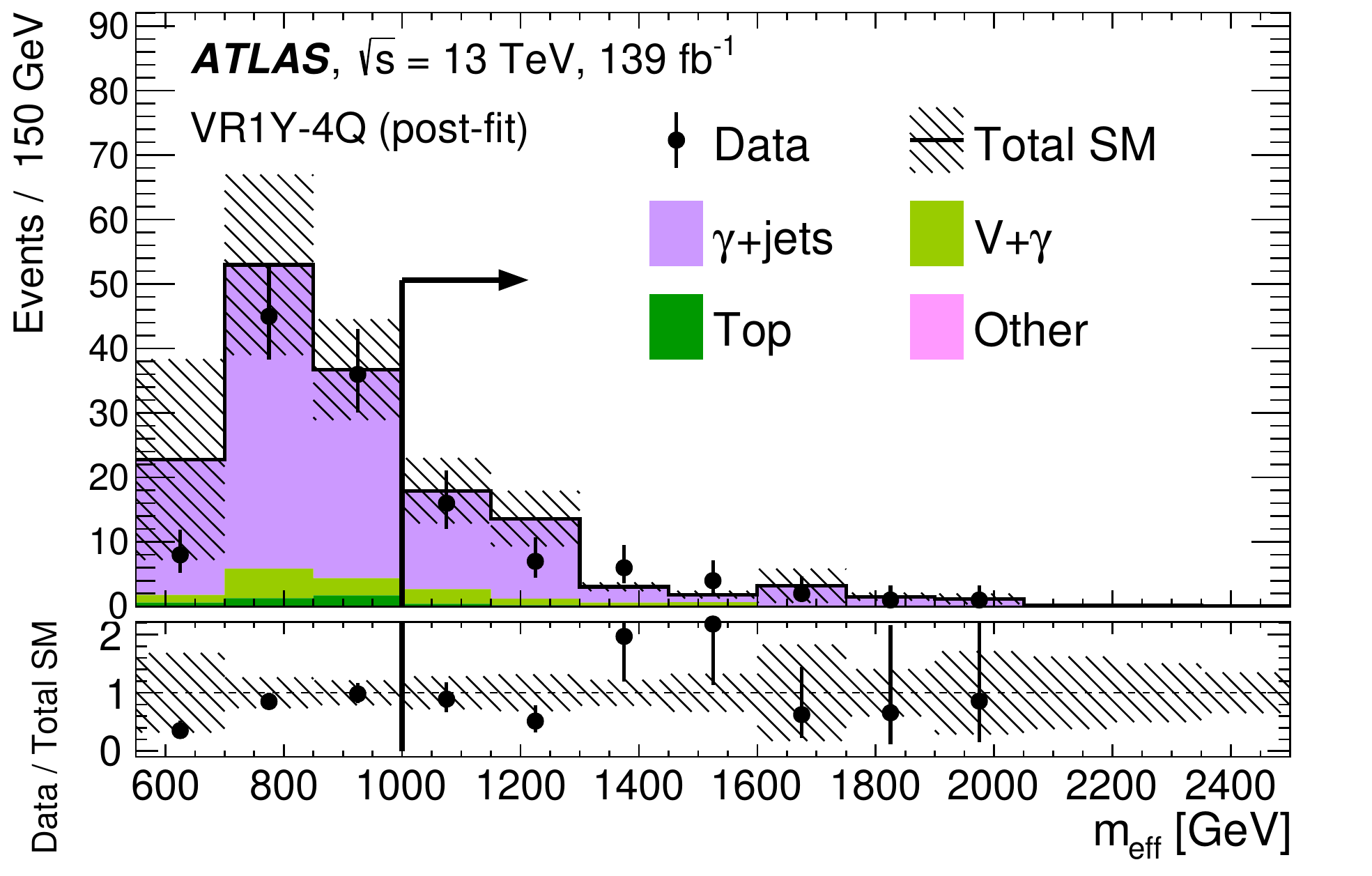}} \\
\subfloat[VR1L-2B2Q]{\includegraphics[width=0.48\textwidth]{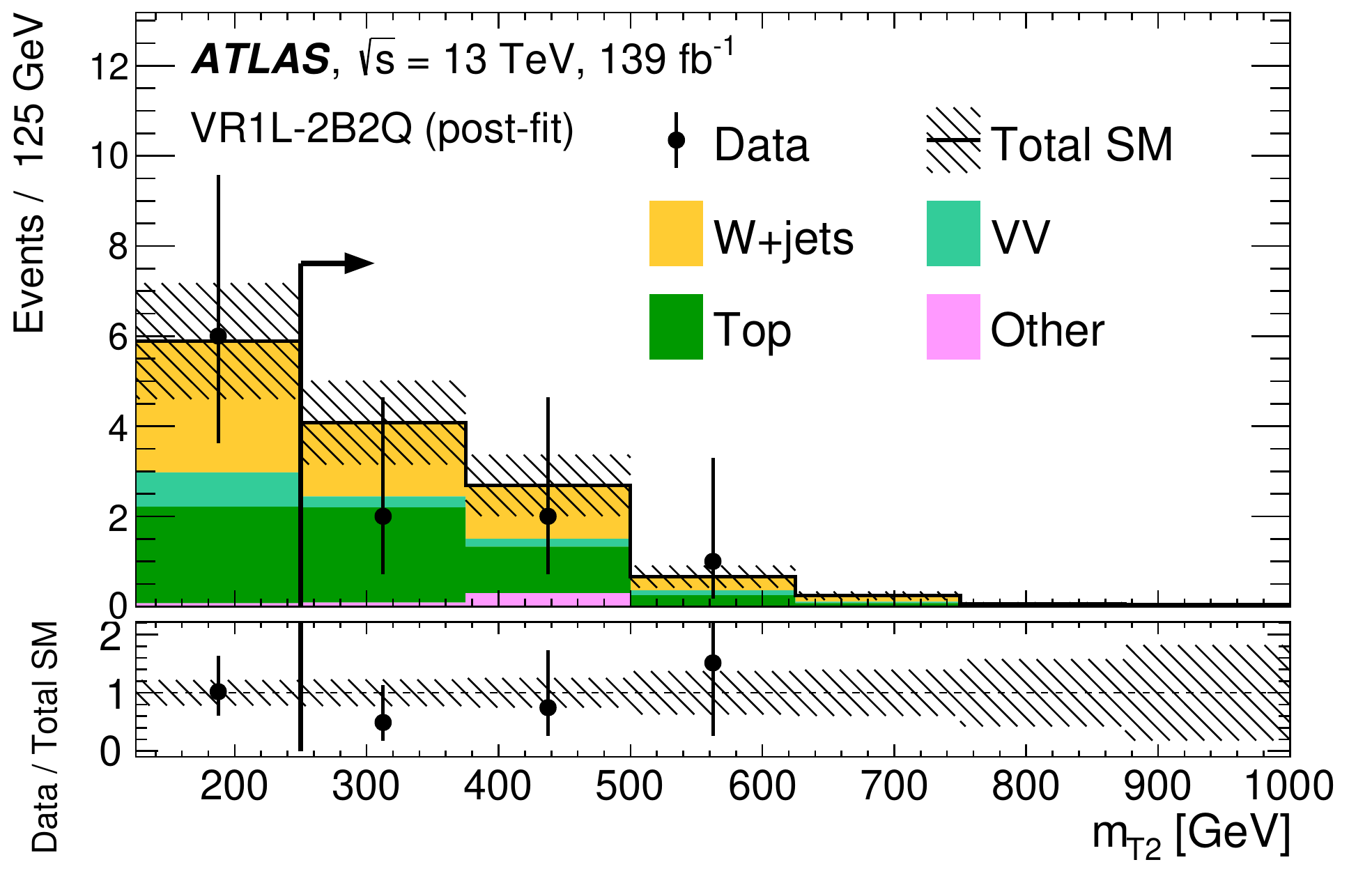}}
\subfloat[VR1Y-2B2Q]{\includegraphics[width=0.48\textwidth]{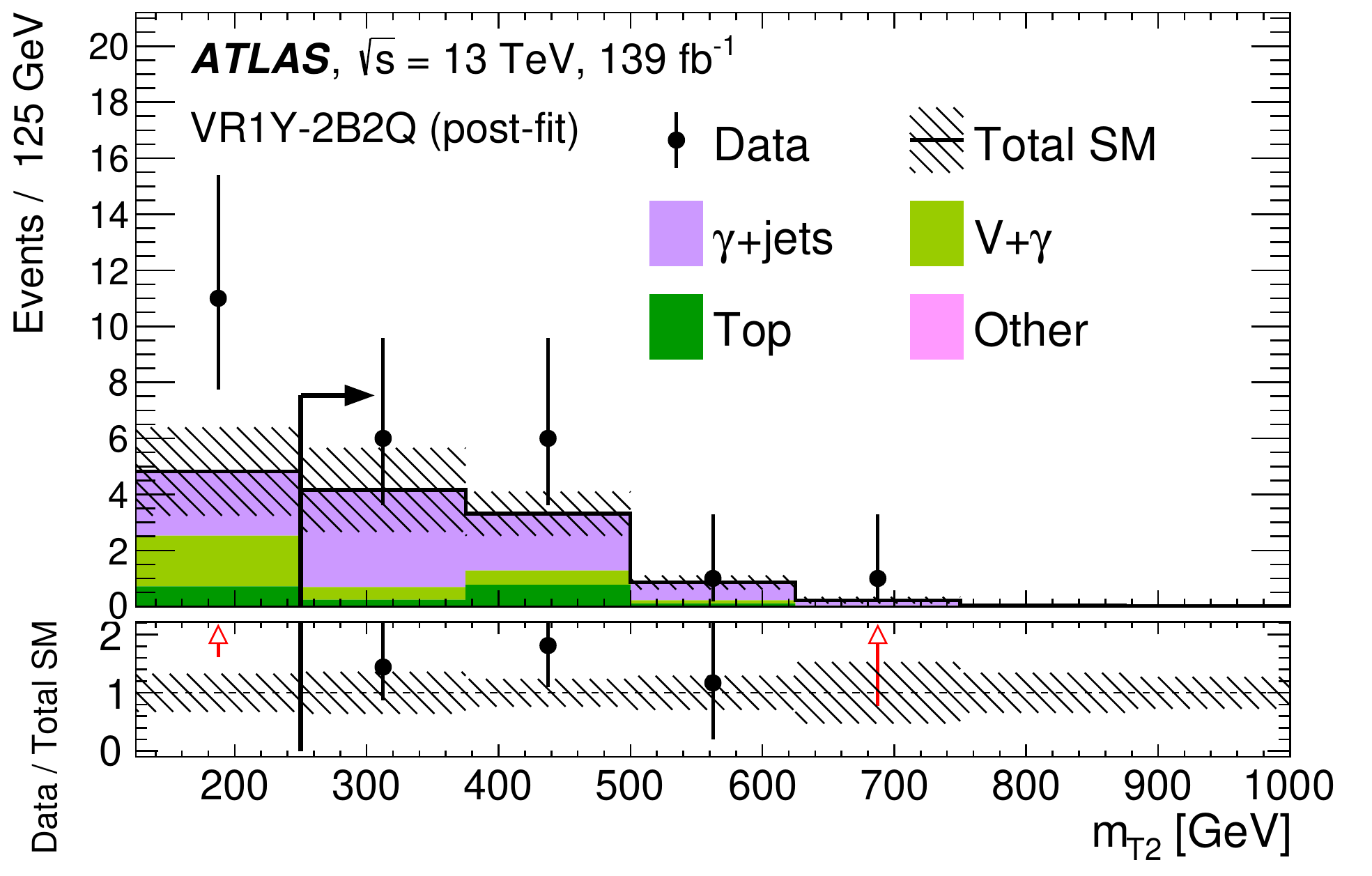}}
\caption{  \meff\ distributions in VR1L-4Q/VR1Y-4Q and \mTtwo\ distributions in VR1L-2B2Q/VR1Y-2B2Q.
The post-fit SM background expectation using the background-only fit is shown in a histogram stack.
``Top'' includes \ttbar, $t{+}X$ and \ttbarX production.
The hatched bands indicate the combined statistical and systematic uncertainty of the background.
The bottom panels show the ratio of the observed data to the background expectation.
The selection criterion for the variable shown by each plot is removed, while the arrow indicates the cut value used to define the region.
}
\label{fig:N-1plots_VR}
\end{figure}

 
\section{Systematic uncertainties}
\label{sec:systematics}
Uncertainties in the expected signal and background yields account for
the statistical uncertainties of the MC samples,
the experimental systematic uncertainties associated with the detector measurements and reconstruction,
and the theoretical systematic uncertainties in the MC simulation modeling.
For the signals and irreducible backgrounds, the uncertainties are assigned directly to the event yields.
For the reducible backgrounds, however, they are assigned to the SR(VR)/CR ratio (``transfer factor'', TF) as a consequence of the normalization performed in the CRs.
The reducible backgrounds are also subject to uncertainties due to the limited data sample size in the CRs.
Each systematic uncertainty is treated as fully correlated across the analysis regions but not across physics processes, unless explicitly stated otherwise.
A summary of the background prediction uncertainties is shown in Figure~\ref{fig:Systematics}.
The post-fit values are quoted after a background-only fit described in Section~\ref{sec:result}.
The MC statistical uncertainties give the largest contribution to the systematic uncertainty, mainly from the limited size of the \Znunu background sample used for the extrapolation.
However, this is not a limiting factor for the analysis sensitivity since the total uncertainty in the SRs is dominated by the statistical uncertainty due to the low number of data events in the SRs.
Details of the experimental and theoretical systematic uncertainties are described in the following subsections.
 
\subsection{Experimental uncertainties}
\label{sec:systematics:exp}
The first class of experimental uncertainties is related to the reconstruction and identification efficiencies for large-$R$ jets, small-$R$ jets, leptons and photons considered in the analysis.
These are assigned as uncertainties in the efficiency correction factors applied to the MC samples,
which correct for discrepancies between the efficiency predicted by MC simulation and the efficiency in data, as measured using dedicated control samples.
 
The uncertainty in $\Wqq(\Zqq)$-tagging contributes the most to this class of efficiency uncertainties.
It originates from the MC modeling uncertainty in the jet substructure variable distributions ($\mJ$, $D_2$ and \ntrk), as well as the precision of the efficiency determination in data.
The MC modeling uncertainty is evaluated by comparing samples from different MC generator configurations~\cite{ATL-PHYS-PUB-2020-017}.
The \Wqq-tagging efficiency in data is measured using $t\bar{t}$ for signal jets (large-$R$ jets that contain the $W \ra qq$ decays)
and QCD multijet/\gammajets for background jets,
following the prescription described in Ref.~\cite{ATL-PHYS-PUB-2020-017}.
The total uncertainty of the \Wqq-tagging efficiency correction factor ranges from 12\% to 23\% for signal jets, and from 7\% to 15\% for background jets, depending on the \pt.
The same efficiency correction factors are applied to the $\Zqq$-tagging selection,
with a simulation-based additional uncertainty of 4\%--5\% to account for differences between $W$ and $Z$ bosons.
 
The uncertainty in the $\Zbb(\hbb)$-tagging is obtained from the $b$-tagging efficiency uncertainty and the \Jbb mass distribution's shape uncertainty.
For the $b$-tagging uncertainty, 1\%--10\%, 15\%--50\%, and 50\%--100\% uncertainties are assigned to the MC correction factor for $b$-jets, $c$-jet mistagging, and light-flavor jet mistagging,
respectively, driven by theoretical uncertainties in the MC-simulated efficiency and the precision of the efficiency measurement in data.
The \Jbb mass scale uncertainty is estimated using the $R_{\textrm{trk}}$ method~\cite{JETM-2018-02},
and a relative scale uncertainty of 2\%--8\% is assigned depending on \ptJ\ and \mJ.
A 20\% relative uncertainty is assigned for the \Jbb mass resolution based on the variation in the simulation~\cite{STDM-2011-19}.
 
Other efficiency uncertainties related to triggering, identification, reconstruction,
and isolation requirements of electrons~\cite{EGAM-2018-01,TRIG-2018-05}, muons~\cite{MUON-2018-03_tmp} and photons~\cite{TRIG-2018-05}, and the jet vertex tagger selection for small-$R$ jets,
are found to be negligible.
 
The second class of experimental uncertainties is related to the energy (or momentum) determination for the reconstructed objects, namely large-$R$ jets~\cite{JETM-2018-02}, small-$R$ jets~\cite{PERF-2016-04},
electrons~\cite{EGAM-2018-01}, muons~\cite{PERF-2015-10} and photons~\cite{EGAM-2018-01}.
These typically come from the precision of simulation-based and in situ calibrations of the energy (or momentum) scale and resolution.
These per-object uncertainties are propagated through the \met calculation,
with additional uncertainties accounting for the scale and resolution of the soft term~\cite{PERF-2016-07}.
 
Additionally, an uncertainty in the integrated luminosity used to normalize the MC samples is considered.
A 1.7\% uncertainty is quoted for the combined 2015--2018 integrated luminosity obtained primarily using the LUCID-2 detector~\cite{Avoni:2018iuv}.
Finally, a pileup modeling uncertainty is assigned to account for the discrepancy between the predicted and measured inelastic cross-sections~\cite{STDM-2015-05}.

\subsection{Theoretical uncertainties}
Theoretical uncertainties in the main reducible backgrounds ($W/Z$+jets, \gammajets and $VV$ production) are estimated with varied generator parameters.
Uncertainties due to the choice of QCD renormalization and factorization scales are evaluated by varying them up and down by a factor of two relative to their nominal values~\cite{ATL-PHYS-PUB-2016-003}.
For $W/Z$+jets, the uncertainties related to the choice of CKKW merging scale are also considered.
These are assessed by shifting the merging scale to 15~\GeV\ or 30~\GeV\ from the default scale of 20~\GeV.
For the \ttbar\ background,
the nominal \POWPY[8] sample is compared with two alternative samples: one from \textsc{MadGraph5\_aMC@NLO} to estimate the hard-scatter modeling uncertainty,
and the other from \POWHEGBOX interfaced to \HERWIG[7.0.4]~\cite{Bellm:2015jjp} and H7UE set of tuned parameters~\cite{Bellm:2015jjp}
to assess the uncertainty due to the choice of parton shower scheme and hadronization model.
Variations in the \ttbar\ initial- and final-state radiation modeling, and renormalization and factorization scales, are also considered following the prescription described in Ref.~\cite{ATL-PHYS-PUB-2018-009}.
Uncertainties related to the choice of \NNPDF[3.0nnlo] PDF sets are assigned to the $W/Z$+jets, \gammajets, $VV$, \ttbar and single-top backgrounds.
These are derived by taking the envelope of the eigenvector variations from 100 propagated uncertainties.

For the reducible background estimation,
an additional uncertainty is assigned for the modeling of the relative background composition.
This is because different physics processes (\Zjets, \Wjets, \ttbar, etc.) are considered as a single component in the fits,
and therefore its composition is predicted solely by the simulation.
While the estimation is insensitive to the composition at first order since the TFs of those physics processes are similar,
the residual TF difference can cause a bias in the estimation
when the composition is significantly mismodeled by the simulation.
The impact of the potential composition mismodeling is evaluated by the variation in the combined TF
when shifting the normalization of each physics process up and down by a factor of two.
An uncertainty of about 2\%--8\% is assigned in the 4Q category,
while 7\%--10\% is assigned for those in the 2B2Q category.
 
For the irreducible backgrounds, cross-section uncertainties are assigned to account for their normalization.
A 50\% uncertainty is quoted for $VVV$ production based on the cross-section measurement by the CMS experiment~\cite{CMS-SMP-19-014}, while a 70\% normalization uncertainty is assigned to \ttbarX
based on the data/MC discrepancy observed in VRTTX, as discussed in Section~\ref{sec:VRTTX}.
 
\begin{figure}[thb!p]
\centering
\includegraphics[width=0.85\textwidth]{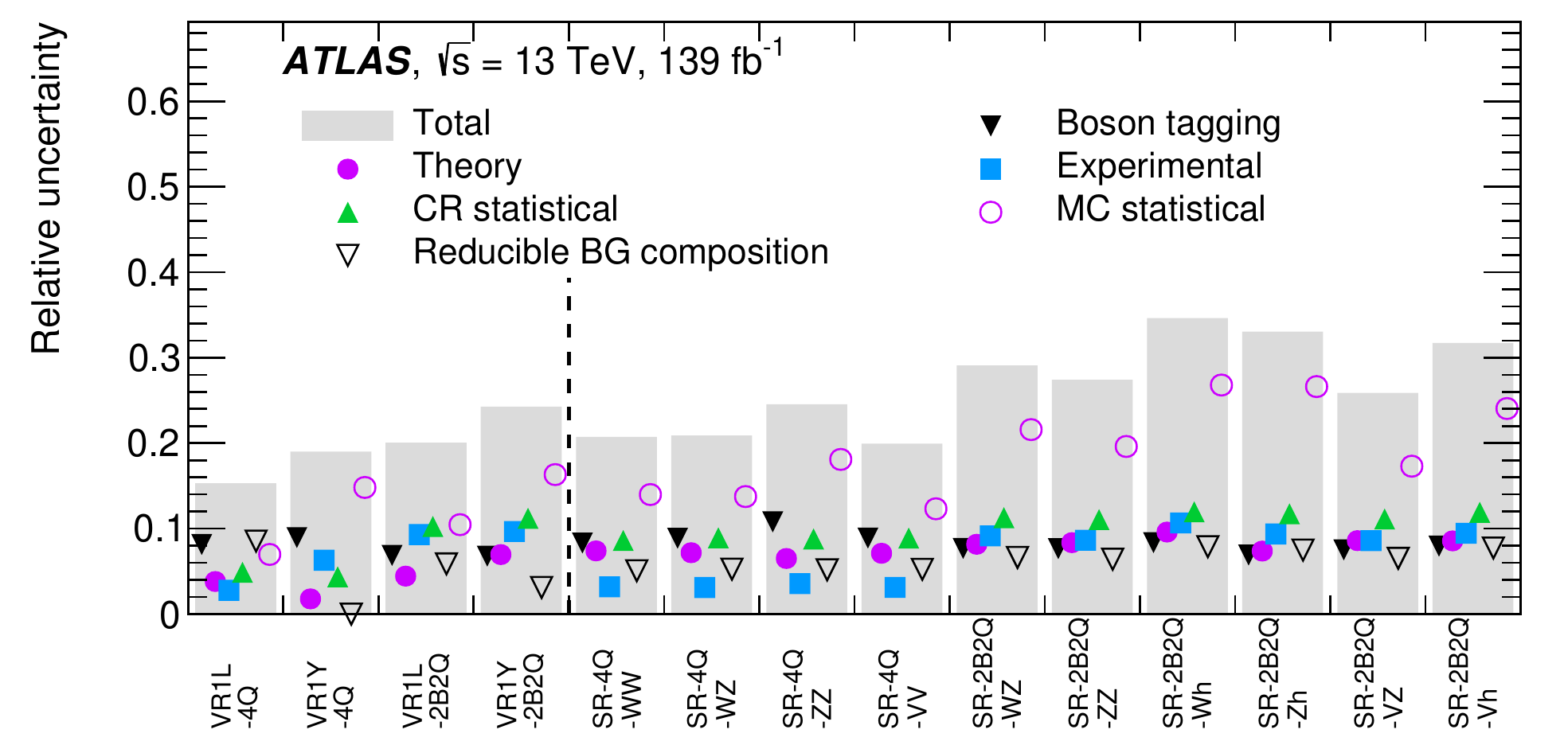}
\caption{The total post-fit uncertainty and the breakdown in each of the SRs and VRs.
``MC statistical'' and ``CR statistical'' present the statistical uncertainty due to the limited sample size either in MC simulation (mainly due to the \Znunu sample size) or in the CR data, respectively.
``Boson tagging'' indicates the uncertainty in the $\Wqq(\Zqq)$-tagging efficiency,
while ``experimental'' shows the contribution from the rest of the experimental uncertainties.
``Reducible BG composition'' refers to the uncertainty in the reducible background estimation due to the relative composition uncertainty.
``Theory'' represents the total uncertainty from all the theoretical uncertainties.}
\label{fig:Systematics}
\end{figure}
 
The uncertainty in the signal yields consists of the cross-section uncertainty and the shape uncertainties.
The cross-section uncertainty ranges from 6\% to 20\% for the production of electroweakinos with masses between $400~\GeV$ and $1~\TeV$,
driven mainly by the PDF uncertainty~\cite{Butterworth:2015oua}.
The shape uncertainties comprise uncertainties in the choice of renormalization/factorization scales and parton shower modeling,
affecting the signal acceptance by 5\%--10\%.
 
\section{Results}
\label{sec:result}
\subsection{Statistical analysis}
\label{sec:stat}
Final background estimates are obtained by performing a profile log-likelihood fit~\cite{Cowan:2010js_paperAndErr}
simultaneously in all CRs and SRs relevant to a given interpretation.
The \textsc{HistFitter}~\cite{Baak:2014wma} framework is employed.
Systematic uncertainties are treated as Gaussian-distributed nuisance parameters in the likelihood, while the statistical
uncertainties of the MC samples are treated as Poisson-distributed nuisance parameters.
 
Three types of fit configurations are used to derive the results.
\begin{itemize}
\item A ``\textit{background-only fit}'' is performed considering only the CRs and assuming no contribution from signals.
The normalization of the total reducible background is allowed to float and is constrained by the fit using the data in the CRs. The normalization factors and nuisance parameters are adjusted by maximizing the likelihood.
Three independent sets of fits are performed in the 0L, 1L, and 1Y categories, respectively.
The 4Q and 2B2Q regions in each category are fitted simultaneously, but with independent normalization factors assigned in 4Q and 2B2Q.
The normalization factors obtained from the fits range from 0.7 to 1.3.
\item A ``\textit{discovery fit}'' performs the hypothesis test for a generic beyond-the-SM (BSM) signal,
setting upper limits on the number of events and visible cross-section for the signal.
The fit uses only a single SR and the associated CR0L bin(s), constraining the backgrounds following the same method as in the background-only fit.
Any contribution from signals is allowed only in the SR, and the signal-strength parameter is defined to be strictly positive.
 
\item An ``\textit{exclusion fit}'' is performed to set the exclusion limit for a given signal model.
The SRs and the corresponding CR0L bins are fit simultaneously to determine the reducible background normalization factors and constrain the systematic uncertainties.
The signal contamination in CR0L is also taken into account according to the model predictions.
\end{itemize}
For each discovery or exclusion fit, the compatibility of the observed data with the background-only or signal-plus-background hypotheses is quantified by calculating a one-sided $p_0$-value with the profile likelihood ratio used as a test statistic~\cite{Cowan:2010js_paperAndErr}.
The upper limits and exclusion are derived using the \CLs prescription~\cite{Read:2002hq} where the 95\% confidence level (CL) exclusion is defined by \CLs $< 0.05$.
 
\subsection{Signal region yields}
The observed data yields in each SR and CR0L together with their SM background expectations are summarized in Table~\ref{tab:SR_yields} and shown in Figure~\ref{fig:SRfit}.
No significant excess is found in any of the SRs.
The distributions of \meff\ in SR-4Q-VV and \mTtwo\ in SR-2B2Q-VZ and SR-2B2Q-Vh are shown in Figure~\ref{fig:N-1plots}
with some representative signal samples overlaid to illustrate the sensitivity.
 
\begin{table}
\caption{Number of observed data events and the SM backgrounds in the SRs and the CR0L bins.
The SM backgrounds are predicted by the background-only fits. Negligible contributions are indicated by ``-''.
Within each CR the reducible backgrounds have the same relative uncertainty in their expected yield because a common normalization factor is assigned to all of them in the fit.
}
\centering{\small
\begin{tabular}{l cccccc}
\toprule
Region   & CR0L-4Q                      &  \multicolumn{1}{c|}{CR0L-2B2Q}        &  SR-4Q-WW       	& SR-4Q-WZ	        & SR-4Q-ZZ	                & SR-4Q-VV	   \\[-0.05cm]
\midrule
Observed & $129$                        &   \multicolumn{1}{c|}{$83$}            &  $2$	        & $3$	                & $1$	                        & $3$	       \\
\midrule
Post-fit & $129\pm 11~~$                &  \multicolumn{1}{c|}{$83\pm 9~~$}      & $1.9 \pm 0.4$      & $3.4\pm 0.7$          & $1.9 \pm 0.5$                 & $3.9 \pm 0.8$              \\
\midrule
 
$W$+jets & $24.2   \pm 2.2~~$           &  \multicolumn{1}{c|}{$16.6\pm 2.0~~$}  & $0.37 \pm 0.08$    & $0.60 \pm 0.13$       & $0.26 \pm 0.07$               & $0.69 \pm 0.15$         	 \\
$Z$+jets & $78     \pm 7~~$             &  \multicolumn{1}{c|}{$44\pm 5~~$}      & $~~1.0  \pm 0.21$  & $1.8 \pm 0.4$         & $1.26 \pm 0.32$               & $2.1 \pm 0.4$         	 \\
$VV$	 & $21.5   \pm 1.9~~$           &  \multicolumn{1}{c|}{$7.1\pm 0.9$}     & $0.35 \pm 0.11$    & $0.73 \pm 0.24$       & $0.26 \pm 0.09$               & $0.79 \pm 0.25$         	 \\
$VVV$	 & $0.9    \pm 0.4$             &  \multicolumn{1}{c|}{$0.10\pm 0.05$}   & $0.17 \pm 0.09$    & $0.19 \pm 0.10$       & $0.11 \pm 0.07$               & $0.23 \pm 0.12$         	 \\
\ttbar	 & $1.38   \pm 0.12$            &  \multicolumn{1}{c|}{$7.8\pm 0.9$}     & $0.039 \pm 0.009$  & $0.060 \pm 0.018$     & $0.025 \pm 0.010$             & $0.063 \pm 0.018$         	 \\
$t{+}X$	 & $1.32   \pm 0.12$            &  \multicolumn{1}{c|}{$2.87\pm 0.34$}   & $0.015 \pm 0.006$  & $0.039 \pm 0.016$     & $0.012 \pm 0.005$             & $0.039 \pm 0.016$         	 \\
\ttbarX	 & $1.3    \pm 0.9$             &  \multicolumn{1}{c|}{$3.7\pm 2.6$}     & -                   	& -         		& -                           & -     	 \\
Other	 & $<0.1$    			&  \multicolumn{1}{c|}{$0.95\pm 0.11$}   & $<0.001$     	& $<0.001$    		& $<0.001$    			& $<0.001$             	 \\
\midrule
\midrule
Region   & SR-2B2Q-WZ	        & SR-2B2Q-Wh	                & SR-2B2Q-ZZ	        & SR-2B2Q-Zh	                & SR-2B2Q-VZ            & SR-2B2Q-Vh  \\[-0.05cm]
\midrule
Observed & 2	                & 0                     	& 2	                & 1                             & 2                     & 1   \\
\midrule
Post-fit & $1.6   \pm 0.4$      & $1.9 \pm 0.7$                 & $1.7 \pm 0.5$         & $1.6 \pm 0.5$                 & $2.2 \pm 0.6$         & $2.5 \pm 0.8$ \\
\midrule
$W$+jets & $0.11  \pm 0.06$     & $0.24 \pm 0.09$               & $0.23 \pm 0.08$       & $0.26 \pm 0.10$               & $0.26 \pm 0.09$       & $0.26 \pm 0.09$    \\
$Z$+jets & $0.84  \pm 0.27$     & $1.3  \pm 0.5$                & $0.78 \pm 0.23$       & $0.66 \pm 0.24$               & $1.15 \pm 0.33$       & $1.4 \pm 0.5$    \\
$VV$	 & $0.33  \pm 0.11$     & $0.09 \pm 0.03$               & $0.32 \pm 0.10$       & $0.085 \pm 0.032$             & $0.37 \pm 0.11$       & $0.085 \pm 0.030$    \\
$VVV$	 & $0.047 \pm 0.027$    & $<0.01$   			& $0.051 \pm 0.032$     & $0.011 \pm 0.007$             & $0.06 \pm 0.04$       & $0.011 \pm 0.007$    \\
\ttbar	 & $0.016 \pm 0.006$    & $0.13 \pm 0.04$               & $0.064 \pm 0.019$     & $0.40 \pm 0.16$               & $0.072 \pm 0.021$     & $0.46 \pm 0.18$   \\
$t{+}X$	 & $0.11  \pm 0.05$     & $0.07 \pm 0.04$               & $0.11 \pm 0.05$       & $0.041 \pm 0.022$             & $0.11 \pm 0.05$       & $0.10 \pm 0.05$       \\
\ttbarX	 & $0.10  \pm 0.08$     & $0.07_{-0.07~~}^{+0.10~~}$    & $0.14 \pm 0.12$       & $0.08_{-0.08~~}^{+0.09~~}$    & $0.18 \pm 0.14$       & $0.10_{-0.10~~}^{+0.11~~}$          \\
Other	 & $<0.01$   		& $0.03 \pm 0.01$               & $<0.01$   		& $0.024 \pm 0.008$    		& $<0.01$   		& $0.037 \pm 0.011$       \\
\bottomrule
\end{tabular}
}
\label{tab:SR_yields}
\end{table}
 
 
\begin{figure}[h]
\centering
\includegraphics[width=0.85\textwidth]{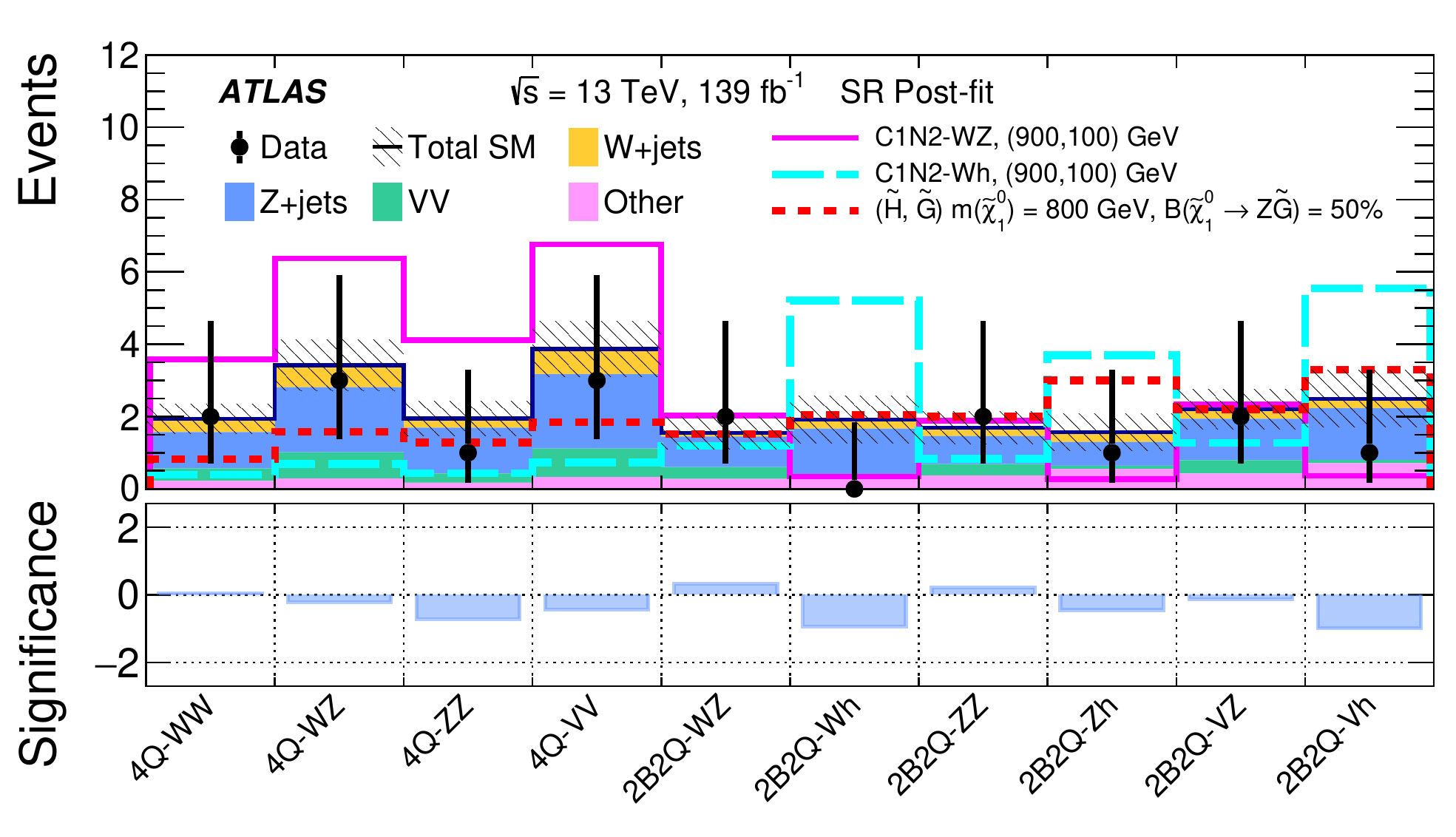}
\caption{Summary of the observed data and predicted SM background in all SRs.
The background prediction in SR-4Q (SR-2B2Q) is obtained by a background-only fit to CR0L-4Q (CR0L-2B2Q).
The total systematic uncertainty in the background prediction is shown by the hatched area.
Distributions of a few representative signals are overlaid.
For the \winoBinoSim models, the label $(900, 100)~\GeV$ indicates $(\mCone, \mNone)$.
The bottom panel shows the statistical significance of the discrepancy between the observed number of events and the SM expectation, following the prescription described in Ref.~\cite{Cousins:2007bmb}.
}
\label{fig:SRfit}
\end{figure}
 
\begin{figure}[h]
\centering
\subfloat[SR-4Q-VV]{\includegraphics[width=0.48\textwidth]{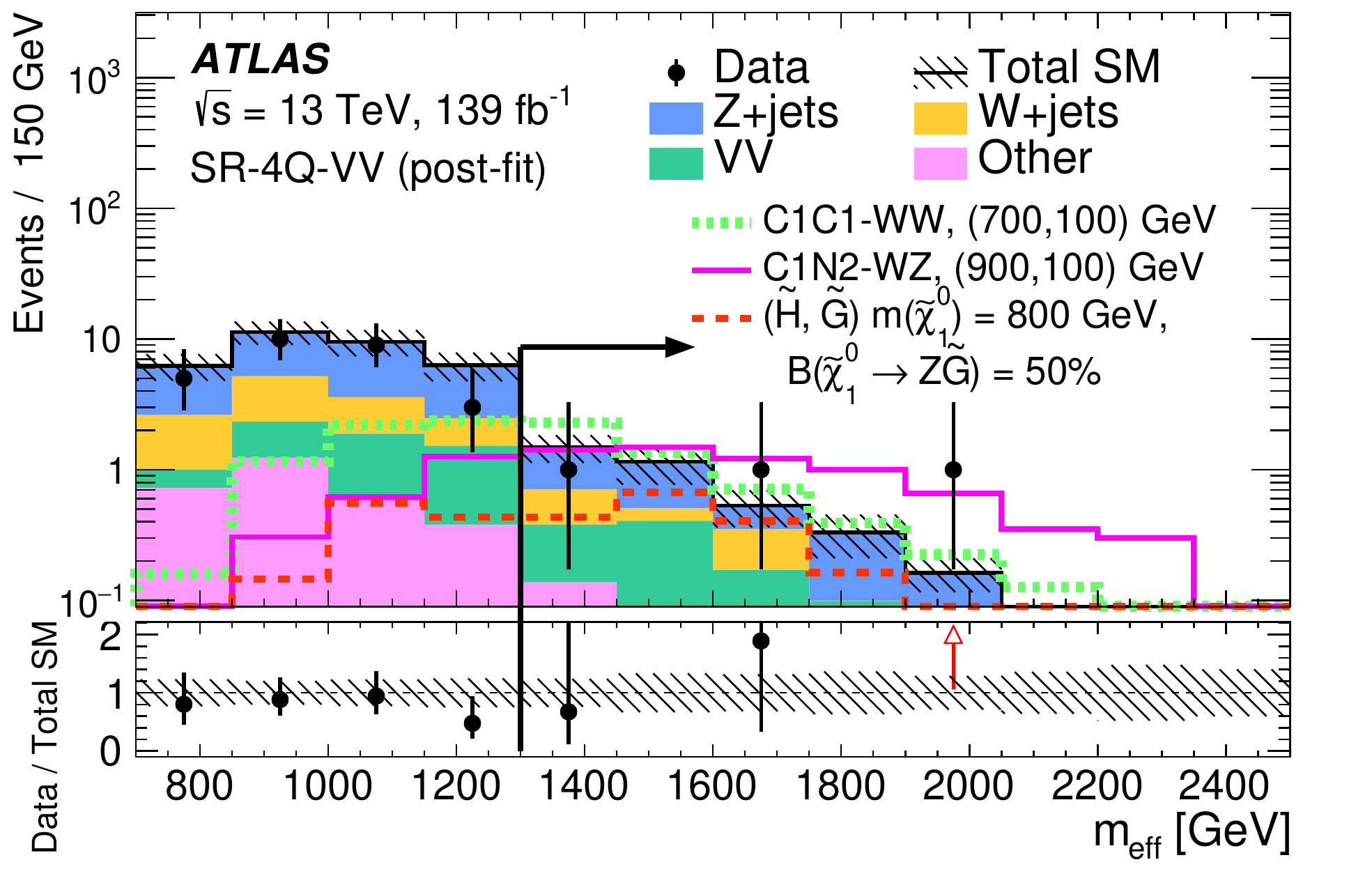}} \\
\subfloat[SR-2B2Q-VZ]{\includegraphics[width=0.48\textwidth]{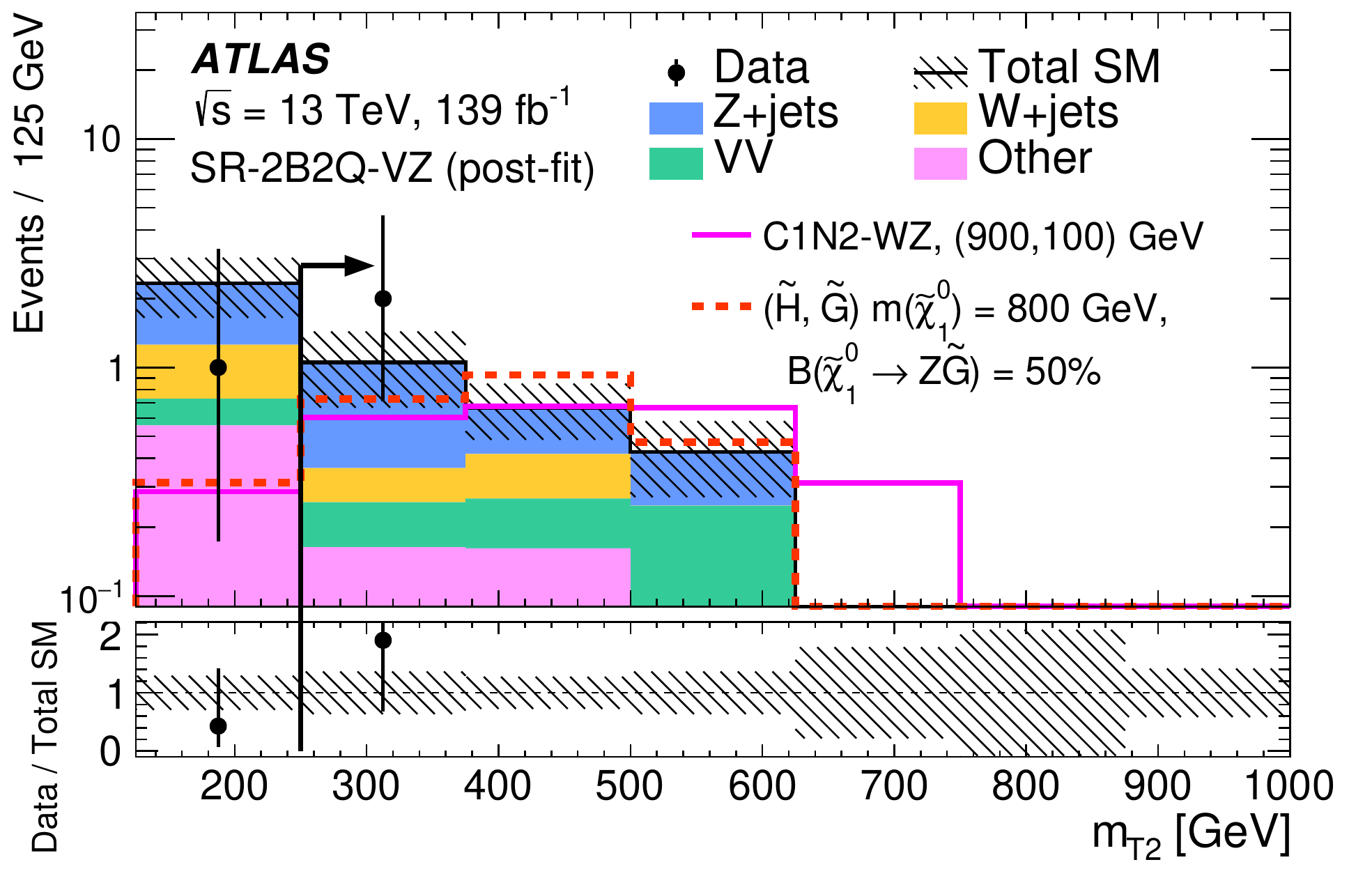}}
\subfloat[SR-2B2Q-Vh]{\includegraphics[width=0.48\textwidth]{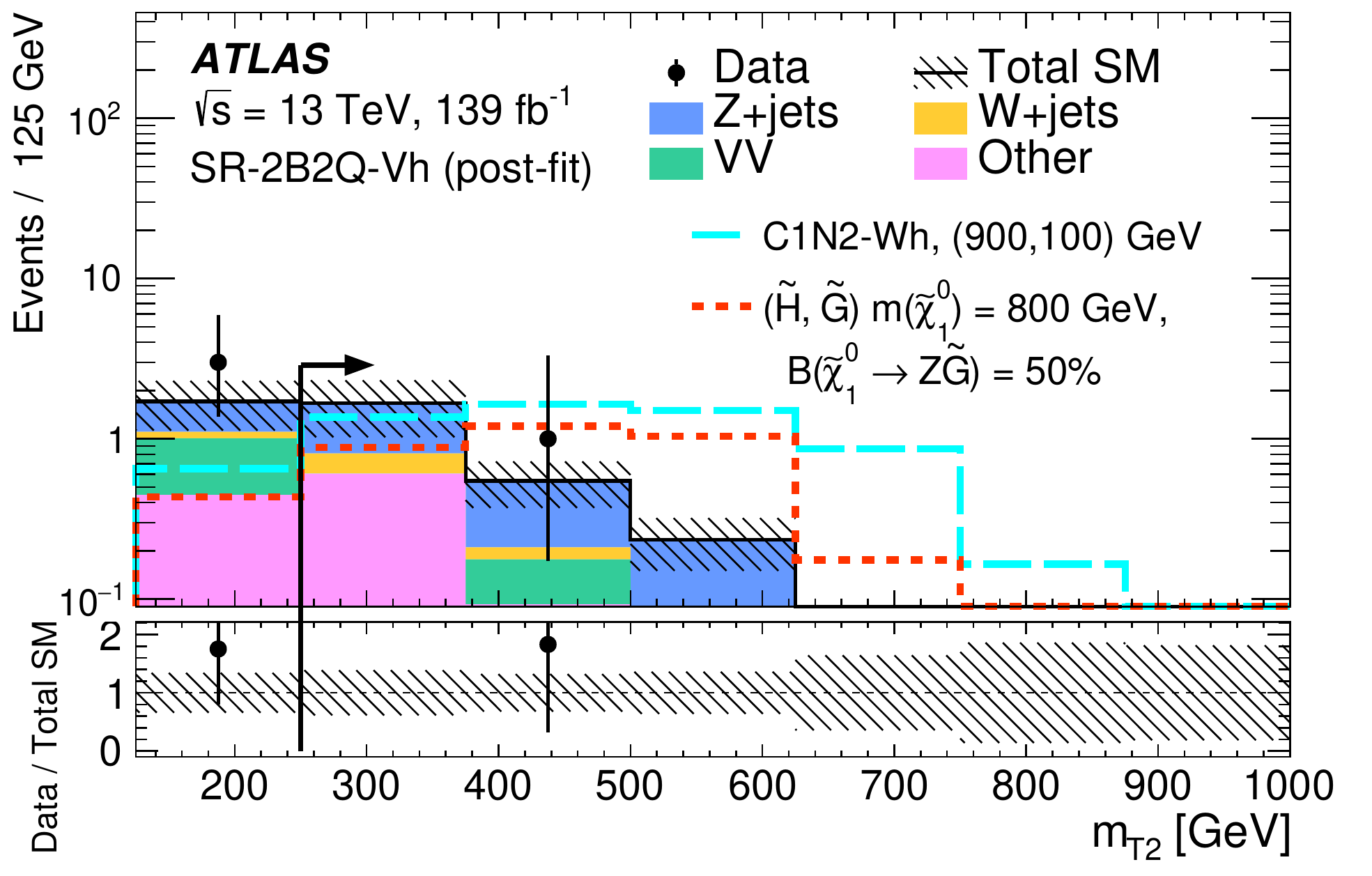}}
\caption{(a) \meff\ distribution in SR-4Q-VV. (b)-(c) \mTtwo\ distributions in SR-2B2Q-VZ and SR-2B2Q-Vh.
The post-fit SM background expectation using the background-only fit is shown in a histogram stack.
Distributions of a few representative signals are overlaid.
The bottom panels show the ratio of the observed data to the background prediction.
The selection criterion for the variable shown by each plot is removed, while the arrow indicates the cut value used to define the region.
For the \winoBinoSim models, the labels $(700, 100)~\GeV$ and $(900, 100)~\GeV$ indicate $(\mCone, \mNone)$.
}
\label{fig:N-1plots}
\end{figure}

\clearpage
\subsection{Model-independent upper limits}
A discovery fit is performed for each SR
to derive the expected and observed 95\% CL upper limits on the number of BSM signal events ($S_{\textrm{exp}}^{95}$ and $S_{\textrm{obs}}^{95}$) as well as the one-sided $p$-value ($p_{0}$) of the background-only hypothesis.
Pseudo-experiments with toy MC are used for the calculation.
An upper limit on the cross-section, $\left<\epsilon \sigma \right>^{95}_{\textrm{obs}}$ where $\epsilon$ represents the efficiency times acceptance of the SR for the given signal,
is obtained by dividing $S_{\textrm{obs}}^{95}$ by the integrated luminosity.
The upper limits and the $p_{0}$ value associated with each SR are summarized in Table~\ref{table:results:exclxsec.pval.upperlimit_after_unblind}.
Two additional ``discovery SRs'' are defined in order to set model-independent upper limits in the inclusive phase space.
First, Disc-SR-2B2Q is defined as the logical union of SR-2B2Q-VZ and SR-2B2Q-Vh,
and then the inclusive discovery signal region Disc-SR-Incl is defined as the logical union of SR-4Q-VV and Disc-SR-2B2Q.
When evaluating Disc-SR-Incl, both CR0L-4Q and CR0L-2B2Q are included in the simultaneous fit and each has
its own floating normalization factor for the reducible backgrounds.
 
\begin{table}
\centering
\setlength{\tabcolsep}{0.0pc}
\caption[Breakdown of upper limits.]{
Left to right: 95\% CL upper limits on the visible cross-section
($\langle\epsilon\sigma\rangle_{\mathrm{obs}}^{95}$).
$S_{\mathrm{obs}}^{95}$ ($S_{\mathrm{exp}}^{95}$) shows the 95\% CL upper limit on the number of
signal events, given the observed number (expected number and $\pm 1\sigma$
excursions) of background events.
The last two columns indicate the \CLb value and the discovery $p$-value ($p(s=0)$) with the corresponding Gaussian significance ($Z$).
\CLb provides a measure of compatibility of the observed data with the 95\% CL signal strength hypothesis relative to fluctuations of the background,
and $p(s=0)$ measures compatibility of the observed data with the background-only (zero signal strength) hypothesis relative to fluctuations of the background.
Larger values indicate greater relative compatibility. $p(s=0)$ is not calculated in signal regions with a deficit with respect to the nominal background prediction.
\label{table:results:exclxsec.pval.upperlimit_after_unblind}}
\begin{tabular*}{\textwidth}{@{\extracolsep{\fill}}lccccc}
\toprule
Signal region                        & $\langle\epsilon{\mathrm \sigma}\rangle_{\mathrm{obs}}^{95}$ [fb]  &  $S_{\mathrm{obs}}^{95}$  & $S_{\mathrm{exp}}^{95}~(\pm 1\sigma)$ & \CLb & $p(s=0)$ ($Z$)  \\
\midrule
SR-4Q-WW    & $0.032$ & $4.5$ & $ { 4.2 }^{ +1.8 }_{ -1.0 }$ & $0.55$ & $ 0.44$~$(0.15)$ \\
SR-4Q-WZ    & $0.036$ & $5.0$ & $ { 5.1 }^{ +2.1 }_{ -1.3 }$ & $0.46$ &  -               \\
SR-4Q-ZZ    & $0.025$ & $3.6$ & $ { 4.1 }^{ +1.8 }_{ -1.0 }$ & $0.30$ &  -               \\
SR-4Q-VV    & $0.034$ & $4.7$ & $ { 5.3 }^{ +2.3 }_{ -1.5 }$ & $0.38$ &  -  \tabularnewline 
\midrule
SR-2B2Q-WZ    & $0.033$ & $4.7$ & $ { 4.0 }^{ +1.7 }_{ -0.7 }$ & $0.66$ & $ 0.33$~$(0.44)$ \\
SR-2B2Q-Wh    & $0.022$ & $3.1$ & $ { 3.9 }^{ +1.3 }_{ -0.7 }$ & $0.28$ &  -               \\
SR-2B2Q-ZZ    & $0.033$ & $4.5$ & $ { 4.1 }^{ +1.7 }_{ -0.9 }$ & $0.63$ & $ 0.37$~$(0.32)$ \\
SR-2B2Q-Zh    & $0.026$ & $3.6$ & $ { 3.9 }^{ +1.4 }_{ -0.7 }$ & $0.38$ &  -               \\
SR-2B2Q-VZ    & $0.032$ & $4.4$ & $ { 4.4 }^{ +1.8 }_{ -1.0 }$ & $0.50$ &  -               \\
SR-2B2Q-Vh    & $0.026$ & $3.6$ & $ { 4.4 }^{ +1.7 }_{ -1.0 }$ & $0.24$ &  -  \tabularnewline 
\midrule
Disc-SR-2B2Q    & $0.034$ &  $4.8$ & $ { 5.6 }^{ +2.4 }_{ -1.6 }$ & $0.30$ &  -  \\
Disc-SR-Incl    & $0.042$ &  $5.9$ & $ { 7.2 }^{ +2.2 }_{ -2.0 }$ & $0.27$ &  -  \\
\bottomrule
\end{tabular*}
\end{table}


\clearpage
\subsection{Model-dependent exclusion limits}
The results are also interpreted in the context of the specific signal models discussed in Section~\ref{sec:models}.
An exclusion fit is performed for each point in the model space,
and a \CLs value is assigned based on the hypothesis test.
The expected and observed 95\% CL exclusion regions correspond to values of \CLs $< 0.05$.
Given the large number of models tested,
an asymptotic approximation~\cite{Cowan:2010js_paperAndErr} is employed in the \CLs calculation
instead of the full calculation using pseudo-experiments.
The validity is checked and the \CLs values of the two methods typically agree within 5\%, and maximally within 10\%.
 
The SRs participating in the simultaneous fit vary with the signal model being tested,
as summarized in Table~\ref{tab:result:models_vs_SRs}.
CR0L-4Q~(CR0L-2B2Q) is included in the fit when at least one SR-4Q~(SR-2B2Q) bin is used in deriving the limit,
while both of the CR0L bins are included when the SR bins from both SR-4Q and SR-2B2Q participate in the fit.

\subsubsection{Exclusion limits on \winoBino, \hinoBino, \winoHino, and \hinoWino models}
\label{sec:excl:winoBinoHino}
The exclusion limits on the \winoBino and \hinoBino models
are summarized in Figure~\ref{fig:binoLSPExcl}.
As discussed in Section~\ref{sec:binoWinoHino},
different $\mathcal{B}(\Ntwo \ra Z\None)$ hypotheses are examined,
and these are shown to result in very similar limits in Figures~\ref{fig:binoLSPExcl:exp} and ~\ref{fig:binoLSPExcl:obs}.
 
The expected and observed limits for a representative slice $\mathcal{B}(\Ntwo \ra Z\None)=50\%$
are shown in Figures~\ref{fig:binoLSPExcl:winoBino} and \ref{fig:binoLSPExcl:hinoBino},
for \winoBino and \hinoBino models respectively.
 
\begin{figure}[h]
\centering
\subfloat[Expected limits with various $\mathcal{B}(\Ntwo \ra Z\None)$]{\includegraphics[width=0.485\textwidth]{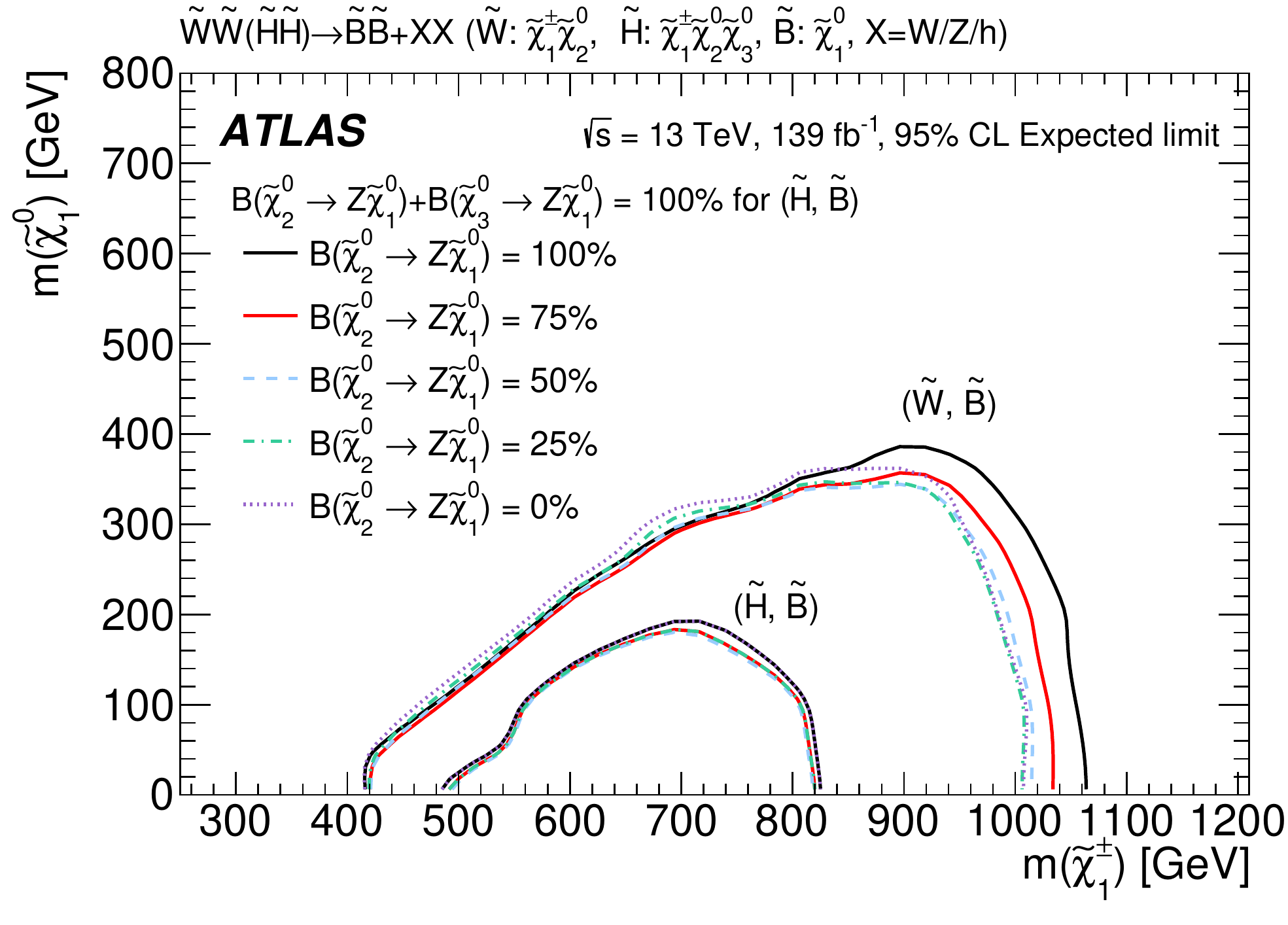} \label{fig:binoLSPExcl:exp}}
\subfloat[Observed limits with various $\mathcal{B}(\Ntwo \ra Z\None)$]{\includegraphics[width=0.485\textwidth]{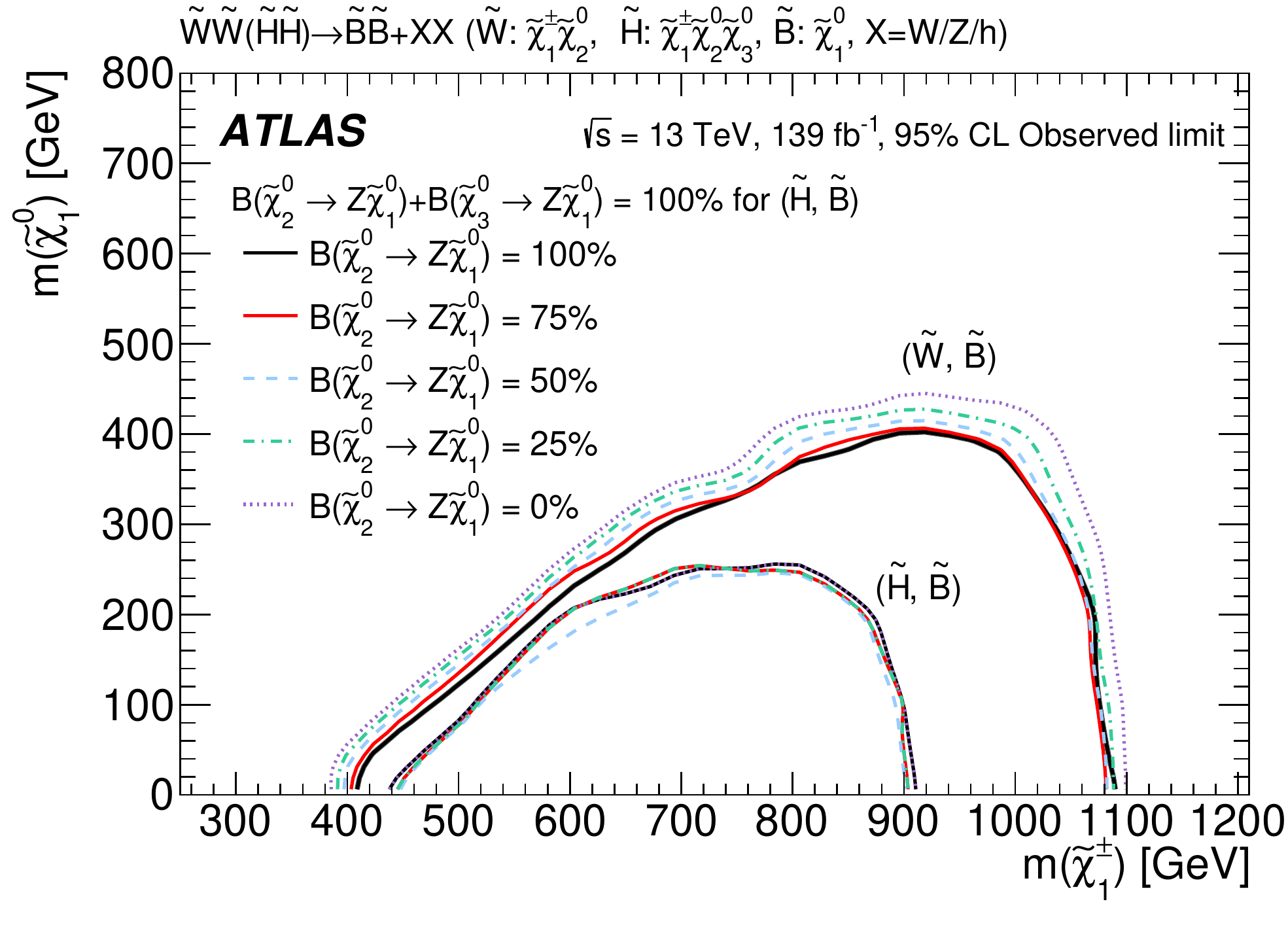} \label{fig:binoLSPExcl:obs}} \\
\subfloat[$(\wino,\bino)$ with $\mathcal{B}(\Ntwo \ra Z\None)=50\%$]{\includegraphics[width=0.485\textwidth]{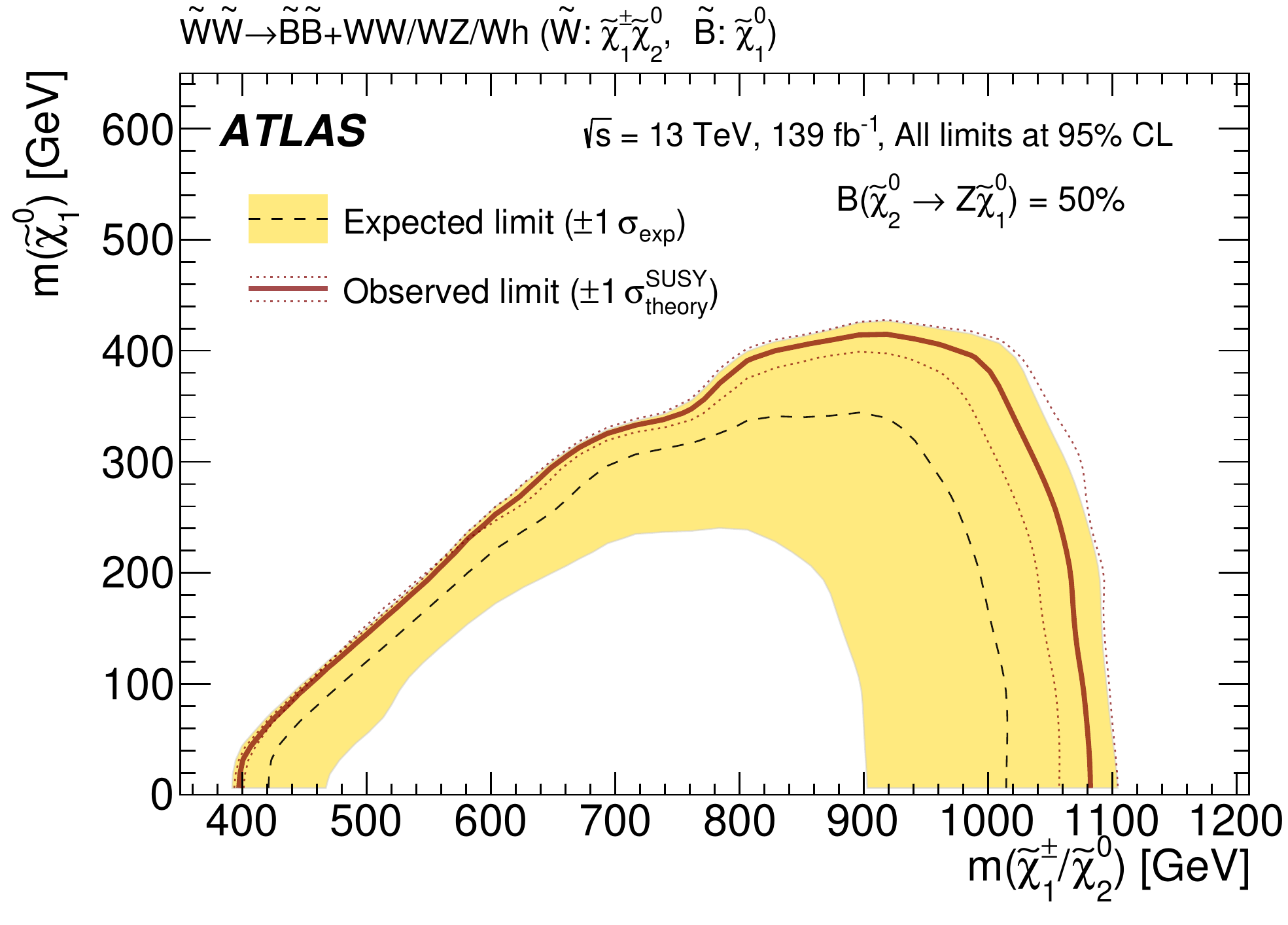} \label{fig:binoLSPExcl:winoBino}}
\subfloat[$(\hino,\bino)$ with $\mathcal{B}(\Ntwo \ra Z\None)=50\%$]{\includegraphics[width=0.485\textwidth]{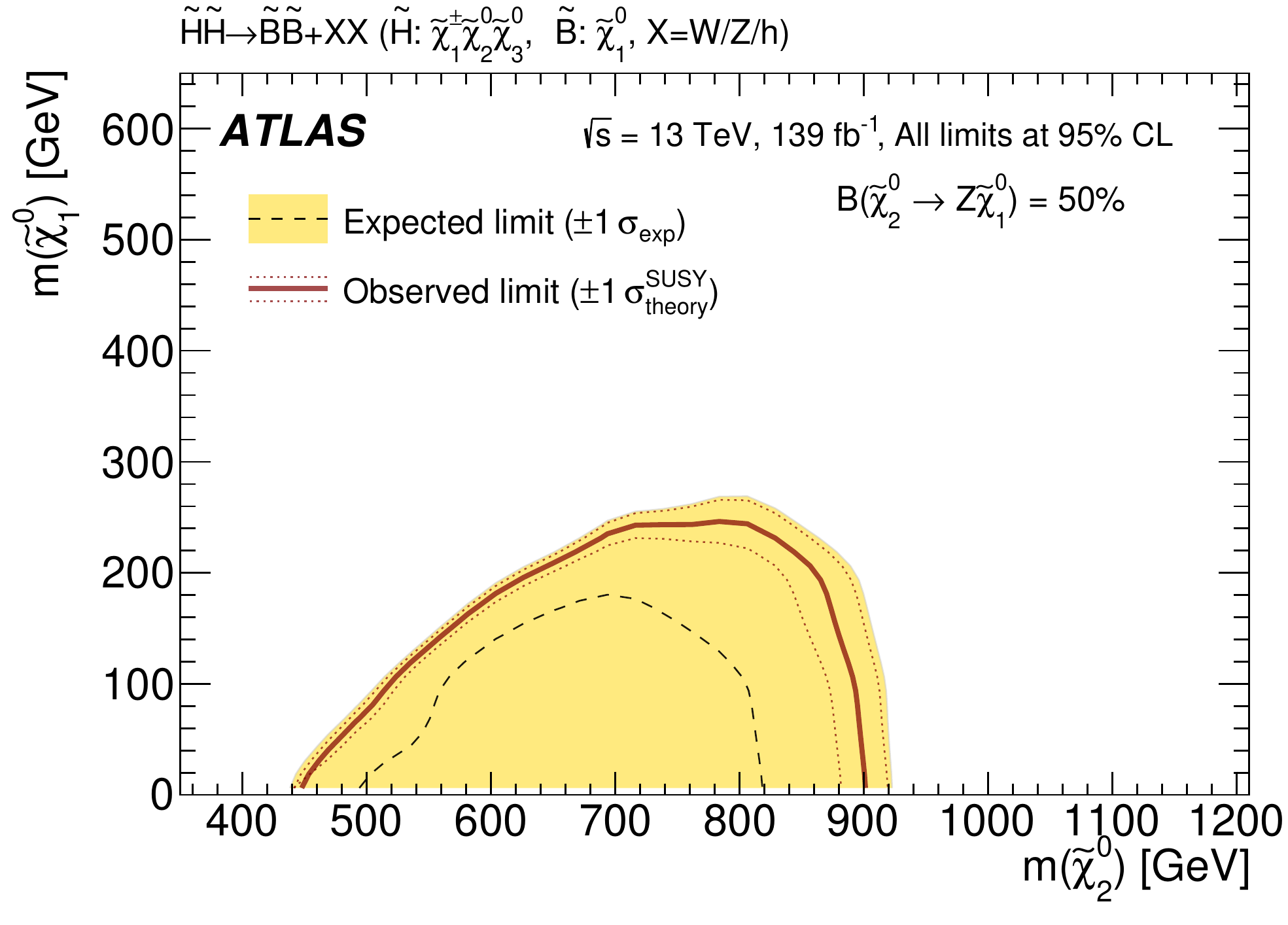} \label{fig:binoLSPExcl:hinoBino}}
\caption{Exclusion limits for the \winoBino and \hinoBino models shown as a function of the wino/higgsino chargino mass \mCone and the bino LSP mass \mNone.
The (a) expected and (b) observed limits for various $\mathcal{B}(\Ntwo \ra Z \None)$ hypotheses are overlaid.
The outer and inner bundles correspond to the limits for the \winoBino and \hinoBino models respectively.
The limits set on the \hinoBino models are highly consistent and thus the contour lines are highly overlapped.
Expected (dashed) and observed (solid red) 95\% CL exclusion limits are shown for the (c) \winoBino and (d) \hinoBino models with a representative branching ratio $\mathcal{B}(\Ntwo \ra Z\None)=50\%$.
}
\label{fig:binoLSPExcl}
\end{figure}
 
The \hinoBino limits are also interpreted for the
$Z/h$-funnel dark matter model described in Section~\ref{sec:binoWinoHino}.
The exclusion limits are shown in Figure~\ref{fig:hinoBino_Zh_funnel},
overlaid with theoretical predictions of higgsino masses (\Ntwo)
reproducing the observed dark matter relic density ($\Omega h^2=0.12$)~\cite{Hamaguchi:2015rxa}.
While the relic density depends drastically on \tanb,
the exclusion limits obtained by the search are assumed to be constant along \tanb since they do not change when varying $\mathcal{B}(\Ntwo \ra Z\None)$,
and are interpreted from the exclusion limits for the $\mathcal{B}(\Ntwo \ra Z\None)=50\%$ hypothesis in Figure~\ref{fig:binoLSPExcl:hinoBino}.
For the $h$-funnel case, where $m(\None)=m_h/2$, the excluded regions are $\tanb>8.5$ for $\mu>0$ and $5.5 < \tanb < 7$ for $\mu<0$.
 
\begin{figure}[h]
\centering
\subfloat[$m(\None)=m_Z/2$~($Z$-funnel)]{\includegraphics[width=0.485\textwidth]{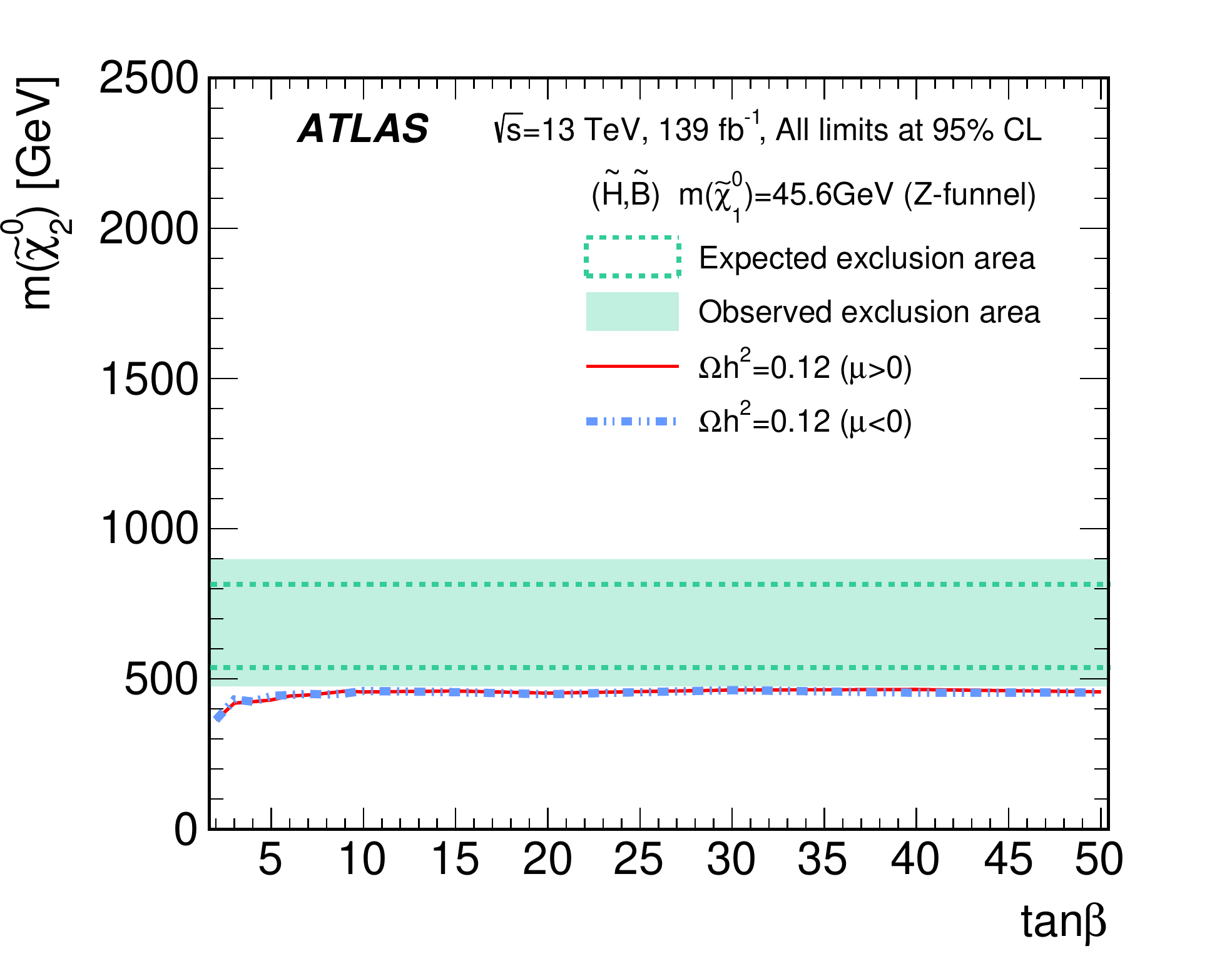}}
\subfloat[$m(\None)=m_h/2$~($h$-funnel)]{\includegraphics[width=0.485\textwidth]{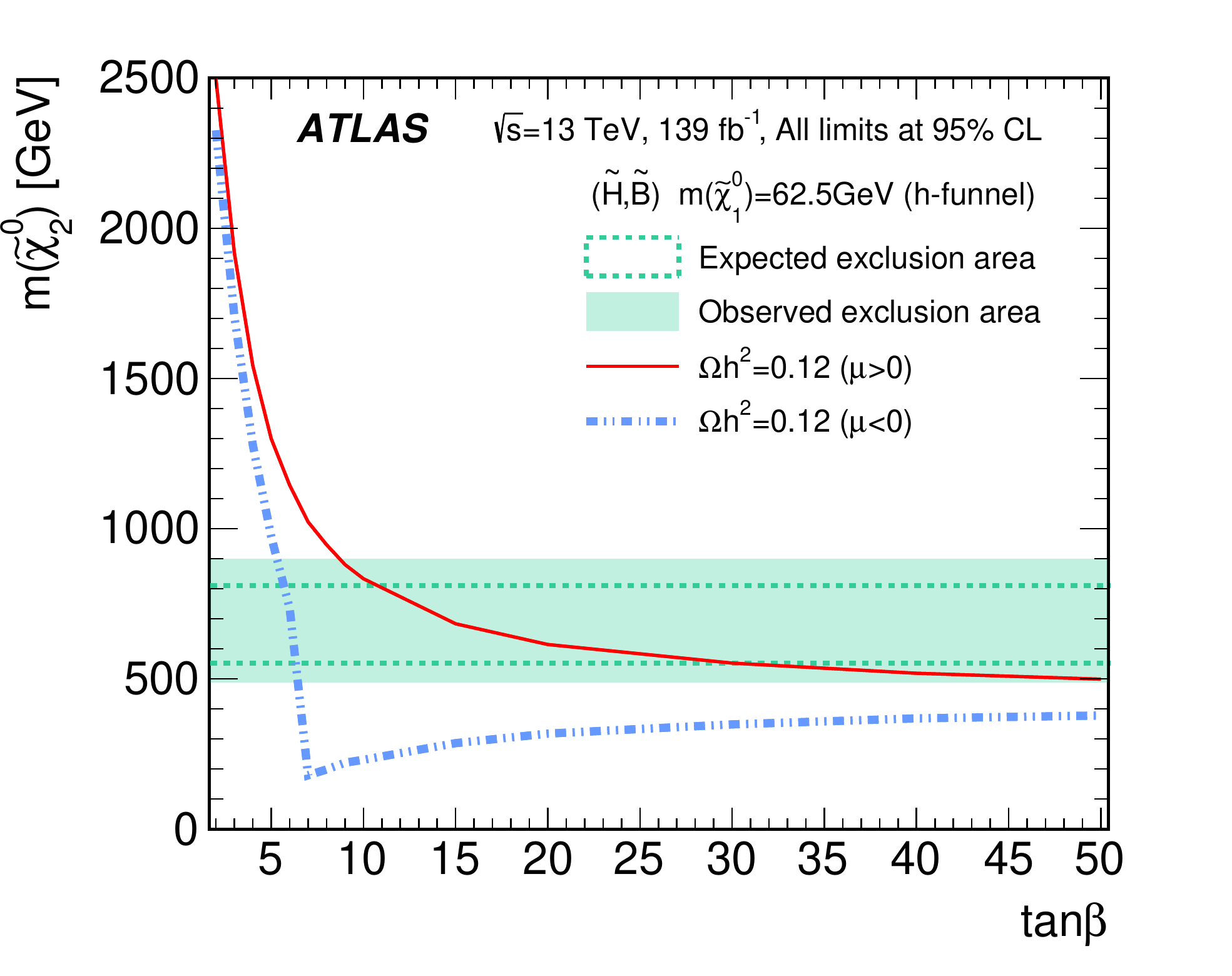}}
\caption{
Expected (dashed green lines) and observed (green band) 95\% CL exclusion limits on the $Z/h$-funnel dark matter model described in Section~\ref{sec:binoWinoHino},
where the mass of bino-like LSP (\None) is (a) half of the $Z$ boson mass~($42.6~\GeV$) or (b) half of the $h$ boson mass~($62.5~\GeV$).
The areas surrounded by the green bands or dashed lines represent the excluded range of \Ntwo.
The overlaid red solid (blue dashed) line indicates for $\mu>0$ ($\mu<0$) the \Ntwo mass that reproduces the observed dark matter relic density ($\Omega h^2=0.12$) as function of \tanb~\cite{Hamaguchi:2015rxa};
below (above) these lines the predicted dark matter relic density is too small (too large).
}
\label{fig:hinoBino_Zh_funnel}
\end{figure}

The exclusion limits set on the \winoHino and \hinoWino models
are evaluated in a three-dimensional model space defined by $(M_2,\mu,\tanb)$.
For each model point, the mass spectra and the branching ratios are determined using the prescription described in Section~\ref{sec:binoWinoHino}.
Figure~\ref{fig:winoHinoExcl:tanb10} shows the expected limit and observed limit as a function of $(M_2,\mu)$ with a fixed $\tanb=10$.
Figure~\ref{fig:winoHinoExcl:expAndObs_M2_mu} shows that varying \tanb or the sign of $\mu$ (\sgnMu) has very little effect on the sensitivity.
The limits are also interpreted as a function of the physical electroweakino masses so that they can be directly compared with the other models.
For a given set of $(\tanb,\sgnMu)$,
a pair of $(M_2,|\mu|)$ can be projected one-to-one to $(m(\Ctwo),m(\None))$
when all the other MSSM parameters are fixed.
Figure~\ref{fig:winoHinoExcl:winoHino} and Figure~\ref{fig:winoHinoExcl:hinoWino} show the limits for the \winoHino model ($M_2>|\mu|$) and \hinoWino model ($M_2<|\mu|$), respectively, assuming $\tanb=10$ and $\mu>0$.
 
\begin{figure}[h]
\centering
\subfloat[Limits vs.\ $(M_2,\mu)$ with $\tanb=10$]{\includegraphics[width=0.485\textwidth]{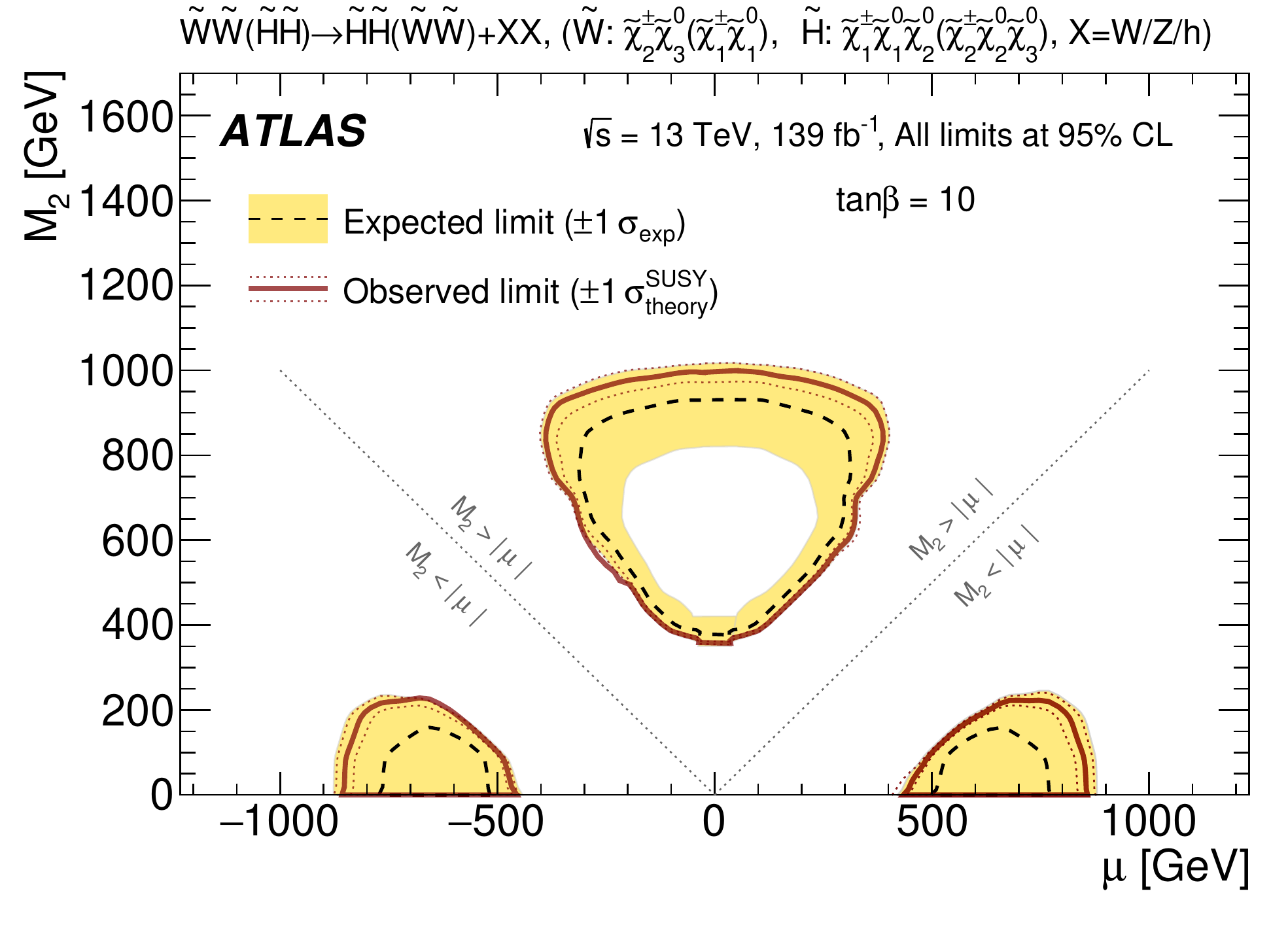} \label{fig:winoHinoExcl:tanb10}}
\subfloat[Limits vs.\ $(M_2,\mu)$ with varying \tanb]{\includegraphics[width=0.485\textwidth]{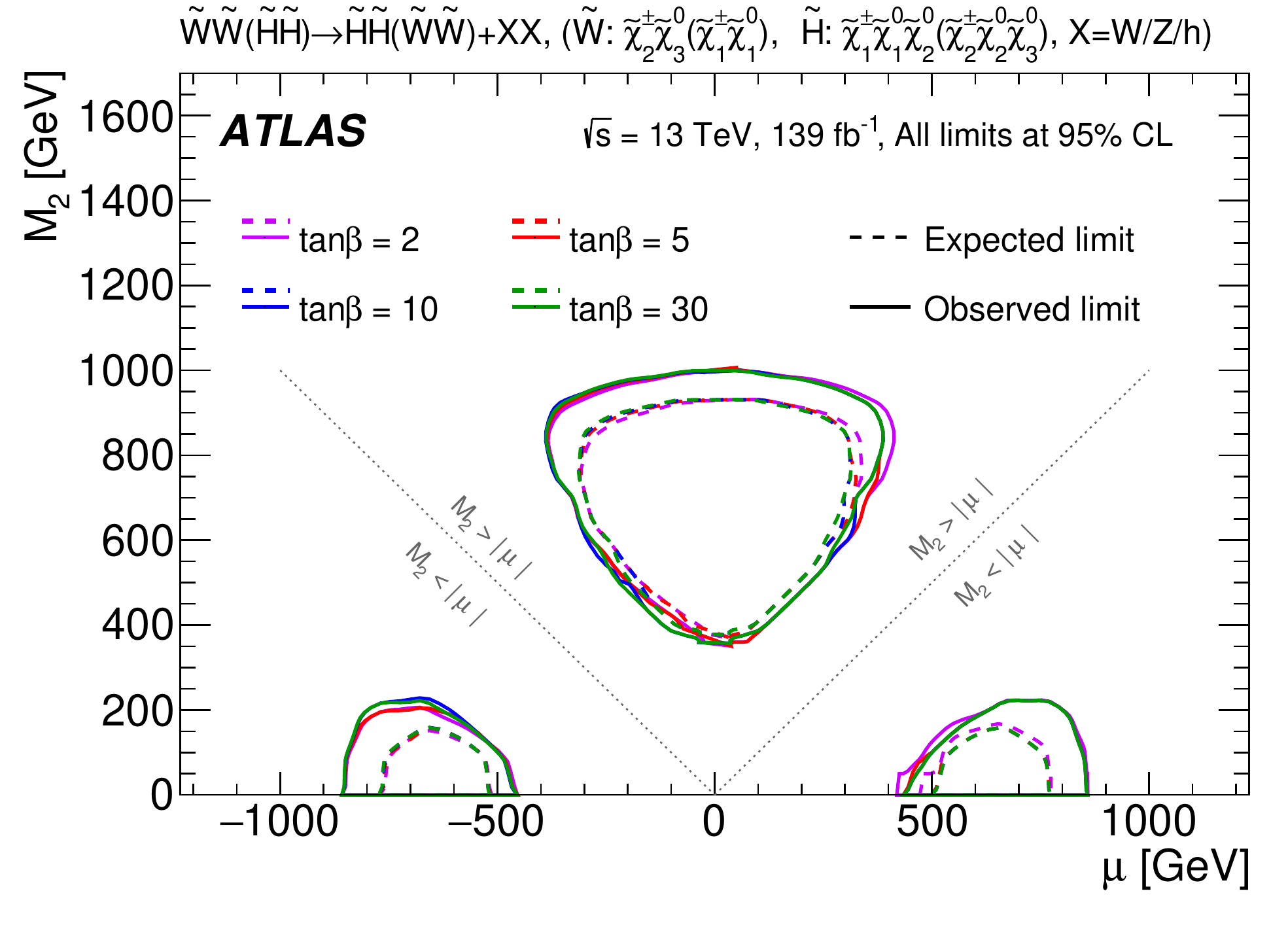} \label{fig:winoHinoExcl:expAndObs_M2_mu}} \\
\subfloat[$(\wino,\hino)$ with $\tanb=10, \mu>0$]{\includegraphics[width=0.485\textwidth]{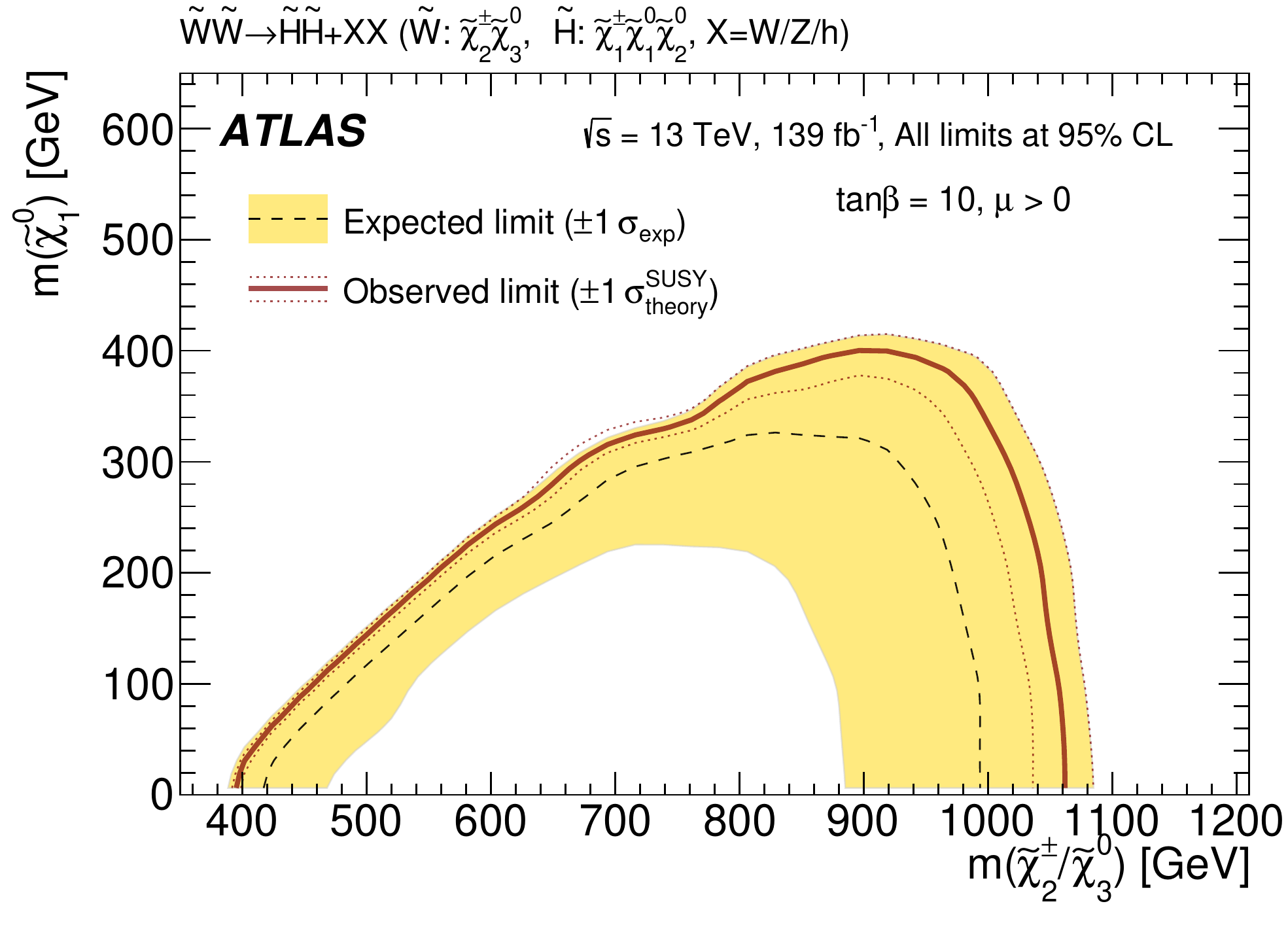} \label{fig:winoHinoExcl:winoHino}}
\subfloat[$(\hino,\wino)$ with $\tanb=10, \mu>0$]{\includegraphics[width=0.485\textwidth]{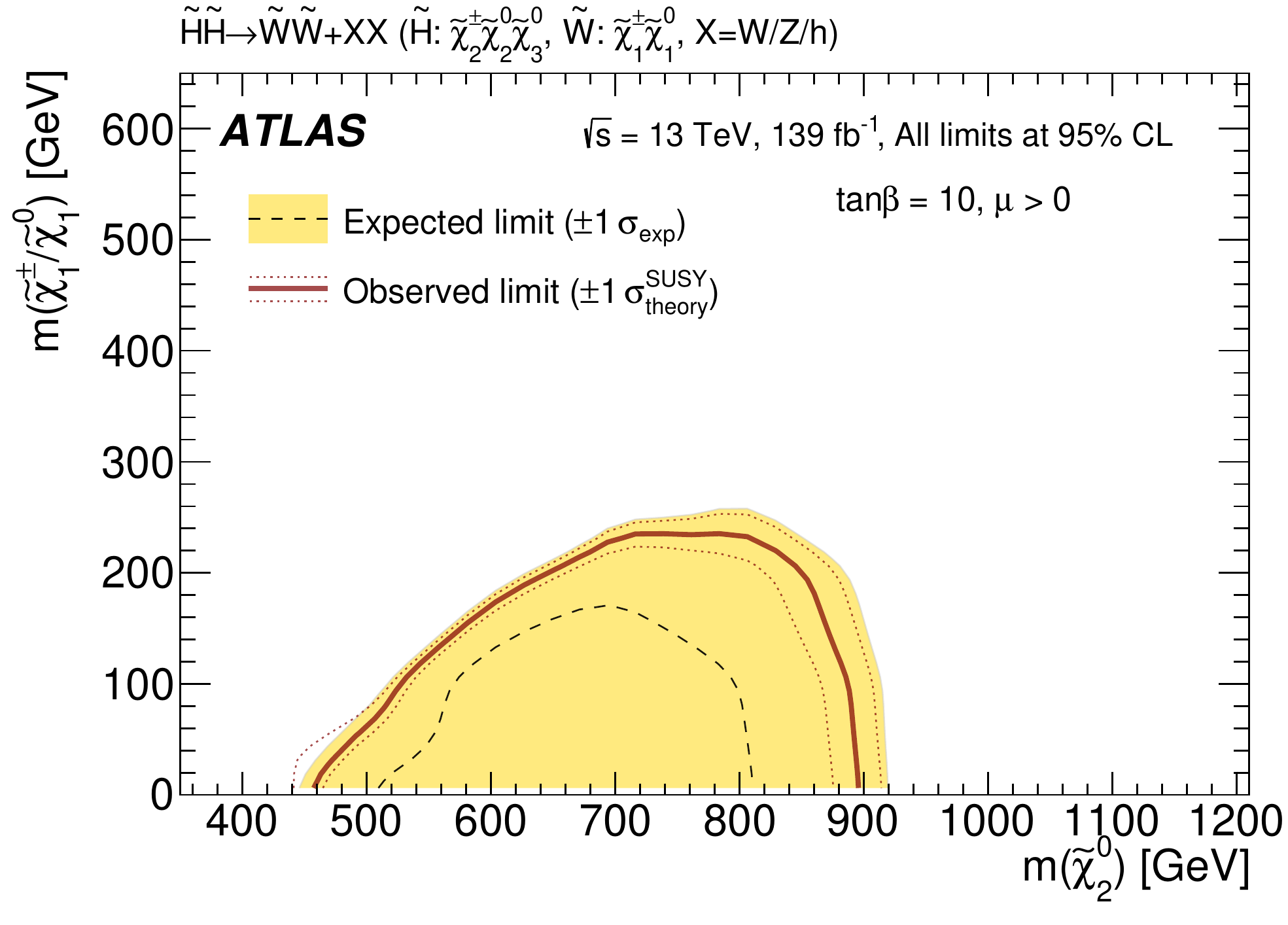} \label{fig:winoHinoExcl:hinoWino}}
\caption{95\% CL exclusion limits for the \winoHino and \hinoWino models.
The limits are projected onto a two-dimensional plane either as a function of the wino/higgsino mass parameters $(M_2,\mu)$ (top figures),
or of the physical electroweakino masses $(m(\Ctwo),m(\None))$ representing $(\mheavy,\mlight)$ (bottom figures).
For the limits shown on the $(M_2,\mu)$ plane,
the excluded regions are indicated by the area inside the contours.
The round excluded area in the top part corresponds to the excluded parameter space in the \winoHino model ($M_2 > |\mu|$),
while the two small areas at the bottom are that in the \hinoWino model ($M_2 < |\mu|$).
}
\end{figure}
 
The mass exclusion limits are shown to be highly stable with respect to the internal variations within each model.
The limits for \winoBino and \winoHino (or for \hinoBino and \hinoWino) are also very similar,
despite the branching ratios of \chiheavy being substantially different between the models.
This model dependency is small mostly due to the statistical combination of SR-4Q and SR-2B2Q,
and the inclusive SR bins (SR-4Q-VV, SR-2B2Q-VZ, and SR-2B2Q-Vh),
which are designed to be agnostic with regard to the difference between $W$ and $Z$ bosons.
 
To summarize, a wino mass between 400~\GeV\ and 1060~\GeV\ is excluded for the wino production models for $\mNone < 400~\GeV$;
and a higgsino mass between 450~\GeV\ and 900~\GeV\ is excluded for the higgsino production models for $\mNone < 240~\GeV$.

\clearpage
\subsubsection{Exclusion limits on \winoBino simplified model: \winoBinoSim}
The exclusion limits for the \winoBino simplified models (\SMWW, \SMWZ and \SMWh) are also derived,
in order to directly compare the search sensitivity with the previous ATLAS analyses.
Figure~\ref{fig:WinoBinoExclusion} shows the obtained exclusion as a function of the produced wino mass \mConeNtwo and the bino LSP mass \mNone.
For \SMWW, \mCone between 630~\GeV\ and 760~\GeV\ is excluded for $\mNone < 80~\GeV$.
For \SMWZ (\SMWh), \mConeNtwo\ between 440~\GeV\ and 960~\GeV\ (400~\GeV\ and 1060~\GeV) is excluded for $\mNone < 300~\GeV~(420~\GeV)$.\footnote{The obtained limits are generally weaker than those set on the \winoBino model shown in Section~\ref{sec:excl:winoBinoHino}. This is because only one production mode is considered in these simplified models, while multiple production modes are included in the signals in the \winoBino model.}
 
The sensitivity to high-mass winos is significantly improved relative to ATLAS searches using the other final states and the same dataset.
For example, the expected limits on \SMWZ (\SMWh) are typically extended by about 300~(140)~\GeV\ in \mConeNtwo,
corresponding to exclusion of signals with a 7.5 (2.4) times smaller production cross-section
than in the search using final states with three leptons~\cite{SUSY-2019-09} (one lepton and two $b$-jets~\cite{SUSY-2019-08}).
This result also sets the most stringent limit on the model to date from the LHC experiments.
 
\begin{figure}[h]
\centering
\subfloat[\SMWW]{\includegraphics[width=0.485\textwidth]{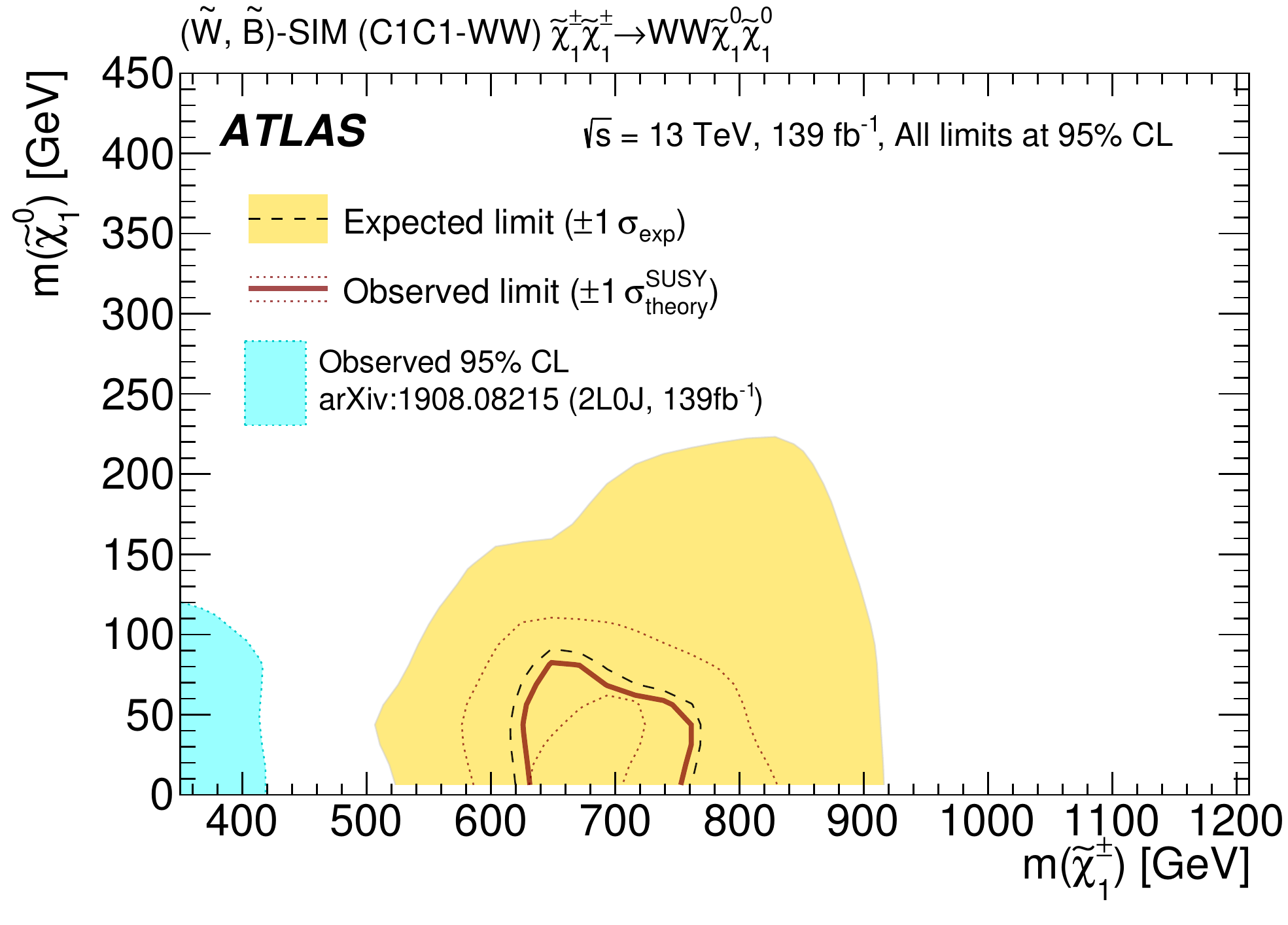}} \\
\subfloat[\SMWZ]{\includegraphics[width=0.485\textwidth]{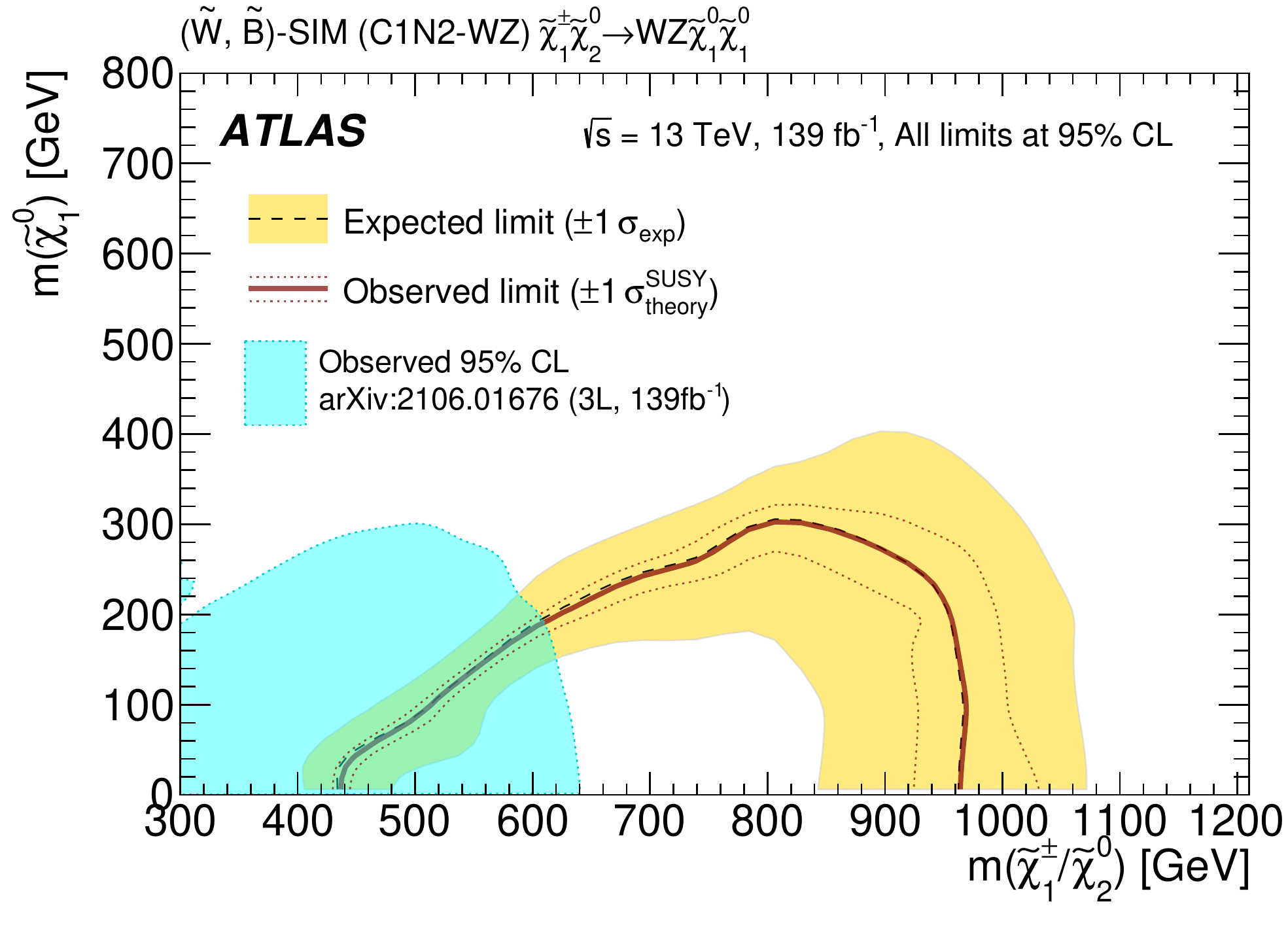}}
\subfloat[\SMWh]{\includegraphics[width=0.485\textwidth]{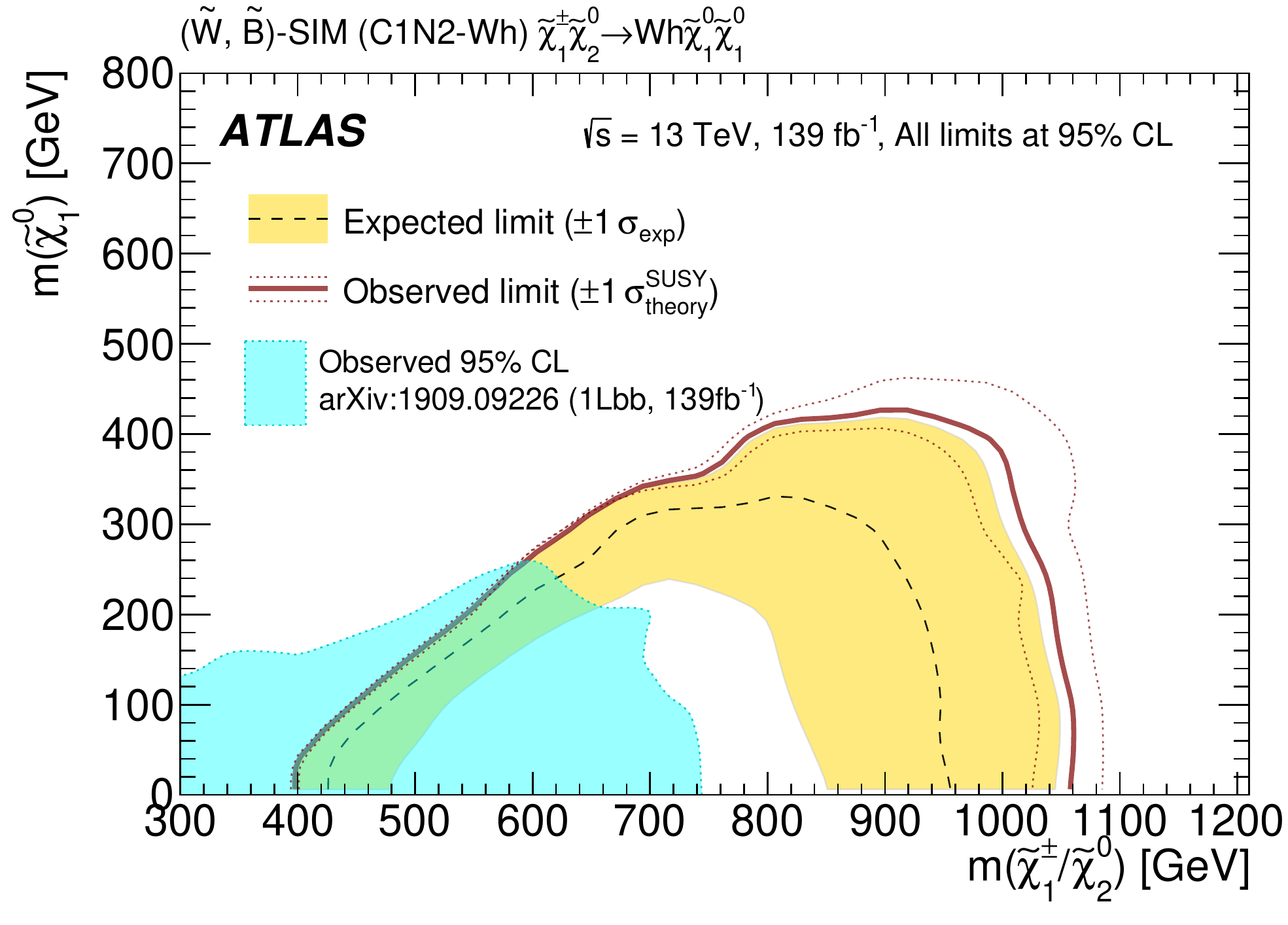}}
\caption{Exclusion limits for \winoBinoSim as a function of the produced wino mass \mConeNtwo and the bino LSP mass \mNone.
Expected (dashed) and observed (solid) 95\% CL exclusion limits on \winoBino simplified models are shown for (a) \SMWW, (b) \SMWZ, and (c) \SMWh.
The limits from the previous ATLAS searches on \SMWW~\cite{SUSY-2018-32},
\SMWZ~\cite{SUSY-2019-09}, and \SMWh~\cite{SUSY-2019-08} are shown by the shaded areas.
}
\label{fig:WinoBinoExclusion}
\end{figure}

\subsubsection{Exclusion limits on \hinoGravitino model}
The exclusion for the \hinoGravitino model is presented in Figure~\ref{fig:GGM_Exclusion},
as the function of the mass of the lightest higgsino, \mNone, and the branching ratio $\mathcal{B}(\None \ra Z\gravitino)$.
For $\mathcal{B}(\None \to Z\gravitino) = 100\%~(50\%)$, \mNone between 450~(500)~\GeV\ and 940~(850)~\GeV\ is excluded,
while the exclusion reaches $\mathcal{B}(\None \to Z\gravitino) \sim 20\%$ for $\mNone \sim 600~\GeV$.
This complements the sensitivity achieved by previous ATLAS searches using 4-lepton final states to target signals with decays into $ZZ$~\cite{SUSY-2018-02} and final states with three or more $b$-jets to target signals with decays into $hh$~\cite{SUSY-2017-02}.
 
\begin{figure}[h]
\centering
\includegraphics[width=0.65\textwidth]{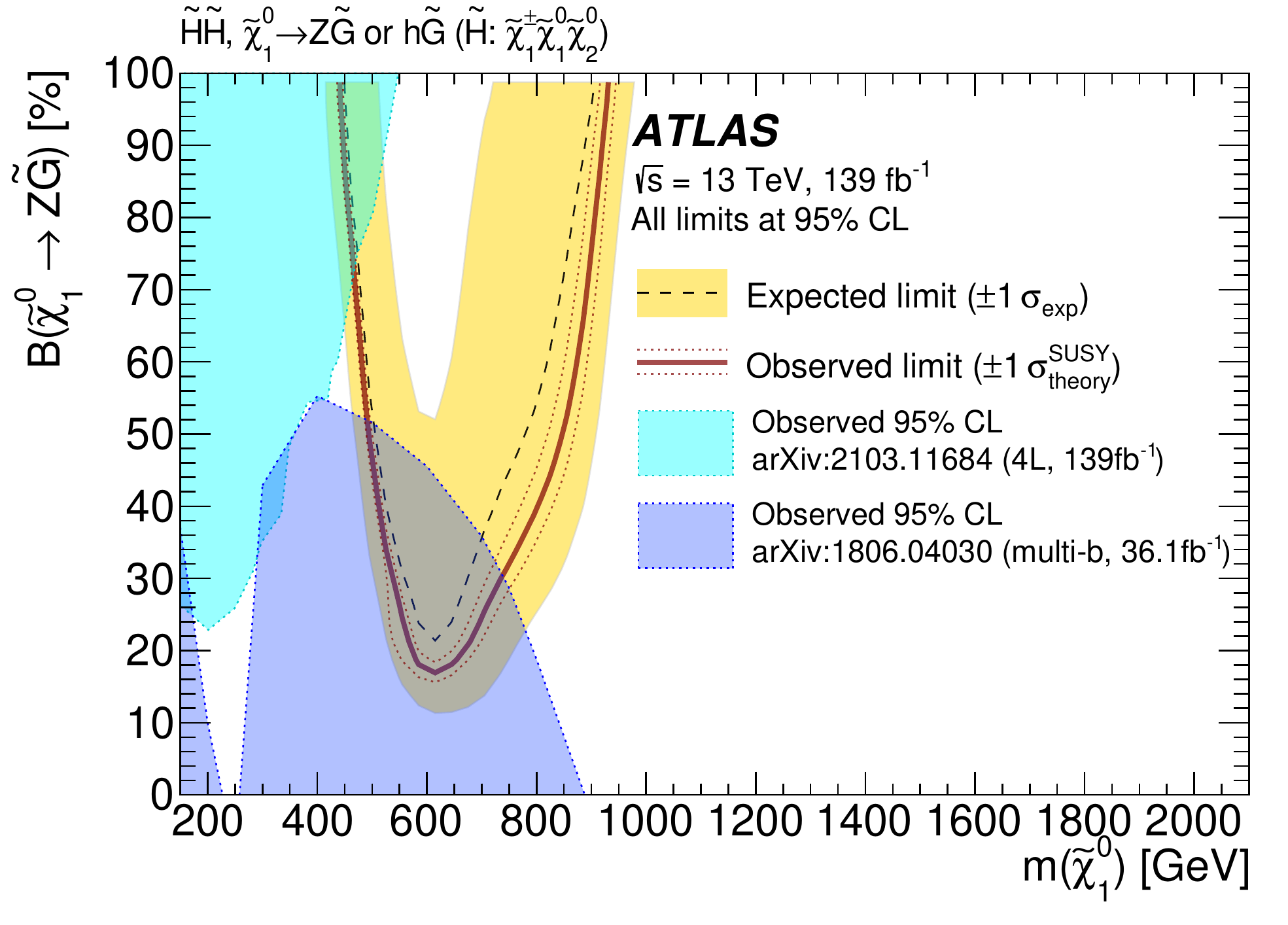}
\caption{Expected (dashed) and observed (solid red) 95\% CL exclusion limits derived for the \hinoGravitino model, as a function of the lightest higgsino mass \mNone and the branching ratio $\mathcal{B}(\None \to Z\gravitino) (=1-\mathcal{B}(\None \to h\gravitino))$.
The excluded region is indicated by the area inside the contour.
The exclusion limits from the previous ATLAS search using 4-lepton final states~\cite{SUSY-2018-02} (cyan, denoted by ``4L''), or final states with three or more $b$-jets~\cite{SUSY-2017-02} (violet, denoted by ``multi-$b$'') are shown by the shaded areas.
}
\label{fig:GGM_Exclusion}
\end{figure}
 
\subsubsection{Exclusion limits on \hinoAxino model}
Lastly, Figure~\ref{fig:Axino_Exclusion} shows the exclusion limits on the \hinoAxino model as the function of the mass of the axino, \mAxino, and the lightest higgsino, \mNone.
Similarly to the \hinoGravitino model, different branching ratio hypotheses of $\mathcal{B}(\None \to Z\axino) (=1-\mathcal{B}(\None \to h\axino))$ are tested.
For $\mathcal{B}(\None \to Z\axino)=100\%$, a higgsino mass between 450~\GeV\ and 940~\GeV\ is excluded when the axino mass is less than 300~\GeV.
The sensitivity decreases with decreasing $\mathcal{B}(\None \to Z\axino)$.
For $\mathcal{B}(\None \to Z\axino)=50\%$, the observed limit excludes higgsino masses of 500--850~\GeV\ when the axino mass is less than $210~\GeV$.
 
\begin{figure}[h]
\centering
\subfloat[$\mathcal{B}(\None \ra Z\axino)=100\%$]{\includegraphics[width=0.485\textwidth]{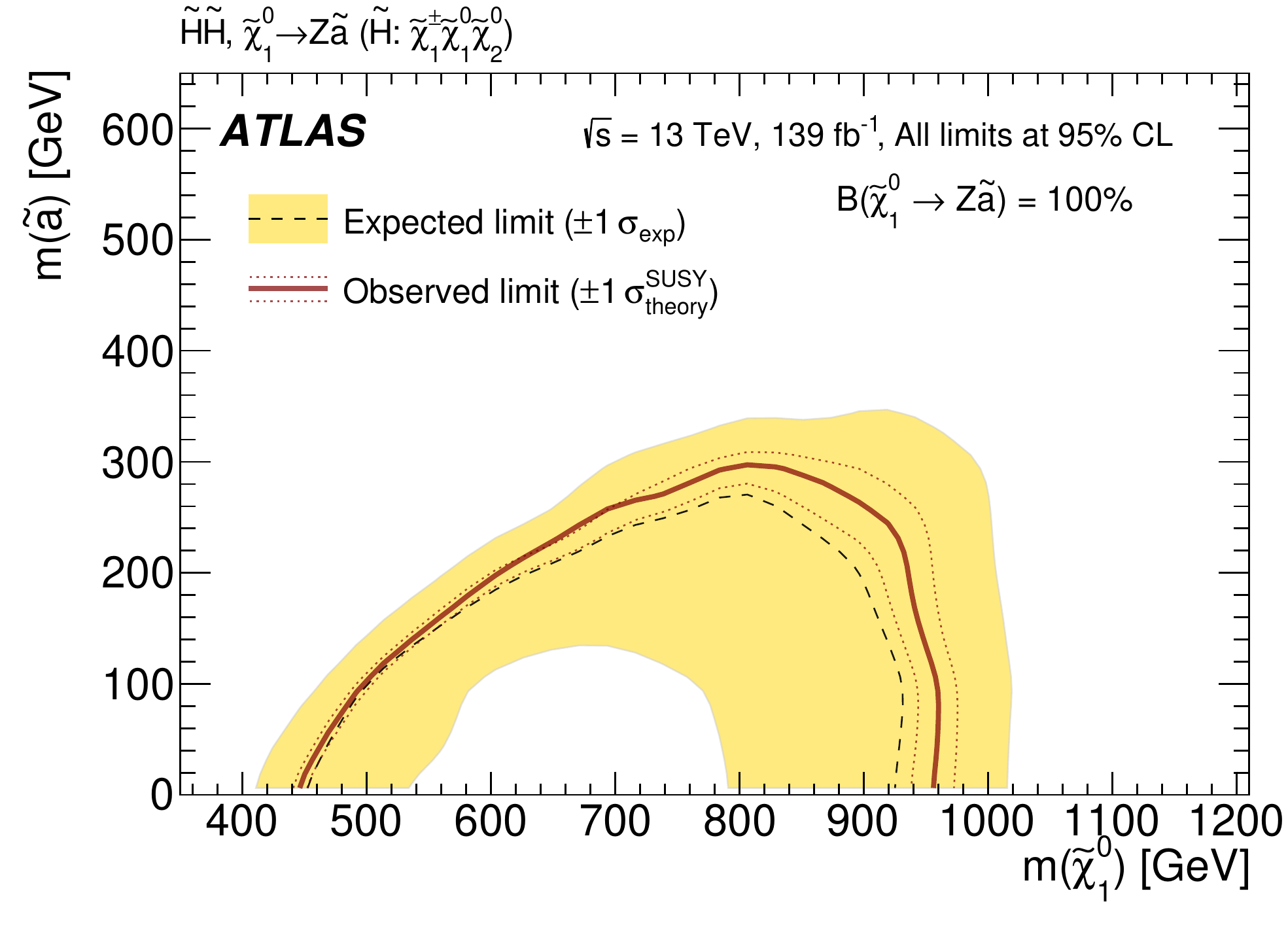}}
\subfloat[Variable $\mathcal{B}(\None \ra Z\axino)$]{\includegraphics[width=0.485\textwidth]{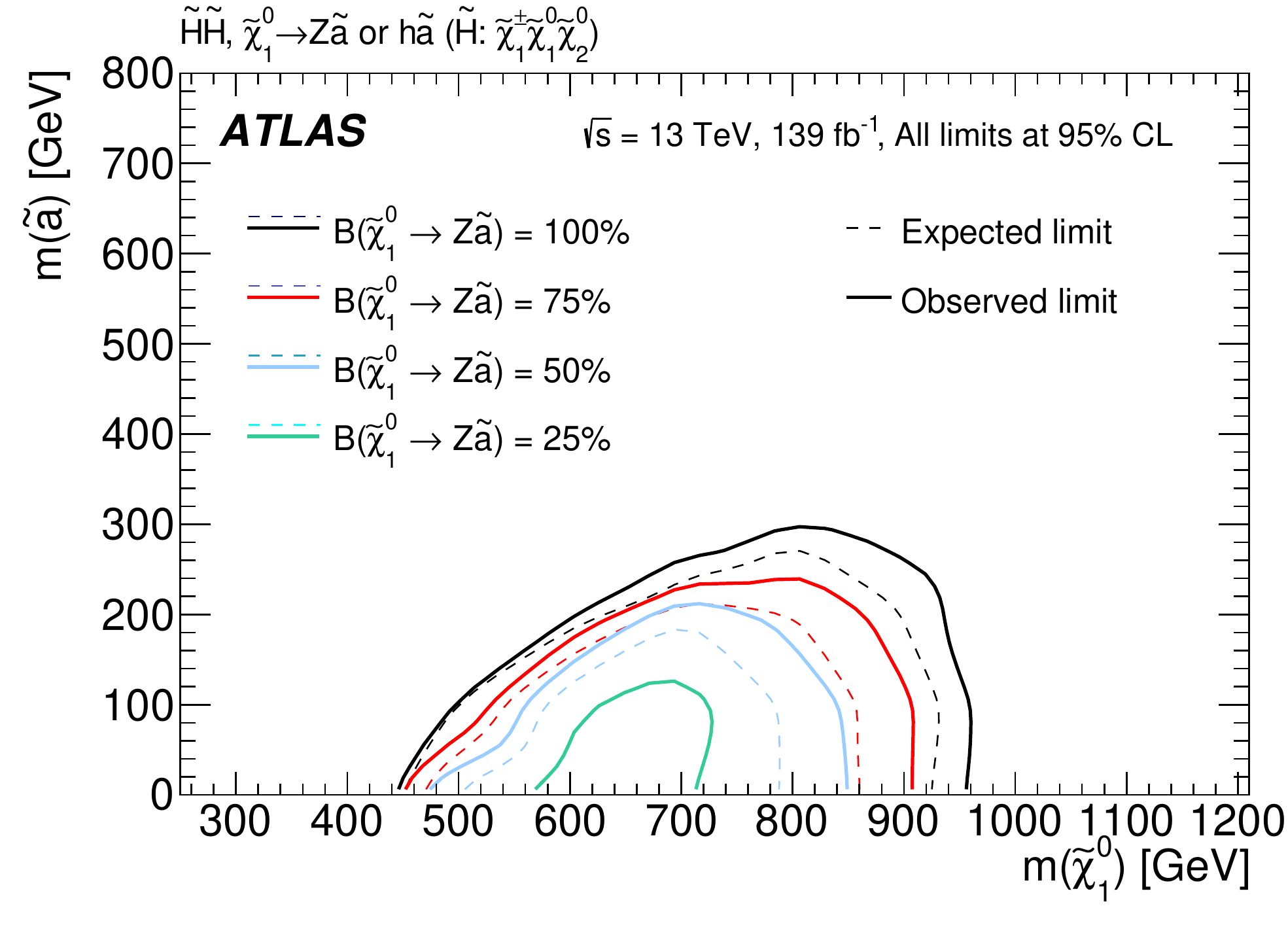}}
\caption{95\% CL exclusion limits for the \hinoAxino model as a function of axino mass \mAxino and lightest higgsino mass \mNone.
(a) Expected (dashed line) and observed (solid red line) limits calculated for $\mathcal{B}(\None \to Z\axino)=100\%$.
(b) Expected (dashed lines) and observed (solid lines) limits with various $\mathcal{B}(\None \to Z\axino) (=1-\mathcal{B}(\None \to h\axino))$ hypotheses.
No expected limit is derived for the case with $\mathcal{B}(\None \to Z\axino)=25\%$ as no mass point on the plane can be excluded.}
\label{fig:Axino_Exclusion}
\end{figure}
 
\clearpage
\section{Conclusion}
A search for electroweakino pair production using final states consisting of \met and two boosted hadronically decaying heavy SM bosons ($W$, $Z$, or $h$) is reported.
Signatures with large mass-splitting between the produced electroweakino and the lightest SUSY particle (LSP) are targeted.
 
The use of fully hadronic final states takes advantage of the large SM boson branching ratios, while the backgrounds are
efficiently suppressed by reconstructing the $W$, $Z$ and $h$ bosons using boosted-boson-tagging techniques.
Using 139~\ifb\ of proton--proton collision data at $\sqrt{s}=13~\TeV$ recorded by the ATLAS detector at the LHC,
this strategy provides unprecedented sensitivity to the production of heavy electroweakinos.
 
No excess over the SM background prediction is observed,
and 95\% CL exclusion limits are set for signal models in various $R$-parity-conserving scenarios.
For wino pair production with direct decays into a bino or higgsino LSP,
wino masses between $400~\GeV$ and $1060~\GeV$ are excluded when the LSP mass is below $400~\GeV$ and the mass splitting is larger than $400~\GeV$.
For higgsino pair production with direct decays into a bino or wino LSP,
higgsino masses between $450~\GeV$ and $900~\GeV$ are excluded when the LSP mass is below $240~\GeV$ and the mass splitting is larger than $450~\GeV$.
The limits are also examined for various wino (higgsino) branching ratio assumptions,
by directly scanning over relevant branching ratios (for the bino LSP models)
or over the MSSM parameters that dictate them ($M_2$, $\mu$, and $\tanb$ for the wino/higgsino LSP models).
The results are shown to be highly consistent for the variations.
 
The results are also interpreted in the context of simplified models of wino production with decays into a bino LSP,
which are more conventionally explored in electroweakino searches at the LHC.
For chargino pair production, with each decaying into a $W$ boson and an LSP,
a chargino mass between $630~\GeV$ and $760~\GeV$ is excluded for an LSP mass below $80~\GeV$.
For chargino--neutralino pair production involving decays into $WZ$ ($Wh$) and two LSPs,
a wino mass between $440~(400)~\GeV$ and $960~(1060)~\GeV$ is excluded for an LSP mass below $300~(420)~\GeV$.
These extend significantly beyond the exclusion limits set by the previous searches at the LHC using different final states.
 
Finally, exclusion limits are set on higgsino production with decays into a massless gravitino LSP or a massless/massive axino LSP,
motivated by the general gauge mediation model (GGM) or new physics models involving the axion, respectively.
A higgsino mass between 450~(500)~\GeV\ and 940~(850)~\GeV\ is excluded for the gravitino LSP model
with $\mathcal{B}(\None \to Z\gravitino) = 100\%~(50\%)$,
and a mass between 450~(500)~\GeV\ and 940~(850)~\GeV\ is excluded for the axino LSP model
with $\mathcal{B}(\None \to Z\gravitino) = 100\%~(50\%)$ when the axino LSP mass is below 300~(210)~\GeV.

 
\section*{Acknowledgments}

We thank CERN for the very successful operation of the LHC, as well as the
support staff from our institutions without whom ATLAS could not be
operated efficiently.
 
We acknowledge the support of
ANPCyT, Argentina;
YerPhI, Armenia;
ARC, Australia;
BMWFW and FWF, Austria;
ANAS, Azerbaijan;
SSTC, Belarus;
CNPq and FAPESP, Brazil;
NSERC, NRC and CFI, Canada;
CERN;
ANID, Chile;
CAS, MOST and NSFC, China;
Minciencias, Colombia;
MSMT CR, MPO CR and VSC CR, Czech Republic;
DNRF and DNSRC, Denmark;
IN2P3-CNRS and CEA-DRF/IRFU, France;
SRNSFG, Georgia;
BMBF, HGF and MPG, Germany;
GSRI, Greece;
RGC and Hong Kong SAR, China;
ISF and Benoziyo Center, Israel;
INFN, Italy;
MEXT and JSPS, Japan;
CNRST, Morocco;
NWO, Netherlands;
RCN, Norway;
MNiSW and NCN, Poland;
FCT, Portugal;
MNE/IFA, Romania;
JINR;
MES of Russia and NRC KI, Russian Federation;
MESTD, Serbia;
MSSR, Slovakia;
ARRS and MIZ\v{S}, Slovenia;
DSI/NRF, South Africa;
MICINN, Spain;
SRC and Wallenberg Foundation, Sweden;
SERI, SNSF and Cantons of Bern and Geneva, Switzerland;
MOST, Taiwan;
TAEK, Turkey;
STFC, United Kingdom;
DOE and NSF, United States of America.
In addition, individual groups and members have received support from
BCKDF, CANARIE, Compute Canada and CRC, Canada;
COST, ERC, ERDF, Horizon 2020 and Marie Sk{\l}odowska-Curie Actions, European Union;
Investissements d'Avenir Labex, Investissements d'Avenir Idex and ANR, France;
DFG and AvH Foundation, Germany;
Herakleitos, Thales and Aristeia programmes co-financed by EU-ESF and the Greek NSRF, Greece;
BSF-NSF and GIF, Israel;
Norwegian Financial Mechanism 2014-2021, Norway;
La Caixa Banking Foundation, CERCA Programme Generalitat de Catalunya and PROMETEO and GenT Programmes Generalitat Valenciana, Spain;
G\"{o}ran Gustafssons Stiftelse, Sweden;
The Royal Society and Leverhulme Trust, United Kingdom.
 
The crucial computing support from all WLCG partners is acknowledged gratefully, in particular from CERN, the ATLAS Tier-1 facilities at TRIUMF (Canada), NDGF (Denmark, Norway, Sweden), CC-IN2P3 (France), KIT/GridKA (Germany), INFN-CNAF (Italy), NL-T1 (Netherlands), PIC (Spain), ASGC (Taiwan), RAL (UK) and BNL (USA), the Tier-2 facilities worldwide and large non-WLCG resource providers. Major contributors of computing resources are listed in Ref.~\cite{ATL-SOFT-PUB-2021-003}.
 

\clearpage
\printbibliography
 
\clearpage 
 
\begin{flushleft}
\hypersetup{urlcolor=black}
{\Large The ATLAS Collaboration}

\bigskip

\AtlasOrcid[0000-0002-6665-4934]{G.~Aad}$^\textrm{\scriptsize 99}$,    
\AtlasOrcid[0000-0002-5888-2734]{B.~Abbott}$^\textrm{\scriptsize 125}$,    
\AtlasOrcid[0000-0002-7248-3203]{D.C.~Abbott}$^\textrm{\scriptsize 100}$,    
\AtlasOrcid[0000-0002-2788-3822]{A.~Abed~Abud}$^\textrm{\scriptsize 34}$,    
\AtlasOrcid[0000-0002-1002-1652]{K.~Abeling}$^\textrm{\scriptsize 51}$,    
\AtlasOrcid[0000-0002-2987-4006]{D.K.~Abhayasinghe}$^\textrm{\scriptsize 91}$,    
\AtlasOrcid[0000-0002-8496-9294]{S.H.~Abidi}$^\textrm{\scriptsize 27}$,    
\AtlasOrcid[0000-0001-5329-6640]{H.~Abramowicz}$^\textrm{\scriptsize 158}$,    
\AtlasOrcid[0000-0002-1599-2896]{H.~Abreu}$^\textrm{\scriptsize 157}$,    
\AtlasOrcid[0000-0003-0403-3697]{Y.~Abulaiti}$^\textrm{\scriptsize 5}$
% ...
\bigskip
\\

% ...
$^{5}$High Energy Physics Division, Argonne National Laboratory, Argonne IL; United States of America.\\
% ...
$^{27}$Physics Department, Brookhaven National Laboratory, Upton NY; United States of America.\\
% ...
$^{34}$CERN, Geneva; Switzerland.\\
% ...
$^{51}$II. Physikalisches Institut, Georg-August-Universit\"{a}t G\"ottingen, G\"ottingen; Germany.\\
% ...
$^{91}$Department of Physics, Royal Holloway University of London, Egham; United Kingdom.\\
% ...
$^{99}$CPPM, Aix-Marseille Universit\'e, CNRS/IN2P3, Marseille; France.\\
$^{100}$Department of Physics, University of Massachusetts, Amherst MA; United States of America.\\
% ...
$^{125}$Homer L. Dodge Department of Physics and Astronomy, University of Oklahoma, Norman OK; United States of America.\\
% ...
$^{157}$Department of Physics, Technion, Israel Institute of Technology, Haifa; Israel.\\
$^{158}$Raymond and Beverly Sackler School of Physics and Astronomy, Tel Aviv University, Tel Aviv; Israel.\\
% ...

\end{flushleft}


\end{document}